\documentclass[12pt]{extarticle}

\usepackage[english]{babel}

\usepackage{graphicx}     
\graphicspath{ {./illustrations/} }
\usepackage[font=small,labelfont=bf]{caption}

\usepackage{hyperref}
\usepackage[super,sort,compress]{cite}

\usepackage{setspace}
\usepackage{amsmath, amssymb}

\usepackage[margin=1.0in]{geometry}

\usepackage{newpxtext}

\usepackage[disable]{todonotes}

\ExplSyntaxOn

\NewDocumentCommand { \ie } { }
  {
    \textit{i.e.}
    \peek_meaning_ignore_spaces:NTF .
      { \skip_horizontal:n { -.3ex } \use_none:n }
      {
        \peek_meaning_ignore_spaces:NF ,
          { \skip_horizontal:n { -.3ex } }
      }
  }
\NewDocumentCommand { \eg } { }
  {
    \textit{e.g.}
    \peek_meaning_ignore_spaces:NTF .
      { \skip_horizontal:n { -.3ex } \use_none:n }
      {
        \peek_meaning_ignore_spaces:NF ,
          { \skip_horizontal:n { -.3ex } }
      }
  }

\ExplSyntaxOff               

\title{Scalable Non-Parametric Bayesian Inference of Free Energy Landscapes and Emission Functions from Observations}

\date{}

\author{Eliane Briand, Helmut Grubm\"uller}

\begin{document}
\maketitle 

\begin{abstract}
Free-energy landscapes govern processes from protein conformational change to climate oscillations, but are typically observed only indirectly, through observables such as FRET intensity or optical-tweezer transmittance.
Neither the landscape shape, nor the emission function mapping latent state to signal, is usually known, and distinct states may produce the same signal.
 We therefore infer landscape and emission function jointly and non-parametrically, using the exact non-linear Bayesian filtering likelihood, to fully use the information contained in observable trajectories that are often noisy, scarce and costly to acquire.
A novel parallel-in-time algorithm makes this tractable: it is up to 27 times faster than sequential filtering on one GPU, scales near-ideally across GPUs, and, given enough resources, has runtime independent of trajectory length.
We demonstrate the method on a single-protein optical tweezer recording, where distinct conformations produce the same signal, and beyond biology on the El Niño–Southern Oscillation and financial market volatility.
\end{abstract}

Many processes across the natural sciences admit as a common description 
an overdamped Brownian dynamics on a free energy landscape.
Examples include protein conformational changes,\cite{frauenfelder1991energy} ion channel gating,\cite{sakmann1995single} and the El Niño–Southern Oscillation (ENSO).\cite{hasselman1976stochastic, penland1996stochastic}
The shape of the landscape encodes the physics of each system, with barrier heights defining transition rates and well depths defining equilibrium populations.

Yet this landscape lives in a latent space, and what is measured is not the latent state itself but a function of it, such as fluorescence intensity, \cite{ha1996probing} ionic conductance,\cite{sakmann1995single} or sea-level pressure differences.\cite{Walker1932}
This mapping, the emission function, is generally non-linear and non-invertible. 
The task is thus to recover the free energy landscape and the emission function jointly, from observed trajectories alone and without assuming parametric forms for either. We will address this problem by Bayesian inference.

Making this inference tractable has so far required additional assumptions on the latent space or the emission function.
The most common is the discrete-state hidden Markov model (HMM).\cite{rabiner1989tutorial, franzke2008hidden}
For continuous systems, like proteins probed by single-molecule FRET,\cite{mckinney2006analysis, noe2008transition, prinz2011markov, chodera2014markov} discretization coarsens the shape of the landscape.
Few states collapse complex transition paths into single rate constants,\cite{sarich2010approximation, schweiger2025cautious} whereas many states turn inference into an ill-posed inverse problem.
Non-parametric HMMs\cite{sgouralis2017introduction} regularize the number of states, but not the smoothness of the landscape, which is the natural prior but difficult to apply on a discrete graph of states.

Continuous diffusion models, in contrast, preserve the landscape but assume the latent coordinate is observed directly, as in single-molecule pulling,\cite{hummer2001free} position-dependent diffusion models,\cite{best2010coordinate} and econometric diffusion estimators.\cite{ait2002maximum}
Bayesian filtering\cite{stratonovich1960application} lifts this assumption but is analytically solvable only for the quadratic landscape of the Kalman--Bucy filter. \cite{kalman1961new} Extended and unscented Kalman filters\cite{julier2004unscented, sarkka2013bayesian} handle non-linear dynamics  but force the latent density to be Gaussian.
Non-parametric landscapes have been inferred jointly with the latent state for single-molecule FRET.\cite{bryan2023learning, saurabh2026resolving, dingeldein2026differentiable}
The non-linear dynamics and the high-dimensional representation of the latent density make such inference costly.
Particle MCMC estimates the exact likelihood by Monte Carlo,\cite{gordon1993novel, andrieu2010particle} whereas simulation-based inference replaces it with a neural surrogate.\cite{cranmer2020frontier, dingeldein2025simulation}

All of these approaches assume a fully specified or parametric emission function.
Jointly inferring the emission allows analysis of systems without a closed-form emission, such as the mapping from a latent climate regime to atmospheric pressures.
It also avoids imposing physical laws that are only approximate in practice, such as the $1/r^6$ law of FRET efficiency,\cite{forster1948zwischenmolekulare} given linker flexibility and fluorophore orientation.
Both gains come at a further computational cost.

With non-linear dynamics and joint landscape and emission inference, the filtering likelihood becomes computationally costly to evaluate.
Worse, its recursion is sequential: each step conditions on the previous one, so runtime grows with trajectory length whatever the parallel resources available.
Parallelizing this recursion requires restructuring it.
Parallel prefix scans do so by recasting it as an associative operation,\cite{sarkka2020temporal, hassan2021temporal} but require prediction and update to compose in closed form, which holds exactly for linear-Gaussian models or transition densities expressible as dense matrices, and only approximately otherwise, through iterated Gaussian linearization.\cite{yaghoobi2025parallel}
A parallel-in-time (PinT) particle smoother exists,\cite{corenflos2022sequentialized} though it draws the particles of each time step independently, resulting in error growing faster with trajectory length than that of sequential smoothers.
Parallel fixed-point solvers sidestep associativity altogether, evaluating general non-linear recursions, MCMC chains included, by Newton or Picard iteration over the whole trajectory.\cite{lim2024parallelizing, gonzalez2024towards, shih2023parallel, zoltowski2026parallelizing, iacob2025parallel}
Under filter stability,\cite{legland2004stability} filters started from different initial densities converge as observations accumulate, leaving distant time points only weakly coupled, the regime in which PinT methods excel.\cite{gander2015fifty, lelarasmee1982waveform}
Yet PinT iteration has not been combined with filter stability to parallelize Bayesian filtering.

Here we infer the landscape and the emission function jointly and non-parametrically, under a prior that imposes a controlled level of smoothness.
The model likelihood is computed by Bayesian filtering, which propagates the non-parametric latent density through the Fokker--Planck equation of overdamped Brownian dynamics.
We remove the sequential bottleneck of this filtering with a parallel-in-time algorithm that splits the trajectory into short blocks, filters them in parallel, and iterates until it reproduces exactly the output of the sequential filter.
For the model studied here, filter stability\cite{legland2004stability} on its compact latent domain makes the iteration count independent of the number of observations $N$.
The span of a likelihood evaluation, its runtime given unlimited parallel resources, therefore falls from $\mathcal{O}(N)$ to $\mathcal{O}(1)$, at the cost of a small multiple of total work.
Because the algorithm treats the filter as a black box, it requires neither the associative reformulation of scan methods nor the Jacobians of fixed-point solvers, and applies to other hidden Markov models, though not necessarily with the same span.

\section{Results}

\subsection{Non-parametric hidden Brownian dynamics model}

\includegraphics[width=0.80\textwidth]{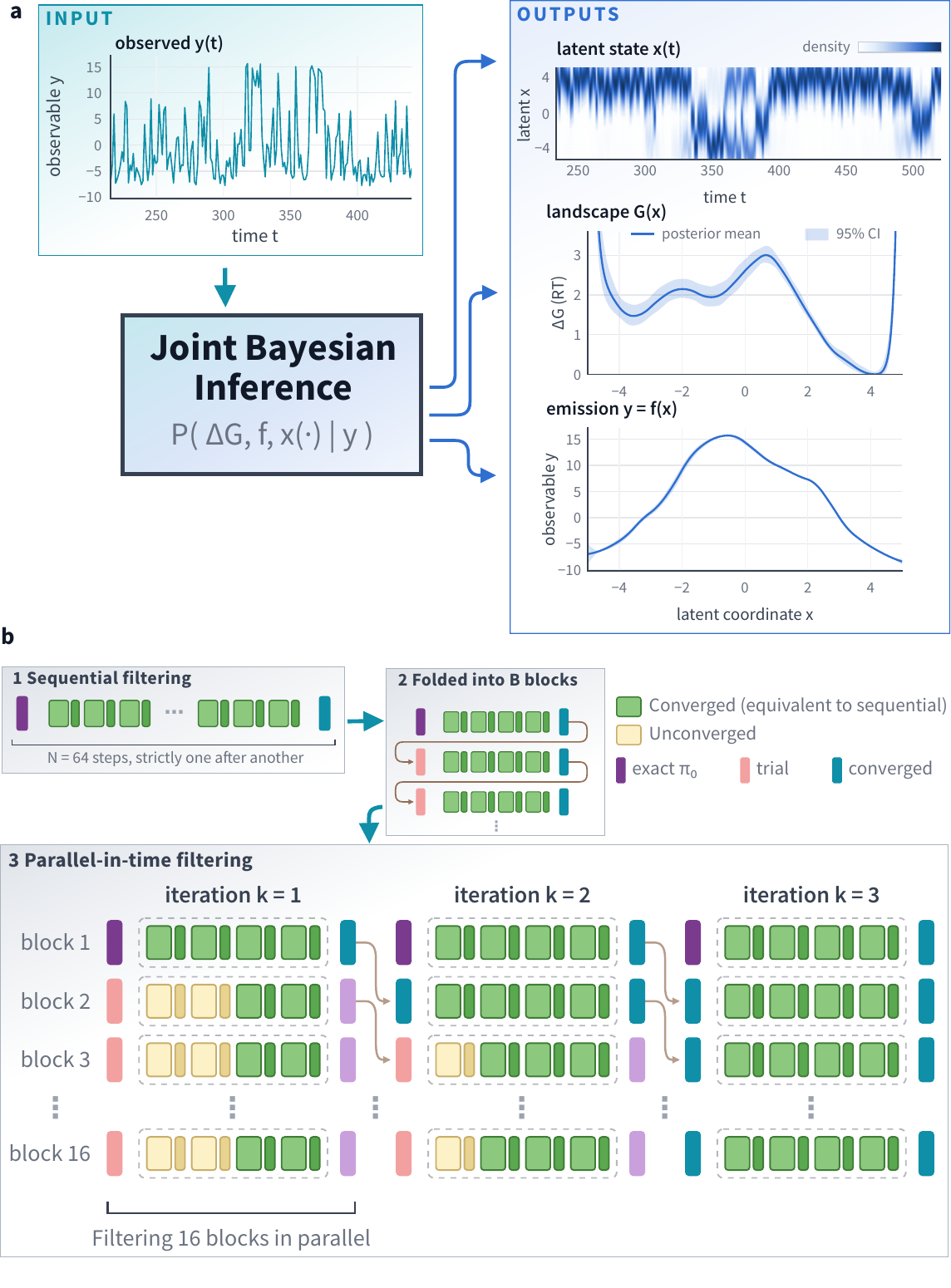}
\captionof{figure}{%
    \textbf{Joint inference of landscape, emission and latent state, and parallel-in-time evaluation of the likelihood.}
    (\textbf{a})~Inference goal: given observable trajectories y(t) (Input panel, teal), we jointly infer the free-energy landscape $G(x)$, the emission function $f$ mapping latent space coordinate $x$ to observable $y = f(x)$ , and the latent state trajectory $x(t)$ (Outputs panel), all three in the form of a Bayesian posterior distribution. (Synthetic data shown). 
    (\textbf{b})~Parallel-in-time (PinT) evaluation of the filtering likelihood. Sequential filtering (1) processes the $N$ steps in order. Folding the trajectory into $B$ blocks (2) exposes parallelism, but each block needs the final density of its predecessor. PinT filtering (3) starts the first block from the exact initial density $\pi_0$ and every other block from a trial density, filters all blocks in parallel, and passes boundary densities on at each iteration until they converge.
    \label{fig:overview_method}
}

We model the latent process as overdamped Brownian dynamics on a free-energy landscape $G(x)$, with $x$ the latent space coordinate, observed through a non-linear emission function $f(x)$ that need not be injective (Figure~\ref{fig:overview_method}a).
Here we focus on a one-dimensional latent coordinate, obeying $dX_t = -\tfrac{D}{k_B T}\nabla G(X_t)\,dt + \sqrt{2D}\,dW_t$.
Observations are drawn conditionally independently given the latent trajectory, from a user-specified distribution whose location parameter is controlled by the emission, $Y_{t_k} \mid X_{t_k} \sim p_{\mathrm{obs}}(\cdot \mid f(X_{t_k}))$.
The dynamics are thus assumed reversible and unforced, with a time-independent landscape and a position- and time-independent diffusion coefficient $D$.
Within these assumptions, $G$ and $f$ are inferred non-parametrically, with a prior on their wavelet coefficients expansion targeting a Besov smoothness class (Methods).\cite{lassas2009discretization}

The likelihood is evaluated by Bayesian filtering,\cite{kushner1964differential, stratonovich1960application, sarkka2013bayesian} which alternates prediction, propagating the latent density by the Fokker--Planck equation, and update, conditioning it on the new observation.
The generator of the Fokker--Planck equation is projected onto a basis and diagonalized, yielding the propagator.
The model likelihood is the product of the normalization factor of the update steps  $\prod_k \int p_{\mathrm{obs}}(y_{t_k} \mid f(x))\, p(x \mid y_{t_1:t_{k-1}})\, \mathrm{d}x$.

Non-injective emissions give rise to local maxima in the posterior, as emissions of different preimage multiplicity are separated by regions of low likelihood.
Our automated inference protocol therefore samples a variety of candidate emission functions of different preimage multiplicities, takes each to their local maximum of the posterior, and selects the top candidate for posterior sampling (Methods).
The method is implemented in Planx, an open-source, GPU-accelerated Python package built on JAX,\cite{jax2018github} whose filtering model can be embedded within larger Bayesian models.

To parallelize filtering, the sequential trajectory (Figure~\ref{fig:overview_method}b, panel 1) is split into $B$ blocks of $L$ time steps that are filtered concurrently (panels 2--3).
Each block needs as its initial density the final density of its predecessor, which is not available \textit{a priori}.
The first block therefore starts from the exact initial density $\pi_0$, and all others from a trial density, here $p_{\mathrm{eq}}$ for both.
Each iteration filters all blocks in parallel, then hands the final density of each block to its successor, until these block-boundary densities stop changing (Methods).
For our model, the number of iterations is set by how quickly the filter forgets an incorrect initial density, not by the trajectory length.
The wall-clock time of a likelihood evaluation therefore does not grow with the number of observations, at the price of filtering each block a few times over.

\subsection{Method validation}

\includegraphics[width=\textwidth]{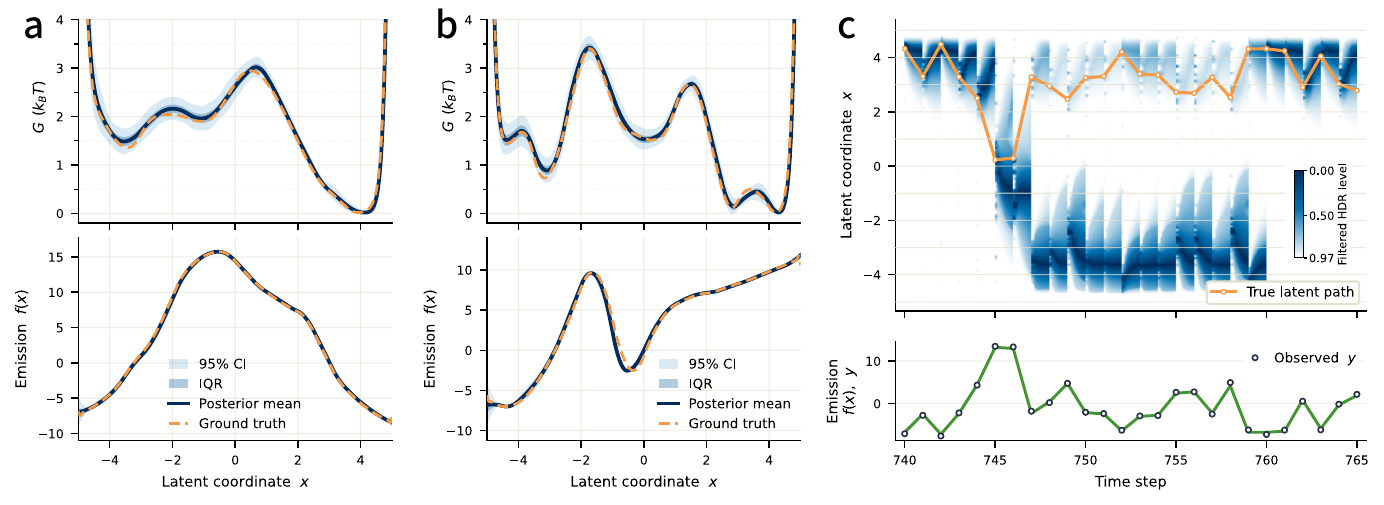}
\captionof{figure}{%
    \textbf{Inference on synthetic data and latent ambiguity resolution.}
    (\textbf{a}), (\textbf{b})~Inferred posterior for the landscape (top) and emission function (bottom) from synthetic trajectories, shown against the ground truth (orange dashed line). The posterior mean (solid blue line) is shown alongside the interquartile range (darker blue band) and the 95\%  credible interval (lighter blue band) quantifying posterior uncertainty.
    (\textbf{c})~Filtered latent density for a trajectory of system~(a), as a function of time (x-axis) and latent coordinate (y-axis).
    The latent density is shown as highest-density-region (HDR) level in shades of blue (darker is more probable), with the true latent trajectory superimposed (orange line).
    The observed value $y$ at each step is shown in the plot below (green line).
    \label{fig:validation_figure}
}

We validated the method on synthetic trajectories generated from known landscapes and emission functions (Figure~\ref{fig:validation_figure}a and b).
As a particular challenge, both emissions are non-injective, a single observation being consistent with up to two latent positions in (a) and three in (b), and both landscapes contain fine structure within their wells.
The data were limited to 32 trajectories of two to ten transitions each, with observation noise narrow ($\sigma = 0.25$) compared to the emission range ($-5$ to $15$), so that any ambiguity in the latent state arises from non-injectivity rather than noise.
Inference followed the protocol of the application cases, without knowledge of the ground truth except for the diffusion coefficient $D = 1$, which only sets the gauge of the latent coordinate (Methods).
The posterior recovers the landscape and the emission quite accurately, including the emission degeneracies and the fine structure of the landscape, with the ground truth well within its credible intervals.

Recovering a landscape despite non-injective emissions requires tracking the latent state through the resulting ambiguities in localization (Figure~\ref{fig:validation_figure}c, one trajectory of system~(a)).
The emission of~(a) is an inverted ``V'', its two branches spanning similar ranges of the observable, so that a single observation cannot resolve on which side of the peak the system is.
When the latent trajectory reaches the peak, around time step 745, the observations become compatible with relaxation into either well, and the filtered density accordingly splits into two modes, at $-4$ and $4$, until later observations fall outside the range of one branch and resolve the ambiguity.
This example shows the advantage of a non-parametric latent density over established Gaussian-latent filters, such as the extended or unscented Kalman filter,\cite{julier2004unscented} which must represent the latent state by a single mode.
Every subsequent observation would then be evaluated against an incorrect latent, biasing the likelihood and thus the inferred landscape and emission.

\subsection{Scaling and benchmark}

Accurate inference of this kind requires many thousands of likelihood-and-gradient evaluations, so its practicality rests on the speed of each evaluation.
We verify empirically the two properties of our PinT filter proven in Supplementary Information~S1, exactness at convergence and, for our model, $\mathcal{O}(1)$ span, then measure its throughput against the sequential filter and on multiple GPUs.

\includegraphics[width=\textwidth]{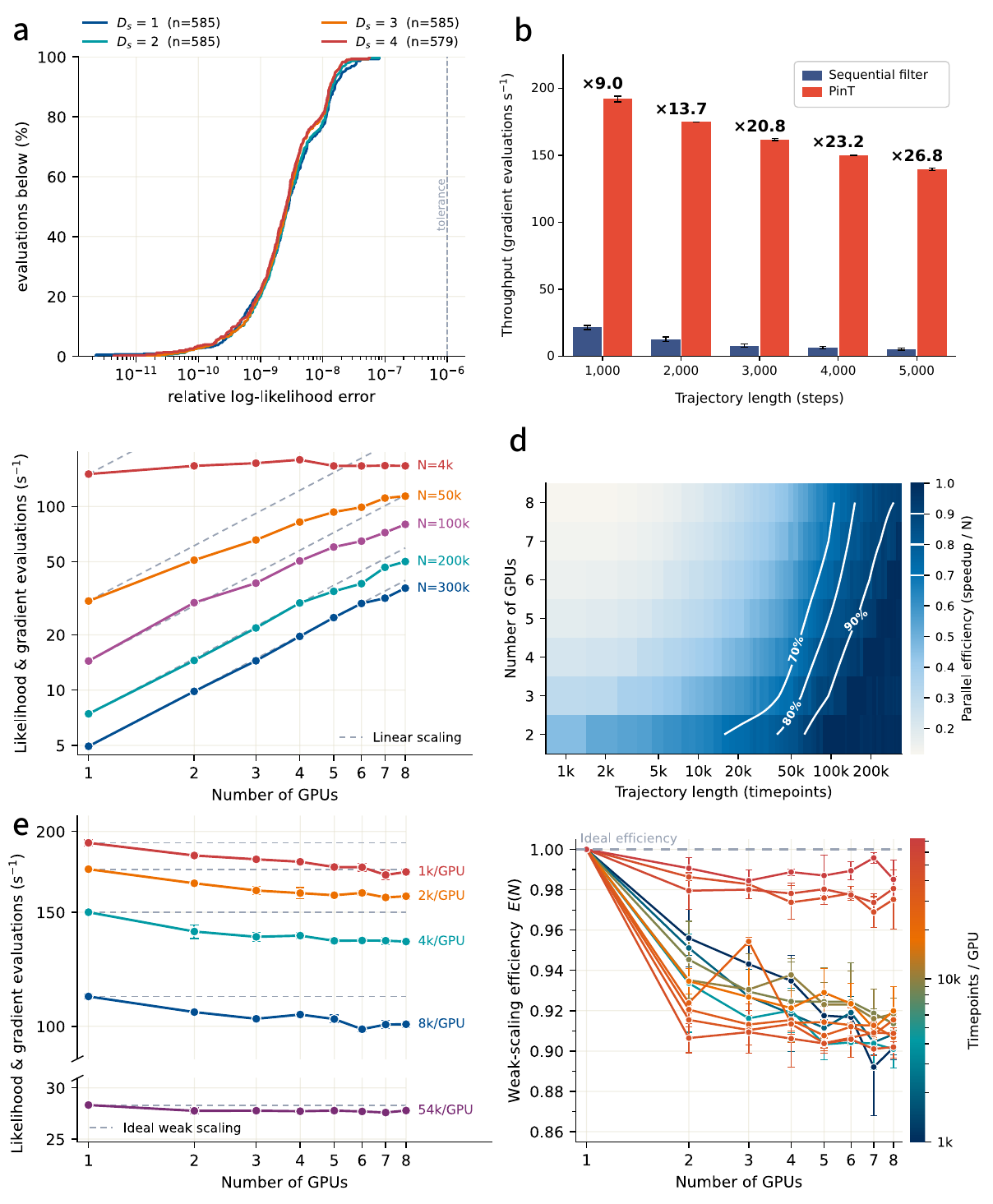}
\captionof{figure}{%
    \textbf{Correctness, speed-up and scaling of PinT filtering.}
    (\textbf{a})~Cumulative distribution of the relative error of the PinT log-likelihood against the sequential filter, for 1 to 4 GPUs (colors), across 14 test systems and trajectory lengths from 400 to 168k time points (Methods). The dashed line marks the accuracy target of $10^{-6}$.
    (\textbf{b})~Throughput, in likelihood-and-gradient evaluations per second within the Hamiltonian Monte Carlo sampler, of the sequential and PinT filters on a single GPU, with the PinT speed-up above each pair of bars. Error bars show the standard deviation across replicate runs.
    (\textbf{c})~Strong scaling: throughput as a function of GPU count, for five trajectory lengths (labeled at right), with ideal linear scaling dashed. Replicate-to-replicate variation is below 1\%, so error bars are omitted.
    (\textbf{d})~Parallel efficiency, the speed-up over a single GPU divided by the GPU count, as a function of trajectory length and GPU count, with the 70, 80 and 90\% iso-efficiency contours in white.
    (\textbf{e})~Weak scaling: throughput as a function of GPU count, for fixed workloads of 1k to 54k time points per GPU (vertical axis broken), with ideal weak scaling dashed. Error bars show 95\% confidence intervals from the inter-replicate variance.
    (\textbf{f})~Weak-scaling efficiency $E(n_\mathrm{GPU})$, the throughput with $n_\mathrm{GPU}$ GPUs and an $n_\mathrm{GPU}$-fold workload relative to a single GPU, as a function of GPU count, for a range of workloads per GPU (color scale). Error bars as in (e).
    \label{fig:benchmark}
}

We computed the model log-likelihood with both the sequential filter and PinT filter on 14 test systems, at parameters both far from and near the high-density region of the posterior (Methods).
The relative error of the PinT log-likelihood, compared to sequential, is typically around $10^{-9}$, and always well below our target of $10^{-6}$, with no increase with trajectory length, no systematic difference between systems, and only a slight increase with GPU count (Figure~\ref{fig:benchmark}a and Supplementary Information~S3).
Our algorithm is therefore numerically equivalent to the sequential filter, to within single-precision round-off.

Near the maximum a posteriori estimate, the iteration count is constant over close to three orders of magnitude of trajectory length, from hundreds to hundreds of thousands of time points, showing a constant span empirically (Supplementary Information~S3).
The iteration count depends on how quickly the filter forgets its initial condition, a property of the landscape and emission function that varies between systems.
Since the iteration count, here 2 to 4, is also the factor by which total work exceeds that of the sequential filter, the speed-ups below cost a corresponding increase in total computation.
Despite this, on a single GPU, PinT filtering evaluates the likelihood and gradient 9 times faster than the sequential filter at 1k time points, and 26.8 times faster at 5k (Figure~\ref{fig:benchmark}b).
This seemingly paradoxical speed-up results from better use of within-GPU parallelism.
The chain of matrix-vector products in the sequential filter is a thin, latency-bound workload, whereas our algorithm uses dense matrix multiplications that use compute and memory bandwidth more efficiently.

Strong scaling, the speed-up from adding GPUs at fixed trajectory length (Figure~\ref{fig:benchmark}c), is limited by an unparallelizable fixed cost of about 4~ms per evaluation, dominated by operator diagonalization, which accounts for more than half of single-GPU runtime below about 6k time points (Supplementary Information~S3).
The 4k-point trajectory accordingly plateaus beyond two GPUs, the 50k-point trajectory speeds up sublinearly up to eight, and trajectories of 200k time points and more scale linearly up to eight.
Ninety percent parallel efficiency is reached at approximately 60k and 100k time points on two and three GPUs, and above 200k on eight (Figure~\ref{fig:benchmark}d), lengths typical of single-molecule recordings.

Weak scaling, the ability to maintain a constant runtime as trajectory length and GPU count grow together, is shown in Figure~\ref{fig:benchmark}e.
Performance drops mainly from one to two GPUs, at the onset of inter-GPU communication.
Each block exchanges only 3 to 6~kB with its neighbors (Methods), so this cost is latency-dominated, at about  0.4~ms per evaluation.
The weak-scaling efficiency stays above 0.90 for workloads above 1k time points per GPU, and approaches 0.98 above 10k (Figure~\ref{fig:benchmark}f).
A trajectory spread over enough GPUs to keep at least 10k time points on each thus scales almost ideally, for instance on two GPUs for the 24k-point optical tweezer recording analyzed below.
Tens to low hundreds of likelihood-and-gradient evaluations per second are therefore readily achievable, even for long trajectories of several hundred thousand time points.
These figures are deliberately conservative, as they were obtained on five-year-old GPUs with inter-GPU transfers staged through host memory (Methods).

\subsection{Application examples}

Next, we applied our method to three representative systems from different fields, a single-molecule optical tweezer trajectory, a century-long climate index, and a daily financial index.

\includegraphics[width=\textwidth]{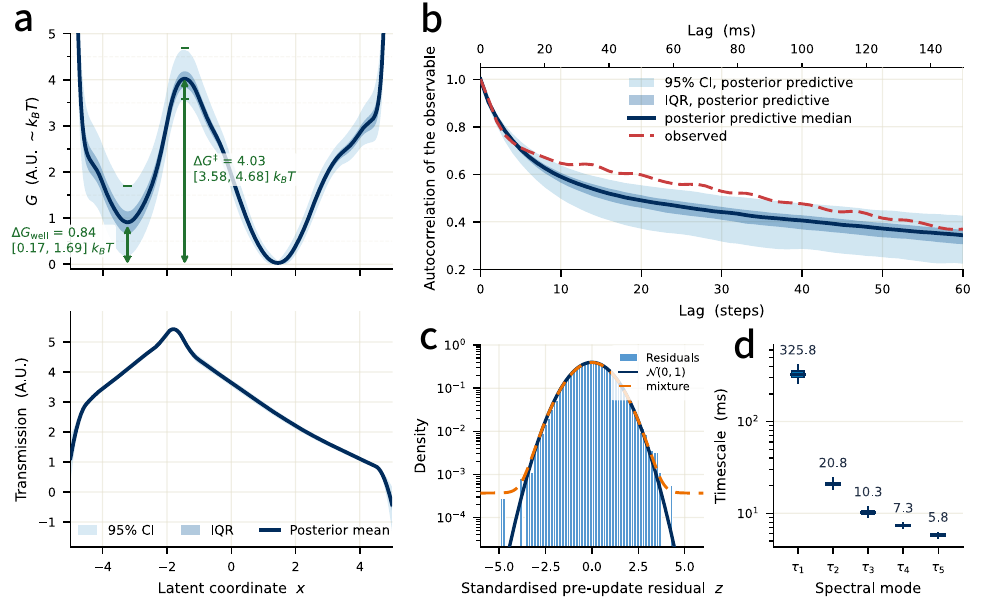}
\captionof{figure}{%
    \textbf{Landscape, emission and kinetics of a single BSA molecule in a nano-aperture optical tweezer.}
    (\textbf{a})~Posterior distribution of the landscape (top, in $k_BT$) and the emission function (bottom, in arbitrary units equal to $1/30$ volt; see Methods), showing the posterior mean (solid blue line), the interquartile range (IQR, dark band) and the 95 percent credible interval (CI95, light band). The free energy difference between the wells $\Delta G$ and the barrier height  $\Delta G^\ddagger$ are marked by arrows. Diffusion $D = 5\cdot10^{-5}$.
    (\textbf{b})~Autocorrelation of the observable, as observed (dashed red line) and under the posterior predictive distribution (dark blue line, with IQR and CI95 as shaded bands). Lag is given in time steps (bottom axis) and in milliseconds (top axis).
    (\textbf{c})~Standardized pre-update residuals, \ie, the difference between each observed value and its one-step-ahead predictive mean, divided by the predictive standard deviation. The observation distribution is a mixture of 99 percent Gaussian and 1 percent uniform; the Gaussian component alone is drawn as a solid dark blue line and the full mixture as a dashed orange line.
    (\textbf{d})~The five slowest relaxation timescales of the inferred landscape, in milliseconds, as a box plot. The posterior median (dark blue line), IQR (box) and CI95 (vertical line) are shown with the median printed above each.
    \label{fig:appl_cases_not}
}

The first application is a single-protein nano-aperture optical tweezer recording of bovine serum albumin (BSA),\cite{peters2024energy} in which the transmission of the trapping laser reports on the conformational state of the protein.
The recovered landscape has two steep-walled wells of uncertain relative depth (Figure~\ref{fig:appl_cases_not}a), with an inter-well relaxation time (325 ms) much slower than all others ($\le 20.8$ ms; Figure~\ref{fig:appl_cases_not}d), justifying a two-state description.
The free energy difference between the wells is 0.84 $k_BT$ (CI95 0.17 to 1.69) and the barrier height 4.03 $k_BT$ (CI95 3.58 to 4.68).
The original non-Bayesian analysis\cite{peters2024energy} reported a  free energy difference near 1.5 $k_BT$ and a barrier around 3 $k_BT$.

The inferred emission function is non-injective, rising at the barrier between wells, so that, as in our most challenging validation example (Figure~\ref{fig:validation_figure}a), a given transmission value is compatible with latent positions on both sides of the barrier.
Unlike FRET efficiency, which follows from the donor--acceptor distance through a known law, optical transmission reports on the polarizability of the protein, a function of its conformation which is well-defined, but whose functional form is system-dependent and difficult to specify.
Such non-injectivity may therefore be frequent in this type of experiment, where it could prevent interpretation by other methods assuming a monotonic emission.

Posterior predictive checks reveal excess autocorrelation at 20 to 135 ms of lag (Figure~\ref{fig:appl_cases_not}b), pointing to a slow degree of freedom hidden by the one-dimensional projection, such as conformational substates within the wells, or to apparatus drift (Supplementary Information~S6.4).
The one-step-ahead residuals also show ten isolated jumps of 4 to 5 standard deviations (Figure~\ref{fig:appl_cases_not}c), which the observation noise model absorbs (Methods).
Both of these findings suggest partial model misspecification, though the inferred landscape and emission are likely still reliable.

\includegraphics[width=\textwidth]{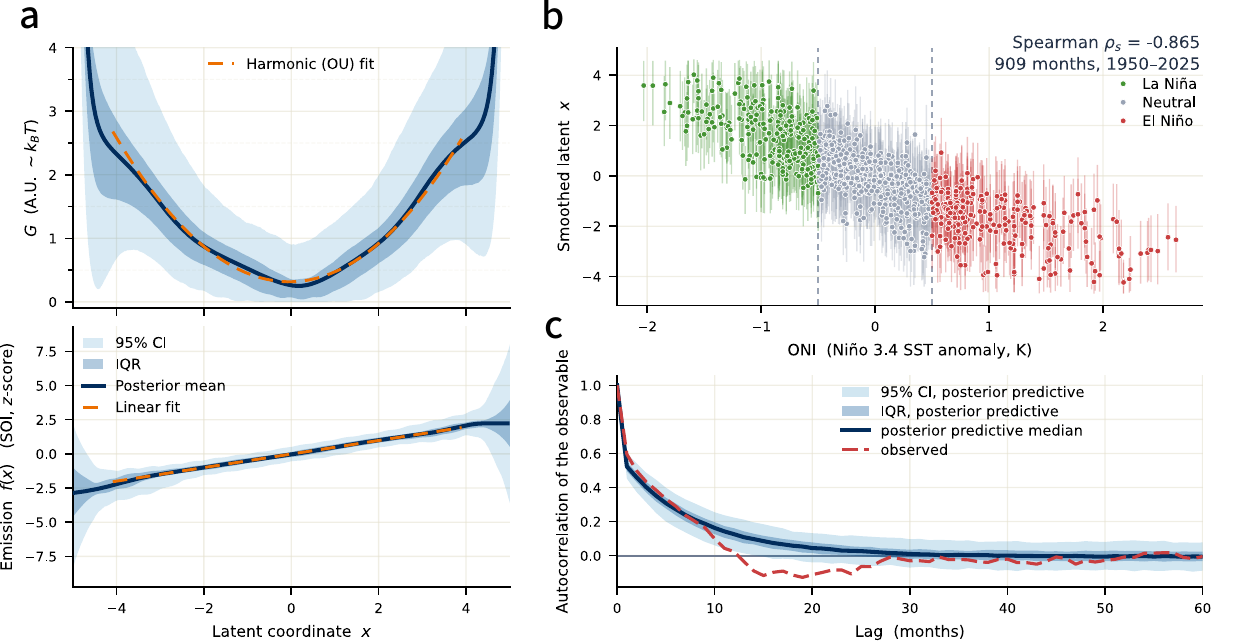}
\captionof{figure}{%
    \textbf{Landscape, emission and external validation of the ENSO latent state.}
    (\textbf{a})~Posterior for the landscape (top) and for the emission function (bottom), showing the posterior mean (solid blue line), the interquartile range (IQR, dark band) and the 95 percent credible interval (CI95, light band). The dashed orange line is a fitted Ornstein--Uhlenbeck model, harmonic in the landscape and linear in the emission. Diffusion $D = 0.35$.
    (\textbf{b})~Smoothed latent coordinate (posterior mean, with CI95 as vertical bars) against the Oceanic Ni\~no Index (ONI), a sea surface temperature index over the Ni\~no~3.4 region that defines the ENSO phase (color). Dashed vertical lines mark the standard NOAA thresholds at $\pm 0.5$~K. 
    (\textbf{c})~Autocorrelation of the observable, as observed (dashed red line) and under the posterior predictive (dark blue line, with IQR and CI95 as shaded bands).
    \label{fig:appl_enso}
}

The second application is the El~Ni\~no--Southern Oscillation (ENSO), the dominant mode of interannual climate variability.
The observable is the monthly Southern Oscillation Index (SOI) from 1866 to 2025, the normalized sea-level pressure difference between Tahiti and Darwin, which measures the atmospheric side of the ENSO.
The inferred landscape is broadly harmonic and the emission linear: an Ornstein--Uhlenbeck model fitted to the posterior mean stays within the 95\% credible interval throughout (Figure~\ref{fig:appl_enso}a, dashed lines).
Our non-parametric approach thus identified this simpler model from the data rather than assuming it.

To test whether the inferred latent coordinate tracks the ENSO state, we compare it with the Oceanic Ni\~no Index (ONI), which was not used during inference (Figure~\ref{fig:appl_enso}b).
The ONI, a sea surface temperature anomaly in the Pacific, measures the oceanic side of ENSO and defines its El~Ni\~no and La~Ni\~na phases.
The smoothed latent coordinate correlates strongly with the ONI (Spearman $\rho_s = -0.865$) and classifies the NOAA phases with a ROC AUC of 0.93 for El~Ni\~no and 0.94 for La~Ni\~na.
Inferred from atmospheric pressure alone, the latent coordinate thus captures the full ENSO state, oceanic as well as atmospheric.

The posterior predictive autocorrelation of the observable matches the observed one over most lags, including the sharp drop from lag zero to one that measures the observation noise (Figure~\ref{fig:appl_enso}c).
The exception is a negative lobe between 14 and 25 months of lag, implying a period of roughly 30 to 38 months.
Reversible dynamics cannot produce a negative autocorrelation at any lag, so this lobe reveals an oscillatory component outside the model class, most likely the quasi-biennial component of ENSO.\cite{rasmusson1990biennial}

\includegraphics[width=\textwidth]{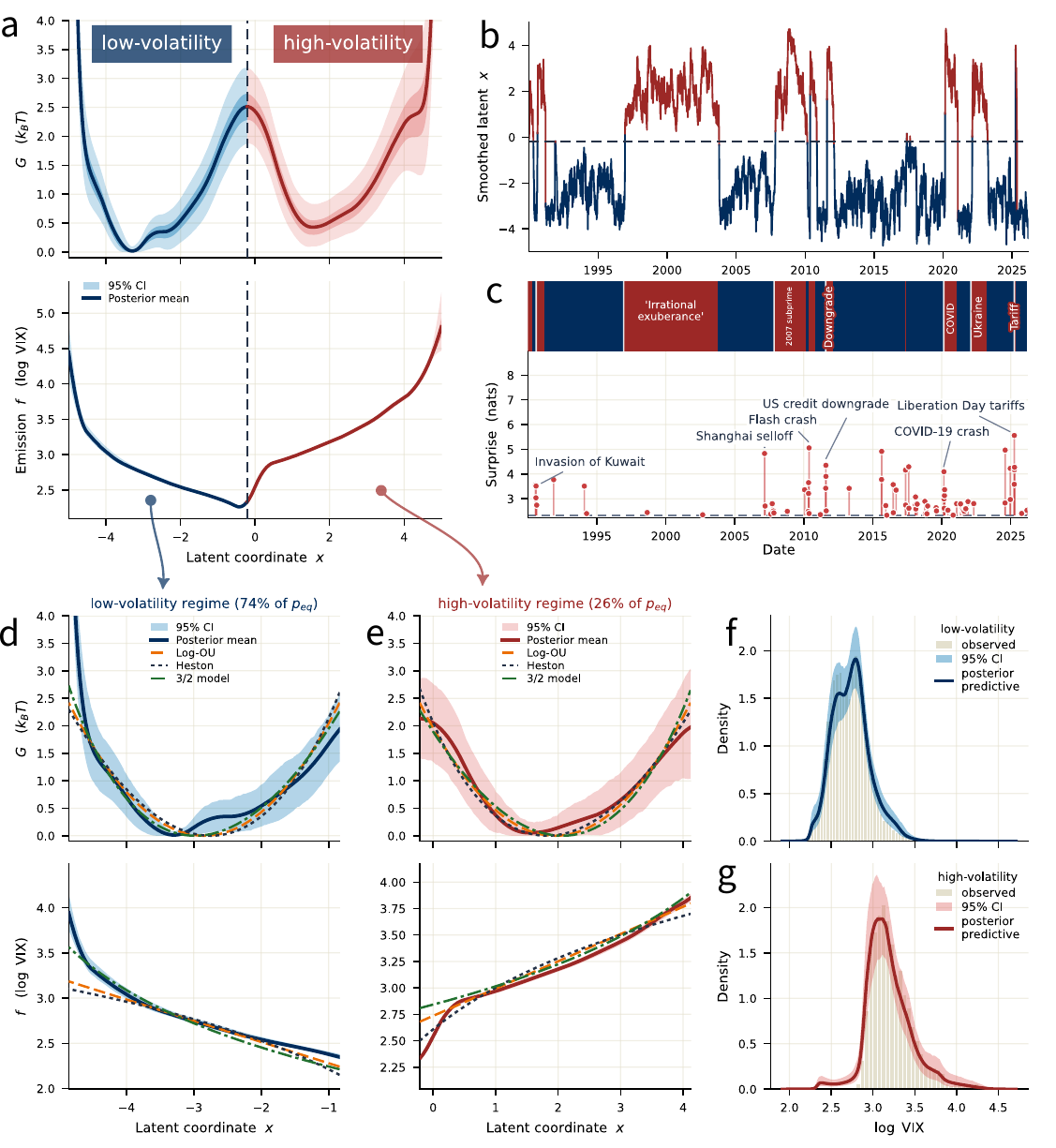}
\captionof{figure}{%
    \textbf{Two volatility regimes in the logVIX record, recovered from a single Brownian dynamics.}
    (\textbf{a})~Posterior for the landscape (top) and for the emission function (bottom), showing the posterior mean (solid line), the interquartile range (IQR, darker band) and the 95 percent credible interval (CI95, lighter band). The dashed vertical line marks the barrier top separating the low-volatility (blue) and high-volatility (red) regimes. Diffusion $D = 0.02$.
    (\textbf{b})~Smoothed posterior mean of the latent coordinate over time. The dashed horizontal line marks the barrier top, and the band between (\textbf{b}) and (\textbf{c}) shows the regime occupied at each time, labeled where a period corresponds to a known episode.
    (\textbf{c})~Model surprise over time: measured as the Kullback--Leibler divergence between the predicted and the filtered latent density at each step,  sharing its time axis with~(\textbf{b}). The dashed horizontal line is the flagging threshold, set at the 95th percentile of that divergence over the whole series. Flagged dates are labeled with the corresponding market event where one is clearly identifiable.    
    (\textbf{d}), (\textbf{e})~Landscape (top) and emission function (bottom) of the low- and high-volatility halves of~(\textbf{a}), the landscape re-referenced to the minimum of each well, with posterior mean and CI95. Panel titles give the equilibrium weight of each regime. Dashed lines show three classical volatility models, log Ornstein--Uhlenbeck, Heston and 3/2, each fitted to the corresponding half.
    (\textbf{f}), (\textbf{g})~Distribution of the observable within the low- and high-volatility regime respectively, as observed (histogram) and under the posterior predictive (solid line, with CI95 as shaded band). Observations are assigned to a regime by the side of the barrier on which the smoothed latent state lies.
    \label{fig:appl_logvix}
}

The third system is financial market volatility, observed through the logarithm of the CBOE Volatility Index (logVIX), recorded daily from 1990 to 2026.
The inferred landscape has two wells, a low- and a high-volatility regime holding 74 and 26 percent of the equilibrium population, clearly separated by a substantial barrier (Figure~\ref{fig:appl_logvix}a).
The emission has one rising branch on each side of the barrier, so that the same logVIX value can arise in either regime (Figure~\ref{fig:appl_logvix}a).
The regimes are thus distinguished by the temporal pattern of fluctuations, which the filter tracks through the latent state, rather than by the level of logVIX alone.

Remarkably, the smoothed latent trajectory shows abrupt regime switches, separated by long stretches, often years, within a single well (Figure~\ref{fig:appl_logvix}b).
High-volatility periods align with episodes of prolonged market stress, from the Asian, Russian and dot-com crises of the late 1990s to the 2007 subprime crisis and the COVID-19 pandemic.
Brief volatility jumps lasting days are flagged by a Kullback--Leibler (KL) surprise metric, the divergence between the predicted and filtered latent densities, which is large when an observation is unlikely under the model (Figure~\ref{fig:appl_logvix}c).
The largest surprises, over 4 nats, coincide with identifiable shocks such as the 2010 Flash Crash or ``Liberation day'' tariffs, though some correspond to no identifiable event.
They occur in both regimes and only some accompany a regime switch, suggesting that short-lived shocks and persistent regime switches are distinct phenomena.

We fitted three classical volatility models, the log Ornstein--Uhlenbeck, Heston and 3/2 models,\cite{goard2013stochastic} to each well (Figure~\ref{fig:appl_logvix}d,e).
All three stay close to the credible interval in the high-volatility regime, but none reproduces local features such as the shoulder in the low-volatility well, nor has any notion of regime switching.
Markov-switching models\cite{hamilton1989new} introduce such switching as explicit jumps between separate regimes.
In our model, by contrast, both regimes and the switches between them emerge from a single Brownian dynamics on the inferred landscape, which also reproduces the observed logVIX distribution within each regime, including their overlap and the heavier tail of the high-volatility regime (Figure~\ref{fig:appl_logvix}f,g).

\section{Discussion}

We have developed a Bayesian method to infer the underlying free energy landscape and emission function jointly and non-parametrically from trajectories of a single observable. 
We have tested the method on three representative systems from different fields such as protein dynamics, climate science and finance.
We have further demonstrated that our method can evaluate the model likelihood for trajectories of hundreds of thousands of points, without approximations such as particle filters, linearization or a discretized latent state, so that the resolution of the inferred landscape and emission is limited only by the data.

Several features distinguish our method from other approaches, each with practical consequences.
First, the emission is inferred rather than specified, and may be non-injective, with several latent states producing the same observed value.
Boltzmann inversion, by contrast, assumes an identity emission, and Gaussian-latent filters such as the extended Kalman filter\cite{julier2004unscented, sarkka2013bayesian}  cannot represent the multimodal latent density that non-injectivity creates.
In contrast, our method can be used for systems for which the emission functional form is not known \textit{a priori}, such as the mapping between transmittance and protein conformation in our optical tweezer application, which moreover was shown to be non-injective.
In the financial market volatility example, non-injectivity is likewise what allows two volatility regimes to emerge.

Second, the landscape is non-parametric and its shape is constrained only by a smoothness prior.
This contrasts with fixed-form landscapes, such as the harmonic potential of the Ornstein–Uhlenbeck process,\cite{uhlenbeck1930theory} and with parametric landscapes, whose number of wells is fixed and only their depths and barriers adjusted.\cite{kwasniok2009deriving}
Using our method, the suitability of a simpler model is therefore a result supported by data, not an \textit{a priori} assumption.
Our ENSO application from climate science illustrated this, with our posterior validating the use of a simpler quadratic-linear model.
In our logVIX application, in contrast, we showed that classical volatility models are severe approximations, and a two-regime model better describe the data.

Third, our model likelihood is exact and explicitly computed,  rather than replaced by a surrogate as in simulation-based inference.\cite{cranmer2020frontier, dingeldein2025simulation}
As a consequence, when the true process is not an overdamped Brownian dynamics, \ie, when the model is misspecified,
the posterior concentrates asymptotically on the overdamped Brownian dynamics closest to the data in Kullback--Leibler divergence.\cite{kleijn2012bernstein} 
Under mild misspecification, the recovered landscape and emission are thus still an informative, albeit inexact, model of the true process, a guarantee of robustness that surrogate likelihoods do not generally share.
Further, misspecification is straightforward to detect, since the observable and statistics predicted under the model can be directly compared to the data.
These checks also make it possible to identify the type of misspecification, such as a hidden slow degree of freedom, external forcing, or non-Brownian jumps, all three demonstrated in our applications.

Finally, our parallel-in-time algorithm makes our likelihood computationally practical to evaluate.
Already faster than sequential filtering on a single GPU, it scales well when using multiple GPUs.
Without our algorithm, inference on the optical tweezer application would have been impractically slow, and our other applications would
require fivefold more runtime.
Unlike parallel prefix scans,\cite{sarkka2020temporal, hassan2021temporal} the algorithm requires no closed-form composition of prediction and update.
It therefore generalizes beyond our model, to hidden Markov models whose filter propagates a probability density, discrete-state models included. 
We conjecture that its span complexity, proven independent of trajectory length for our model, would grow only logarithmically in the general case.
More broadly, our parallel-in-time algorithm reduces the need to trade off model sophistication and data volume in likelihood-based inference for such models.

The following limitations of our approach point to directions for future work. 
It currently supports only a single observable and a one-dimensional latent space, whereas many systems are probed through several observables simultaneously.
Further observables would enter as additional factors in the observation likelihood, providing the information needed for higher-dimensional latent inference.
However, the explicit basis we use to represent the latent density, and with it the computational cost, grows exponentially with the latent dimension, becoming prohibitive beyond two dimensions.
Particle filters could avoid this cost, scaling instead with the number of particles, but whether or not our parallel-in-time algorithm could be adapted to them is an open question.
If such adaptation were possible can, our approach would extend to models whose transitions can be simulated but not evaluated, a class currently addressed mainly by simulation-based inference.\cite{cranmer2020frontier, dingeldein2025simulation}

\section{Methods}
Our inference problem combines three components: a model for the latent dynamics, a model for the emission function, and Bayesian filtering to compute the likelihood of observed trajectories given both.
We define each in turn, then the wavelet representation with its smoothness priors.
The filtering likelihood follows, with the parallel-in-time algorithm that evaluates it, the sampling and benchmarking protocols, and the synthetic and application cases.
The convergence proof of the parallel-in-time algorithm, complete parametrizations, and implementation details are given in the Supplementary Information (S1, S6 and S2, respectively).

\subsection{Model definition}

We consider a continuous--discrete filtering model: a continuous-time, continuous-state latent process, observed at discrete times through an emission function that is non-linear and need not be injective.

The latent process $\{X_t\}_{t \geq 0}$ takes values in the state space $\mathcal{X} = [a, b]^n \subset \mathbb{R}^{n}$ and satisfies the overdamped Langevin equation

\begin{equation}
dX_t = -\frac{D}{k_B T} \nabla G_{\theta_{\mathrm{land}}}(X_t)\, dt + \sqrt{2D}\, dW_t,
\end{equation}

where $D > 0$ is the diffusion coefficient and $\{W_t\}_{t \geq 0}$ the standard Wiener process.
The equation holds for overdamped dynamics, in which momentum relaxes fast compared with the observation interval.
The free energy landscape $G_{\theta_{\mathrm{land}}}: \mathcal{X} \to \mathbb{R}$ decomposes as $G_{\theta_{\mathrm{land}}} = G_{\mathrm{interior},\, \theta_{\mathrm{land}}} + G_{\mathrm{wall}}$, the first term parameterized by $\theta_{\mathrm{land}} \in \Theta_{\mathrm{land}}$, the second fixed and known, rising steeply near $\partial\mathcal{X}$ (Supplementary Information~S2).
Both terms are assumed continuously differentiable with bounded second derivative on $\mathcal{X}$, and a formal zero-flux condition at $\partial\mathcal{X}$ closes the domain.
The stationary density is then $p_{\mathrm{eq}}(x) = Z^{-1} e^{-\beta G_{\theta_{\mathrm{land}}}(x)}$, with $\beta = 1/k_B T$ and $Z$ the normalizing constant.

The observed process $\{Y_{t_k}\}_{k=1}^{N}$ takes values in the observation space $\mathcal{Y} \subseteq \mathbb{R}$, sampled at times $t_k = k \Delta t$ for simplicity, though the method allows non-uniform observation times.
Conditional on the latent state, observations are generated via the emission model

\begin{equation}
Y_{t_k} \mid X_{t_k} \sim p_{\mathrm{obs}}(\cdot \mid f_{\theta_{\mathrm{em}}}(X_{t_k}))
\end{equation}

where $f_{\theta_{\mathrm{em}}}: \mathcal{X} \to \mathcal{S}$ is the emission function parameterized by $\theta_{\mathrm{em}} \in \Theta_{\mathrm{em}}$, assumed continuously differentiable with bounded second derivative.
Here $\mathcal{S}$ is the parameter space of the observation (noise) distribution and $p_{\mathrm{obs}}(\cdot \mid s)$ is a probability density on $\mathcal{Y}$ indexed by $s \in \mathcal{S}$.
If $p_{\mathrm{obs}}$ is Gaussian with fixed $\sigma$ and variable mean, for instance, then $\mathcal{S} = \mathbb{R}$ and $s$ is that mean.
The observation distribution is specified by the user rather than inferred, subject only to $f_{\theta_{\mathrm{em}}}$ controlling a location parameter of it.
Its identifiability jointly with the landscape and emission is untested.
In the application cases, we chose it empirically from the one-step-ahead residuals of a preliminary run.
Given the trajectory $\{X_t\}$, the observations $\{Y_{t_k}\}$ are conditionally independent.

The model is identified only up to gauge freedoms of the latent coordinate: isometries such as reflections and rotations, and its scale set by the diffusion coefficient $D$ (Supplementary Information~S4).
We break the isometries with a weak prior and fix $D$ rather than infer it.
Its value comes from a preliminary run in which $D$ is free, fixed within the range the posterior supports and reported alongside the posteriors (gauge-fixing procedure, Supplementary Information~S4).

\subsection{Representation and regularization}

Recovering $G$ and $f$ from finite, noisy trajectories is an inverse problem, and unregularized solutions to such problems fit detail that the data cannot resolve.
Bayesian inference mitigates this by returning a posterior rather than a point estimate, so that unsupported detail appears as uncertainty.
We regularize nonetheless on two grounds.
First, as an inductive prior: a projection onto a few collective coordinates averages over many degrees of freedom, so that the resulting landscape and emission are generally smooth.
Second, for efficiency, as restricting the search space reduces the sampling needed to reach a given accuracy.
Besov spaces $B^s_{p,p}$ comprise the functions of a prescribed regularity, indexed by a smoothness order $s$ and a sparsity type $p$.
We target this class by expanding $G$ and $f$ in a wavelet basis and placing the corresponding prior on the coefficients, which shrinks them geometrically with resolution level and is discretization-invariant in the sense of Lassas \textit{et al}.\cite{lassas2009discretization}
We use the biorthogonal 3.3 synthesis wavelet, a compactly supported quadratic spline, continuously differentiable with bounded second derivative, so that finite expansions meet the regularity assumed above.
Defaults are $s = 2$, the lowest order that penalizes curvature, which enters the Fokker--Planck operator in the symmetrized form we diagonalize, and $p = 2$, which makes the coefficient prior Gaussian.
Both are user-settable (Supplementary Information~S5).
The posterior is robust to the prior scales and to the weight of the isometry-breaking prior, more sensitive to $s$, and concentrates on the ground truth as the trajectory lengthens (Supplementary Information~S6.6).

The wavelet basis carries the prior, but filtering is computed in a second basis.
We parametrize $G$ and $f$ in the wavelet basis, 24 coefficients each, and compute in a basis of Gaussian functions on a regular grid, $N_g = 96$ per dimension (256 for the logVIX case), giving $N_b = N_g^n$ basis functions.
A linear change of basis maps between the two.
Gaussians are chosen for quadrature, the non-linear operations of filtering being evaluated at Gauss-Hermite nodes and projected back by the quadrature weights, efficiently on GPU (Supplementary Information~S2).
Computation is mixed precision: filtering runs in single precision, while operator assembly, diagonalization and the log-likelihood reduction run in double.

\subsection{Filtering and smoothing}

The central computational task is evaluating the likelihood $p(\{Y_{t_k}\} \mid \theta)$, which Bayesian filtering decomposes into predict--update iterations.\cite{sarkka2013bayesian}
Prediction propagates the density between observations, $p^{\mathrm{pred}}_{t_{k}} = e^{\mathcal{L}\Delta t} p_{t_{k-1}}$, with $\mathcal{L}$ the Fokker--Planck operator, projected onto the computation basis and diagonalized to form the propagator.
The update multiplies by the observation likelihood and renormalizes, with

\begin{equation}
\label{eq:likelihood_integral}
Z_{t_{k}} = \int_\mathcal{X} 
  p_{\mathrm{obs}}(Y_{t_k} \mid f_{\theta_{\mathrm{em}}}(x)) 
  \cdot p^{\mathrm{pred}}_{t_{k}}(x) \; dx,
\end{equation}

so that the total log-likelihood is $\sum_k \log Z_{t_{k}}$.
Because the update takes place in $\mathcal{X}$, no Jacobian correction is needed despite the non-invertible emission.
The recursion is initialized with $p_{t_0} = p_{\mathrm{eq}}$, and the backward Kitagawa smoother gives exact non-linear smoothing for latent trajectory reconstruction.
Implementation details are given in Supplementary Information~S2.

\subsection{Parallel In Time (PinT) filtering}

The filtering recursion is sequential: the density at step $k$ depends on that at step $k-1$, giving a span complexity of $\mathcal{O}(N)$ in the number of time steps.

We convert this span into parallel work with an iterative algorithm in the manner of waveform relaxation\cite{lelarasmee1982waveform} (Figure~\ref{fig:overview_method}b).
Each trajectory of $N$ time steps is partitioned into $B = \lceil N/L \rceil$ contiguous blocks of length $L$, distributed across the available devices.
Splitting the trajectory this way requires the latent density at each block boundary, which is not available in advance.
The leading block therefore starts from the model's initial density, $p_{\mathrm{eq}}$ here, and every other from a trial density, for which we also use $p_{\mathrm{eq}}$.
Each iteration filters all blocks concurrently, accumulating their likelihood contributions, then passes the terminal density of block $b-1$ to the start of block $b$.
At the fixed point of this iteration, the boundary densities and the log-likelihood agree exactly with those of the sequential filter (Supplementary Information, Proposition~P1).
Blocks communicate only through these boundary densities, so the inner filter enters as a black box: any predict--update recursion can be substituted for ours.
Exactness at the fixed point follows for any such substitution; the iteration count established below does not, resting on the forgetting rate of whichever filter is used.
Pseudocode and convergence proofs, with partial Lean formalization, are given in Supplementary Information~S1.

Convergence is measured by the change in the block-boundary densities between successive iterations, in the weighted root-mean-square (WRMS) norm standard in adaptive ODE error control (Supplementary Information~S1.1).
The norm scales each basis coefficient by a mixed absolute and relative tolerance, so that it is dimensionless and stays well defined for coefficients approaching zero.
Because the sequential filter is reproduced exactly only where the boundary densities stop changing altogether, the iteration stops once this norm falls to one or below, with $\mathrm{atol} = \mathrm{rtol} = 10^{-6}$, about ten times the machine epsilon of the single precision in which the filter runs, so that the fixed point is reached to within floating-point noise.

Convergence is guaranteed in at most $B$ iterations and without approximation, since exact boundary conditions propagate across one further block per iteration.
This guarantees convergence even for a filter that never forgets its initialization, though of course the speed-up is then nil.
A stronger result holds for our model: the iteration count does not grow with $N$, giving a span of $\mathcal{O}(N_{\mathrm{iter}} L)$ rather than $\mathcal{O}(N)$.
At iteration $n$, block $b$ has been conditioned on $nL$ consecutive filtering steps.
By filter stability\cite{legland2004stability} a misspecified initialization is forgotten after some number of steps $k_{\mathrm{f}}$, so once $nL > k_{\mathrm{f}}$ every block has converged, and $N_{\mathrm{iter}} = \lceil k_{\mathrm{f}}/L \rceil$.
It is the particularly strong forgetting rate of our model that allows an $\mathcal{O}(1)$ span, as it provides a global $k_{\mathrm{f}}$ derived from the compactness of the latent domain or the spectral gap of the propagator.
For models with neither property, and hence without such a global forgetting time, we expect instead $\mathcal{O}(\log N)$ span.
Total work is $\mathcal{O}(N_{\mathrm{iter}} N)$, a small multiple of that of the sequential filter, the price of the reduced span.

HMC also requires the gradient of the likelihood, which we obtain without differentiating through the iterations.
Writing that iteration as $P^{(n+1)} = \Phi(P^{(n)}, \theta)$, with $P$ the block boundary densities, reverse-mode gradients follow from the implicit function theorem.\cite{christianson1994reverse}
At convergence the gradient of the likelihood with respect to $\theta$ solves the adjoint equation $u = \bar P + (\partial \Phi/\partial P)^\top u$, itself a contractive fixed-point iteration.
Differentiating the converged solution rather than the iterations keeps the gradient exact and its memory cost independent of the iteration count.

\subsection{Inference protocol}

Reaching the region of highest posterior density is difficult: emission functions differing in preimage multiplicity, the number of latent values mapping to a given observation, occupy separate basins that a chain rarely crosses even when the destination is better.
Our protocol therefore covers the basins by construction, rather than requiring any chain to cross between them.

We fan out over a library of candidate parameter sets, 25 to 150 per chain, each pairing a flat landscape with an emission function drawn as a piecewise-linear path through a coarse grid of nodes.
Each candidate is taken to a local maximum of the log-posterior by L-BFGS, then explored by HMC, both under a tempered posterior $\pi_\gamma(\theta) \propto [\pi(\theta)\,p(Y \mid \theta)]^{\gamma}$ with $\gamma$ starting at $\gamma_0 = 0.8$ for the application cases and $0.1$ for the validation cases.
The observation distribution is tempered in its tails as well, mixed with a broad density (triangular, or uniform where $p_{\mathrm{obs}}$ already has a uniform component) spanning the observed range, $p^{(\varepsilon)}_{\mathrm{obs}} = (1-\varepsilon)\,p_{\mathrm{obs}} + \varepsilon\,\mathrm{Tri}(y_{\min}, y_{\max})$, with $\varepsilon$ between 0.05 and 0.3 (Supplementary Information~S6), adjusted empirically upward if divergences are encountered during HMC.
Far from the true landscape, a predicted density and an observation can overlap negligibly, so the likelihood and its gradient reflect almost nothing about how far apart they are.
The mixture restores that signal, acting in one dimension much like a Wasserstein distance between the two densities.
Annealing then proceeds in two stages, first raising $\gamma$ to one, then lowering $\varepsilon$ to a residual 0.5 percent that keeps the likelihood positive for every $\theta$, after which a short untempered HMC phase settles each candidate into its local optimum.
The Hessian at the L-BFGS optimum sets the initial diagonal mass matrix, which is then adjusted, together with the step size, by warmup-style adaptation throughout the fan-out.

One candidate is retained by importance sampling on its posterior probability at the end of that phase.
The best candidate typically has overwhelmingly better posterior probability than the other candidates: the other basins are local optima carrying negligible posterior mass, so that retaining one candidate discards no appreciable part of the posterior.
Step size and a diagonal mass matrix are then re-adapted for the production run by the Stan window algorithm,\cite{carpenter2017stan} and samples are collected by HMC.\cite{neal2011mcmc}
The whole protocol, fan-out included, is run independently for each of four chains, which are therefore independent in the strict sense.
Convergence is assessed by split $\hat{r}$, Bayesian fraction of missing information, and effective sample size.
We use static HMC with a fixed number of leapfrog steps and large sample counts rather than adaptive path lengths (as in NUTS), since GPU-accelerated likelihood evaluation makes over-sampling inexpensive.
Sample counts and remaining settings are given in Supplementary Information~S6.

\subsection{Benchmarks}

All benchmarks ran on a single compute node with eight NVIDIA A4000 GPUs, on one landscape (double well) and one emission function (linear).
The measured quantity is the time to collect 10 posterior samples by HMC with 8 leapfrog steps.
The accuracy comparison (Figure~\ref{fig:benchmark}a) used 14 test systems: the validation and application systems of this work, further applications not reported here, and additional validation runs such as the prior sensitivity scans (Supplementary Information~S6).
Each was evaluated at the initial flat landscape, at the ground-truth parameters where available, and at posterior samples.
The benchmark therefore carries every overhead of practical use, likelihood accumulation and the backward pass for gradients included.
Block length was scanned from 3 to 20 time steps per block, and the best performance for each problem size and GPU count is reported.
Each configuration was replicated five times, re-running the whole sampling loop without recompilation, to give confidence intervals on the timings.
Sequential timings were collected by the same protocol.
Non-monotonic speed-ups beyond run-to-run noise reflect JAX emitting different machine code for each GPU count and problem size, a known limitation of the framework.

Communication between blocks, and hence between GPUs, is nearest-neighbor: each block receives one boundary density per trajectory from its predecessor and sends one to its successor.
The volume is \textit{de minimis}, a density of $N_b$ coefficients, 3 to 6~kB.
The algorithm is therefore latency-bound in its communication, so we deliberately used the slowest available transfer, a staging buffer in host memory, to give worst-case conditions for within-node parallelization.

\subsection{Validation examples}

Ground-truth landscapes and emission functions were drawn from the prior ($s = 2$, $p = 2$), then modified to add multiple wells and two- and three-way non-injectivity.
From 10 such draws we retained the two most challenging, those with the most overlapping emission branches, on the reasoning that a method handling those handles easier cases.

Trajectories were generated by Euler--Maruyama integration at a step roughly one hundred times smaller than the observation interval, reduced further so that $\Delta t$ falls on an exact integer multiple of it.
Integration error is therefore negligible at the scale of the observation interval, so that the validation tests the method essentially against the continuous-time process.
Observations were drawn from a Gaussian observation distribution with $\sigma = 0.25$.

The diffusion coefficient was set to $D = 1$ for both generation and inference.
The gauge is therefore fixed by construction rather than by the preliminary run used for the application cases.
Inference then followed the protocol above unchanged, starting from the candidate library with no knowledge of the ground truth, exactly as for the application cases.
Posteriors are shown as produced, with no alignment to the ground truth except for fixing the minimum free energy of each sample to zero, the dynamics being invariant to an offset of the whole landscape.

Prior sensitivity was assessed by re-inferring the landscape of validation system~(b) across smoothness orders $s \in \{1, 1.5, 2, 2.5, 3\}$ at $p = 2$ and isometry-breaking weights $w_{\mathrm{sym}} \in \{1, 4, 8, 10, 20\}$, and that of system~(a) across prior scales $\sigma_0 \in \{10, 20, 40\}$ and multipliers $C_s \in \{2, 5, 20\}$, with $C_s\sigma_0$ the scale of the scaling-function prior.
Together, $\sigma_0$ and $C_s$ set our prior belief in the overall free-energy variation.
Data sensitivity was assessed on a separate double-well landscape with linear emission, for trajectories of $N = 10\,000$ to $50\,000$ time steps.
Trajectory counts and lengths, integrator settings, and the remaining parameters are given in Supplementary Information~S6.1, S6.2 and S6.6.

\subsection{El Niño (ENSO) case study}

We model ENSO as a stationary diffusion over the 160-year observational record, the same assumption underlying autoregressive models routinely fitted to climate indices over comparable windows.
Slow non-stationarity from volcanic, orbital, and anthropogenic forcing is absorbed into a single time-averaged potential.
The inferred landscape is the dimensionless effective potential $G(x)$, defined by $p_\text{eq} \propto e^{-G}$; no thermal interpretation is needed, though $G$ directly determines transition statistics (\eg, occupation ratios, mean residence times).

Observable: the Southern Oscillation Index (SOI), the standardized pressure differential between Tahiti and Darwin, recorded monthly from January 1866 to December 2025 (1920 observations).\cite{ropelewski1987extension, CruSOITimeseries}
As independent validation, we compare inferred ENSO phase assignments against the Sea Surface Temperature-based Oceanic Niño Index (ONI),\cite{NOAA_CPC_ONI, huang2017extended} available for 909 of these months, using standard NOAA thresholds ($\pm 0.5$~K) to derive El~Niño, La~Niña, and neutral labels.
The ONI plays no role in the inference.
The observation distribution is a Student-$t$ ($\nu = 9.5$, $\sigma = 0.60$), chosen empirically from the one-step-ahead residuals of a preliminary run, and the diffusion coefficient was fixed at $D = 0.35$ after a short run with free $D$.
See Supplementary Information~S6.3 for detailed parametrization.

\subsection{Nano-aperture optical tweezer application}

Observable: the transmission of the trapping laser through a double nano-hole aperture, recorded as an avalanche photodiode voltage, reporting on the conformational state of a single bovine serum albumin molecule held at 37\,$^\circ$C.\cite{peters2024energy}
We use 24 thousand points, sampled every 2.5~ms, of the single trajectory available as open data.

The recorded voltage was offset by its minimum and rescaled by a factor of 30 to improve numerical conditioning; the emission function is reported in these arbitrary units, which correspond to $1/30$ volt.
The observation distribution is a mixture, 99 percent Gaussian ($\sigma=0.10$) and 1 percent uniform over the observable range, the uniform component serving only to absorb ten isolated frame-to-frame jumps of 4 to 5 standard deviations, incompatible with Brownian dynamics (Figure~\ref{fig:appl_cases_not}c).
As in the other case studies, the diffusion coefficient was fixed, here at $D = 5 \cdot 10^{-5}$, after a short run with free $D$.

See Supplementary Information~S6.4 for detailed parametrization.

\subsection{Implied volatility case study}

Observable: option-implied, forward-looking volatility of the S\&P~500 as logVIX (daily close, 2 Jan 1990 – 13 Mar 2026, Chicago Board Options Exchange; 9142 observations).\cite{cboe_vix_historical} 
The index was introduced in 1993, and its methodology revised in September 2003, when it moved from S\&P~100 to S\&P~500 options and to a wider range of strikes.
Values before that date were computed retroactively by the exchange under the revised methodology.
As in the ENSO study, the inferred landscape is the dimensionless effective potential $G(x)$.
For the observation distribution we use an empirically determined Student-$t$ ($\nu=2.6$, $\sigma=0.023$), based on residual fitting after an initial inference.
The diffusion coefficient was fixed at $D = 0.02$ after a short run with free $D$.

See Supplementary Information~S6.5 for detailed parametrization.

\subsection{Data availability}

The datasets, input files, and scripts used for the validation and application cases presented here are available at Zenodo (\href{https://doi.org/10.5281/zenodo.24311789}{DOI 10.5281/zenodo.24311789}), including
final posterior data.

\subsection{Code availability}
The Planx package implementing the described method, and used to produce the application cases, is available under
the GPLv3 license at \url{https://gitlab.com/ElianeBriand/planx}.

\subsection{Use of AI statement}
The large language model Claude (Anthropic) was used to help write the software package and the plotting code, and to edit this manuscript.

\subsection*{Appendix}

\subsection*{Acknowledgments}

This work was funded by the German Federal Ministry of Education and Research (BMBF)
as part of its SCALEXA (New Methods and Technologies for Exascale Computing) initiative.

\bibliography{bayesian_filtering}

\end{document}


\begin{center}
  {\LARGE\bfseries Supplementary Information}\\[0.5em]
  {\large Scalable Non-Parametric Bayesian Inference of Free Energy Landscapes and Emission Functions from Observations}\\[1em]
  Eliane Briand, Helmut Grubm\"uller
\end{center}

\tableofcontents
\newpage

\section{Convergence of PinT filtering algorithm}
\label{sec:conv_pint}

\noindent\small
The grey boxes documents how the proof various proposition are expressed the Lean~4 / Mathlib
formalization, \ie{} the computer-verified version of the proof. Readers not interested in this
formalization may skip these boxes.
\normalsize

\subsection{Setting and scope}\label{sec:pint_setting}

\paragraph{Algorithms under consideration.}

The following three algorithms/subroutines are the subjects of the proof.

\begin{quote}\small
\textsc{BlockFilter}$(p_{\mathrm{in}}, Y_{\mathrm{seg}}, \theta)$:\\
\hspace*{1.2em}$p \leftarrow p_{\mathrm{in}}$; \quad $\ell \leftarrow 0$\\
\hspace*{1.2em}\textbf{for each} $Y$ \textbf{in} $Y_{\mathrm{seg}}$:
  \hfill\emph{$L$ steps in a block}\\
\hspace*{2.4em}$p_{\mathrm{pred}} \leftarrow e^{\mathcal{L}\Delta t} p$   \hfill\emph{predict}\\
\hspace*{2.4em}$Z \leftarrow \int p_{\mathrm{obs}}(Y \mid f_\theta(x))\,p_{\mathrm{pred}}(x)\,dx$  \hfill\emph{normalizer  = likelihood}\\
\hspace*{2.4em}$p \leftarrow p_{\mathrm{obs}}(Y \mid f_\theta(\cdot))\,p_{\mathrm{pred}} / Z$ \hfill\emph{update} \\
\hspace*{2.4em}$\ell \leftarrow \ell + \log Z$\\
\hspace*{1.2em}\textbf{return} $p, \ell$

\end{quote}

\textsc{BlockFilter} is a generic, density-based non-linear Bayesian filter operating on a block of $L$ time steps with observations $Y_{\mathrm{seg}} = \{ Y_{i=1..L} \}$, indexed locally within the block, and Fokker--Planck operator $\mathcal{L}$.
In addition to the latent densities $p$, the model log-likelihood is also returned in $\ell$.

\begin{quote}\small
\textsc{SequentialFilter}$(Y, \theta)$:\\
\hspace*{1.2em}$p_0 \leftarrow \pi_0$\\
\hspace*{1.2em}\textbf{for} $\beta = 0,\dots,B-1$:
  \hfill\emph{strictly serial, span $\mathcal{O}(N)$}\\
\hspace*{2.4em}$p_{\beta+1}, \ell_\beta \leftarrow \textsc{BlockFilter}(p_\beta, Y_{\mathcal{I}_\beta}, \theta)$\\
\hspace*{1.2em}\textbf{return} $\sum_\beta \ell_\beta, \{p_\beta\}$
\end{quote}

\textsc{SequentialFilter} is the sequential Bayesian filter, the naive algorithm corresponding to sequential application of the filtering recursion.
We express it equivalently as a sequential iteration on $B = N/L$  blocks of $L$ time steps, given a trajectory of $N$ total time steps, with repeated application of \textsc{BlockFilter}, passing the boundary density resulting from each application to the next one.

\begin{quote}\small

\medskip
\textsc{PintFilter}$(Y, \theta, \pi_0, \bar\pi_0)$:\\
\hspace*{1.2em}$p^{(0)}_0 \leftarrow \pi_0$;    \hfill\emph{True initial condition} \\
\hspace*{1.2em} $p^{(0)}_\beta \leftarrow \bar\pi_0$ for $\beta = 1,\dots,B-1$ \hfill\emph{Trial densities as initial condition for each block}\\
\hspace*{1.2em}\textbf{for} $j = 1,\dots,B$:     \hfill\emph{PinT iteration (at most $B$ iterations)} \\
\hspace*{2.4em}\textbf{parallel for} $\beta = 0,\dots,B-1$:\\
\hspace*{3.6em}$q^{(j)}_\beta, \ell^{(j)}_\beta \leftarrow
  \textsc{BlockFilter}(p^{(j-1)}_\beta, Y_{\mathcal{I}_\beta}, \theta)$ \hfill\emph{Filter each block in parallel} \\
\hspace*{2.4em}$p^{(j)}_0 \leftarrow \pi_0$
  \hfill\emph{Leading boundary pinned}\\
\hspace*{2.4em}$p^{(j)}_\beta \leftarrow q^{(j)}_{\beta-1}$ for $\beta = 1,\dots,B-1$ \hfill\emph{Pass densities tail-to-head to next block}\\
\hspace*{2.4em}$\delta^{(j)} \leftarrow \max_{\beta\geq1}
  \bigl\|\mathbf{c}[p^{(j)}_\beta] - \mathbf{c}[p^{(j-1)}_\beta]\bigr\|_{\mathrm{WRMS}}$  \hfill\emph{Convergence criterion in density}\\
\hspace*{2.4em}\textbf{if} $\delta^{(j)} \leq 1$: \textbf{break}\\
\hspace*{1.2em}\textbf{return} $\sum_\beta \ell^{(j)}_\beta, \{p^{(j)}_\beta\}$
\end{quote}

\medskip

\textsc{PintFilter} is our parallel-in-time filter.  
As in \textsc{SequentialFilter}, each trajectory of $N$ time steps is partitioned into $B = N/L$ contiguous blocks of $L$ steps.
Filtering the blocks in parallel requires the latent density at each block boundary, a density that only the chain of preceding blocks can produce, and which is thus not available in advance.
We therefore start the leading block from the true initial density $\pi_0$, and other blocks from a trial density $\bar\pi_0$.
In each iteration $j$, all blocks are concurrently filtered, accumulating their log-likelihood contributions, and the terminal density of block $\beta-1$ is then passed on to block $\beta$, as its next initial condition.
To detect convergence, we measure, for each block, how much its initial condition has changed compared to the previous iteration,
\[
  \delta^{(j)}_\beta \;=\; \bigl\|\mathbf{c}[p^{(j)}_\beta] - \mathbf{c}[p^{(j-1)}_\beta]\bigr\|_{\mathrm{WRMS}},
\]
and take the maximum over blocks, $\delta^{(j)} = \max_{\beta\geq1} \delta^{(j)}_\beta$, which we
call the \emph{increment} at iteration $j$.
When no initial condition changes are detected, the algorithm has converged, since the block filter is deterministic and identical inputs yield identical outputs.
In exact arithmetic, the test would be $\delta^{(j)} = 0$.
Floating-point arithmetic makes successive passes boundary densities differ slightly, so the implementation uses a Weighted Root Mean Square (WRMS) norm with absolute and relative tolerances, both at the floating-point floor (\eg, $10^{-6}$ in single precision).
The log-likelihoods $\ell^{(j)}_\beta$ and boundary densities $p^{(j)}_\beta$ returned on exit both come from the $j$-th pass, whose inputs are the boundary densities $p^{(j-1)}_\beta$ of the previous iteration.

The weighted root-mean-square (WRMS) norm is defined as
 \begin{equation}
   \delta^{(n)} = \max_b \big\| p^{(n)}_{0,b} - p^{(n-1)}_{0,b} \big\|_{\mathrm{WRMS}}, \qquad
   \| v \|_{\mathrm{WRMS}} = \sqrt{\frac{1}{N_b}\sum_{i=1}^{N_b} \left(\frac{v_i}{\mathrm{atol} + \mathrm{rtol}\cdot|u_i|}\right)^{2}},
 \end{equation}
 where $u_i$ denotes coefficient $i$ of $p^{(n-1)}_{0,b}$ and $N_b$ is the basis dimension.

\medskip

 \includegraphics[width=0.80\textwidth]{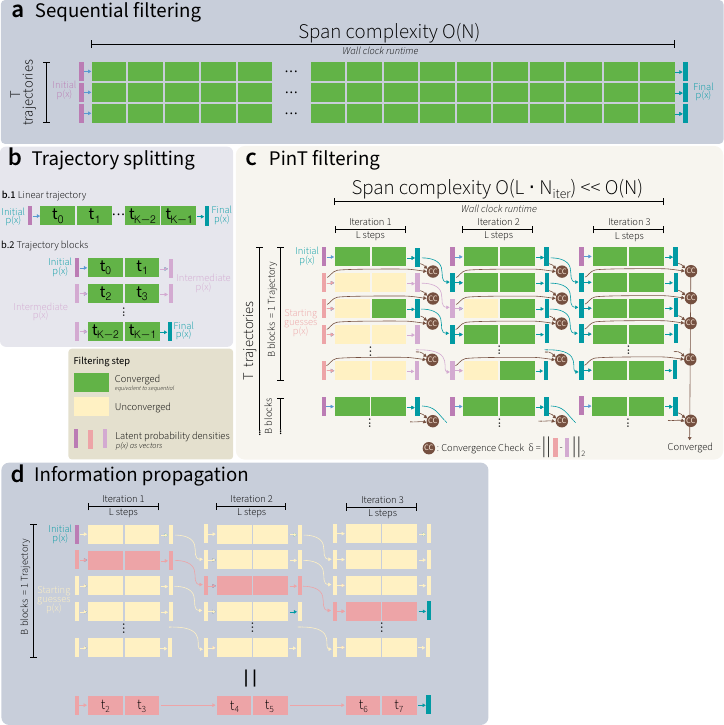}
 \captionof{figure}{\textbf{Sequential versus PinT filtering.}
 \label{fig:waveform_relax_explain}
 \textbf{a},~Sequential filtering processes each of $N_\mathrm{traj}$ trajectories as a chain of $N$ filtering steps; trajectories are independent and thus parallelizable, but within each every step depends on its predecessor, yielding a span complexity of $\mathcal{O}(N)$.
 \textbf{b},~Splitting each trajectory into $B = N/L$ contiguous blocks of length $L \ll N$, with the associated intermediate $p(x)$ densities.
 \textbf{c},~Parallel-in-Time filtering: at the first iteration, the leading block is initialized with the true initial density while all subsequent blocks receive a starting guess $p_{\mathrm{eq}}(x)$ (purple and teal tall rectangles) in place of the intermediate densities, which are not yet available.
 All blocks are then filtered concurrently.
 Between iterations, the terminal intermediate density of each block is passed forward as the initial condition of its successor (arrows), and a convergence check (CC) is applied at each block boundary.
 Blocks whose initial condition has converged (green) produce the same filtered densities as the sequential algorithm.
  \textbf{d}, Information propagation with iteration: each block has its history prolonged backward by the boundary density propagation, one block at every iteration, hence the relevance of the filter stability results.
 }

\medskip

Let us discuss informally some properties of this algorithm to contextualize the proofs stated in the following sections:

\paragraph{Remark (cost of speed-up).} We will show below that, for our particular model, \textsc{PintFilter} has $\mathcal{O}(1)$ span complexity: for a fixed block length $L$, its span does not grow with the trajectory length $N$, whereas that of the sequential filter is $\mathcal{O}(N)$.
On parallel hardware, this results in a shorter wall-clock runtime.
The speed-up comes at the cost of more total computation.
In each iteration, all $B$ blocks of $L$ steps are filtered, which costs as much as the whole sequential filter, so the total work is the number of iterations times the work of the sequential filter.

\paragraph{Remark (the algorithm always terminates).} After $B-1$ iterations, the true initial condition has propagated through all blocks, and the boundary densities are exactly those of the sequential filter (Corollary~S-9).
The increment at the next iteration is then zero, so the algorithm stops after at most $B$ iterations.
In this worst case, however, nothing is gained: the span is $B \cdot L = N$ steps, the same as that of the sequential filter, and the total work is $B$ times that of the sequential filter, so that $B-1$ full passes of computation are wasted.

\paragraph{Remark (trial densities).} In our implementation, the trial density $\bar\pi_0$ is the equilibrium density $p_\mathrm{eq}$ of the model, which is also the true initial density $\pi_0$.
Any other trial density is admissible.
The number of iterations to convergence would change for different trial densities, but not the converged result, since the fixed point is the sequential filter whatever $\bar\pi_0$ is (Theorem~S-12).
The $p_\mathrm{eq}$ density is a natural choice as the latent marginal, likely the best among trial densities that ignore the observations.

\paragraph{Remark (iterations as extending history).} The iterations can also be read as extending, for every block, the history it conditions on.
At iteration $j$, the initial condition of block $\beta$ is the result of filtering the $j$ preceding blocks, starting from the trial density (Lemma~S-7).
Each iteration adds one more block of past observations.
This continues until either the history reaches the true initial condition, or it is long enough for filter stability to make the density forget the misspecified trial density.
From that iteration on, the density at the block boundary changes only negligibly, whatever density it was started from, \ie convergence.

\paragraph{Remark (iterations as a Picard iteration).} The algorithm is also a Picard iteration for a fixed-point problem.
Collect the boundary densities of iteration $j$ into $P^{(j)} = (p^{(j)}_\beta)_\beta$.
One iteration, which filters all blocks in parallel and passes each terminal density on, is the map $P^{(j)} = F(P^{(j-1)})$, with $F$ the boundary-update map defined in Section~\ref{sec:algotheoremsection3}.
The iteration stops when the boundary densities no longer change, that is, at a fixed point of $F$, and this fixed point is the sequential filter (Theorem~S-12).

\paragraph{Remark (relation to other PinT algorithms).} The algorithm is somewhat similar to waveform relaxation~\cite{lelarasmee1982waveform}, with its splitting of the time domain and passing of boundary values between iterations.
It is also a special case of Parareal~\cite{lions2001resolution}.
Parareal updates each boundary value as
\[
  U^{(j)}_{\beta+1} \;=\; \mathcal{G}\bigl(U^{(j)}_\beta\bigr) + \mathcal{F}\bigl(U^{(j-1)}_\beta\bigr) - \mathcal{G}\bigl(U^{(j-1)}_\beta\bigr),
\]
with a fine propagator $\mathcal{F}$ and a coarse propagator $\mathcal{G}$.
With a coarse propagator ignoring its input and always returning $\bar\pi_0$, the two $\mathcal{G}$ terms cancel, and the update becomes $U^{(j)}_{\beta+1} = \mathcal{F}(U^{(j-1)}_\beta)$.
This is our iteration, with $\mathcal{F} = \textsc{BlockFilter}$.
Like Parareal, and as discussed above, it therefore converges exactly, in the worst case, once the initial condition has propagated through all blocks.

\paragraph{Remark (discretization error).} We want to prove that \textsc{PintFilter} reproduces the sequential filter as implemented, not that either of them is close to the exact, idealized filter.
To that end, both algorithms call the same implementation of \textsc{BlockFilter}, so they share the same discretization error, up to the stopping tolerance.
We do not attempt to bound the discretization error itself.

\subsection{Notation}

The following notation is used throughout the proof.

\paragraph{Our model.} As described in the Methods section:
\begin{itemize}\itemsep2pt
\item the latent domain is a cube $\mathcal{X} = [a,b]^n$, with a reflecting zero-flux
  boundary.
\item the free energy landscape $G \in W^{2,\infty}(\mathcal{X})$, continuously differentiable with bounded second derivative, is bounded with $p_{\mathrm{eq}} \propto
  e^{-G/k_BT}$ its equilibrium density, and its equilibrium measure $P_{\mathrm{eq}}(dy) =
  p_{\mathrm{eq}}(y)\,dy$.
\item $Q_t(x,dy)$ is the transition kernel of the latent dynamics over a time $t$, and $\kappa_t$
  its density with respect to $P_{\mathrm{eq}}$: $Q_t(x,dy) = \kappa_t(x,y)\,P_{\mathrm{eq}}(dy)$.
\item observations $Y_{t_1},\dots,Y_{t_N}$ are taken at the times $t_k = k\Delta t$, and $N$ is the
  trajectory length.
\item the latent state $x$ is observed through the emission $f_\theta : \mathcal{X} \to
  \mathcal{S}$ and the observation density $p_{\mathrm{obs}}(\cdot \mid s)$. The likelihood of the
  $k$-th observation, as a function of the latent state, is $\Psi_k(x) := p_{\mathrm{obs}}(Y_{t_k}
  \mid f_\theta(x))$.
\end{itemize}

\paragraph{Blocks and iterates.} As in the algorithms above:
\begin{itemize}\itemsep2pt
\item $L$ divides $N$, so each trajectory is split into $B = N/L$ blocks. Block $\beta$
  update based on the observations $Y_{\mathcal{I}_\beta}$ with indices $\mathcal{I}_\beta =
  \{\beta L+1,\dots,(\beta+1)L\}$.
\item $p^{(j)}_\beta$ is the density at the start of block $\beta$ after iteration $j$, \ie, the
  initial condition of that block in the next pass. $\bar\pi_0$ is the trial density used for
  every block $\beta \geq 1$ at iteration $0$.
\item $p^{\mathrm{seq}}_\beta$ is the sequential filter's density at the start of block $\beta$.
\item $\mathcal{P}(\mathcal{X})$ is the set of probability measures on $\mathcal{X}$. We identify a
  density $p$ with the measure $p(x)\,dx$, so the filter states $p^{(j)}_\beta$ and
  $p^{\mathrm{seq}}_\beta$ are elements of $\mathcal{P}(\mathcal{X})$.
\end{itemize}

\paragraph{Norms.}
\begin{itemize}\itemsep2pt
\item $\|\cdot\|$ is the total variation norm, normalized as in Le~Gland--Oudjane, \ie,
  $\|\mu - \mu'\| \leq 2$ for probability measures. We write $\|\cdot\|_{\mathrm{TV}}$ when confusion is possible.
\item $\|\cdot\|_{\mathrm{WRMS}}$ is the weighted root-mean-square norm that the implementation
  applies to the coefficient vectors $\mathbf{c}[p] \in \mathbb{R}^{N_b}$ of densities. Both are
  defined in Section~\ref{sec:algotheoremsection3}.
\end{itemize}

\subsection{Assumptions}

We assume the following for the proof.

\paragraph{(A1) Nondegenerate channel.}
For every $s \in \mathcal{S}$, $p_{\mathrm{obs}}(\cdot \mid s)$ admits a Lebesgue density on
$\mathcal{Y}$; hence $\Psi_k$ is a well-defined non-negative measurable function on
$\mathcal{X}$.

\paragraph{(A2) Non-vanishing, integrable likelihood.}
For the observed record, $0 < \int_\mathcal{X} \Psi_k\,dP_{\mathrm{eq}} < \infty$ for every $k$.

\medskip

\paragraph{(A3) Source of mixing.} One of the following assumptions, (A3a) or (A3b), must hold, to provide mixing condition \eqref{eq:E}.
(A3a) is the default route: it holds for every $\Delta t$ and gives the best possible constant
in~\eqref{eq:E}, but only as the minimum and maximum of the transition density, which are not known
in closed form. (A3b) applies in the sparse-sampling regime, where it gives a constructive lower
bound on that constant, computable from the spectral gap $\lambda_1$ and the diagonal values of the
transition density. (Further
variants used to establish~\eqref{eq:E} could be added without modifying the proof apart from
Section~\ref{sec:algotheoremsection1}.)

\paragraph{(A3a) Compact positivity.}
The transition density $\kappa_{\Delta t}$ is continuous on $\mathcal{X}\times\mathcal{X}$, strictly
positive, and satisfies
\[
  \int \kappa_{\Delta t}(x,y)\,P_{\mathrm{eq}}(dy) = 1 \qquad\text{for every } x.
\]

\paragraph{(A3b) Spectral gap, with sparse sampling.}
For every $t > 0$, the transition density $\kappa_t$ is continuous on
$\mathcal{X}\times\mathcal{X}$ and symmetric, $\kappa_t(x,y) = \kappa_t(y,x)$
(reversibility). Moreover, the dynamics has a spectral gap $\lambda_1 > 0$: for every $t > 0$
and every $f \in L^2(P_{\mathrm{eq}})$ with $\int f\,dP_{\mathrm{eq}} = 0$,
\[
  \|P_t f\|_{L^2(P_{\mathrm{eq}})} \;\leq\; e^{-\lambda_1 t}\,\|f\|_{L^2(P_{\mathrm{eq}})},
  \qquad P_t f(x) := \int_\mathcal{X}\kappa_t(x,y)\,f(y)\,P_{\mathrm{eq}}(dy).
\]
Finally, the sampling is sparse enough: for some $t_0 \in (0, \Delta t/2)$,
\[
  e^{-\lambda_1(\Delta t - 2t_0)}\,\sup_x\bigl(\kappa_{2t_0}(x,x) - 1\bigr) \;<\; 1 .
\]

\paragraph{(A4) Finite-dimensional representation.}
The states of the filter are densities in a finite-dimensional subspace $V \subset
L^1(\mathcal{X})$ with a basis $e_1,\dots,e_{N_b}$, and $\mathbf{c} : V \to \mathbb{R}^{N_b}$ is
the coordinate map in this basis, $p = \sum_i c_i[p]\,e_i$. In particular, two states with
equal coefficient vectors are equal. We write
\[
  M_e \;:=\; \sup_{0 \neq p \in V}\, \frac{\max_i |c_i[p]|}{\|p\|_{\mathrm{TV}}}
\]
for the norm of $\mathbf{c}$ from the total variation norm to the maximum norm. It is finite
because $V$ is finite-dimensional, and $\max_i |c_i[p]| \leq M_e\,\|p\|_{\mathrm{TV}}$ for every
$p \in V$.

\paragraph{Remark (on (A2)).}
The integral in (A2) is the equilibrium observation marginal at the observed value,
\[
  \int_\mathcal{X}\Psi_k\,dP_{\mathrm{eq}}
  \;=\; \int_\mathcal{X} p_{\mathrm{obs}}(Y_{t_k} \mid f_\theta(x))\,p_{\mathrm{eq}}(x)\,dx
  \;=\; p_Y(Y_{t_k}),
\]
so (A2) says that $p_Y(Y_{t_k}) \in (0,\infty)$. When the data are generated by the model
itself, at its true parameters, this holds almost surely given (A1). The latent chain starts at
$\pi_0 = p_{\mathrm{eq}}$ and is stationary, so each $Y_{t_k}$ has density $p_Y$, and a density
is almost surely positive and finite under its own law. Since there are countably many $k$, (A2)
then holds for almost every observation record, and so does Proposition~P1.

If the model is misspecified, or at parameters $\theta$ far from the bulk of the posterior, as in
the early phases of inference, (A2) is a genuine assumption. It holds for every $\theta$ when
$p_{\mathrm{obs}} > 0$ everywhere, as with Gaussian observation noise, since then $\Psi_k > 0$.
When $p_{\mathrm{obs}}(\cdot \mid s)$ has bounded support, by contrast, an observed $Y_{t_k}$ can
fall outside the support of every emission distribution $p_{\mathrm{obs}}(\cdot \mid
f_\theta(x))$. Then $\Psi_k \equiv 0$, the normalizer $Z$ vanishes, the log-likelihood is
$-\infty$, and the update step is undefined. A practical remedy is to implement the observation
model as a technical mixture of $p_{\mathrm{obs}}$ with a uniform component of weight $1\%$, which
keeps $\Psi_k > 0$, and hence (A2) for every $\theta$. One then checks \textit{a posteriori} on
the inferred posterior whether some observations are explained only by the uniform component.

\paragraph{Remark (on (A3)).}
(A3) is the most stringent assumption on the model. In either variant, it makes the propagator
mix uniformly, through compactness and positivity in (A3a) and through a spectral gap in (A3b).
This is what gives the algorithm its $\mathcal{O}(1)$ span complexity.
In practice, if $\Delta t$ is longer than the relaxation timescale $1/\lambda_1$ (sparse-observation
regime), then (A3b) is satisfied, and it gives an explicit value for the constant in~\eqref{eq:E},
hence an explicit bound on the iteration count. Under (A3a) alone, the constant exists but its value
is not known, so the iteration count is shown to be finite and independent of $N$, without an
explicit bound. When sampling is fast, the constant is moreover small, whichever route is used. The
resulting bound is then still sufficient for the span complexity result, but far above the iteration
counts observed in practice, so the proof gives no insight into the actual mechanism of
convergence.

\subsection{Statement of the result (Proposition P1)}

\begin{restatable}[Constant span complexity]{propPone}{PoneStatement}
Assume (A1)--(A4), and let $\varepsilon \in (0,1)$ satisfy~\eqref{eq:E}, as provided by
Section~\ref{sec:algotheoremsection1}. Run \textsc{PintFilter} with the WRMS stopping rule. Then
the following hold.
\begin{enumerate}\itemsep2pt
\item \emph{Equivalence.} The iteration has a unique fixed point, the sequential filter: at it,
  every filtered density and the total log-likelihood agree exactly with those of
  \textsc{SequentialFilter} (Theorem~S-12), and an exactly zero increment verifies that it has
  been reached (Corollary~S-13).
\item \emph{Convergence, at an $N$-independent rate.} At iteration $j$, the total-variation error
  between the densities of \textsc{PintFilter} and \textsc{SequentialFilter},
  $E^{(j)}_\beta := \|p^{(j)}_\beta - p^{\mathrm{seq}}_\beta\|_{\mathrm{TV}}$, satisfies
  $\max_\beta E^{(j)}_\beta \leq \min\{2, C_0\tau^{jL}\}$, where $\tau < 1$ and $C_0$ depend only
  on $\varepsilon$ (Proposition~S-10). Hence any target accuracy
  $\delta_{\mathrm{TV}}$ is reached within $N_{\mathrm{iter}}(\delta_{\mathrm{TV}})$ iterations,
  which depends only on $(\varepsilon, L, \delta_{\mathrm{TV}})$ (Theorem~S-11).
\item \emph{Span complexity.} The WRMS rule stops the algorithm within
  $N_{\mathrm{stop}}$ iterations, where $N_{\mathrm{stop}}$ depends only on $(\varepsilon, L,
  M_e, \mathrm{atol})$ (Corollary~S-17). The span complexity is therefore
  $\mathcal{O}(N_{\mathrm{stop}}\cdot L)$, independent of $N$.
\item \emph{A posteriori certificate.} At iteration $j$, the total-variation error, which the
  program cannot observe, is bounded by the WRMS increment it computes: $\max_\beta E^{(j)}_\beta \leq
  A\,\Gamma\,\delta^{(j)}$, with constants $A = A(\varepsilon, L)$ and $\Gamma$, which depends
  only on the basis and the tolerances (Proposition~S-15, Lemma~S-16). In particular, when the
  rule $\delta^{(j)} \leq 1$ stops the algorithm, this error is at most $A\,\Gamma$,
  independently of $N$.
\end{enumerate}
\end{restatable}

\begin{leanbox}[Lean: Proposition P1 --- \texttt{WaveformRelax.converges\_to\_sequential}]
Top-level proposition in \verb|PlanxAlgorithm/Relaxation.lean|. 

The names in the Lean statement are mapped to the human-readable proof as follows:
\begin{itemize}\itemsep1pt
\item \texttt{S} is the set of states, here the probability measures on $\mathcal{X}$, and
  \texttt{dist} is the total variation distance on it.
\item \texttt{$\Phi$ b} is the block map $\Phi_\beta$ of block $\beta$ = \texttt{b}; \texttt{$\pi_0$}
  and \texttt{$\pi_0$'} are the prior $\pi_0$ and the trial density $\bar\pi_0$.
\item \texttt{iter $\Phi$ $\pi_0$ $\pi_0$' n b} is $p^{(j)}_\beta$, with $j$ = \texttt{n};
  \texttt{seq $\Phi$ $\pi_0$ b} is $p^{\mathrm{seq}}_\beta$; \texttt{IsFixedPoint $\Phi$ $\pi_0$ P}
  says that $P$ is a fixed point of the boundary-update map; \texttt{iterCount} is
  $N_{\mathrm{iter}}$.
\item \texttt{L} is the block length, and the hypothesis \texttt{hL : 0 < L} says that blocks are
  not empty. It is used only to obtain $\tau^L < 1$, for the iteration count and for the geometric
  series in the a posteriori bound.
\item \texttt{F : BlockForgetting $\Phi$ L} is the hypothesis that the block maps satisfy the
  contraction bounds~\eqref{eq:Fabs} and~\eqref{eq:Flip}, the output of
  Section~\ref{sec:algotheoremsection2}. It bundles the constants \texttt{F.$\tau$},
  \texttt{F.C$_0$} and \texttt{F.C$_1$}, which are $\tau$, $C_0$ and $C_1$, together with the two
  bounds. Despite the letter, it is unrelated to the boundary-update map $F$ of the text.
\end{itemize}
\begin{leancode}
theorem converges_to_sequential {L : ℕ} (F : BlockForgetting Φ L) (hL : 0 < L) :
    (∀ P : ℕ → S, IsFixedPoint Φ π₀ P ↔ P = seq Φ π₀) ∧
    (∀ n b, dist (iter Φ π₀ π₀' n b) (seq Φ π₀ b) ≤ F.C₀ * F.τ ^ (n * L)) ∧
    (∀ δ : ℝ, 0 < δ → ∀ n, iterCount F.C₀ F.τ L δ ≤ n →
      ∀ b, dist (iter Φ π₀ π₀' n b) (seq Φ π₀ b) ≤ δ) ∧
    (∀ d : ℕ → ℝ,
      (∀ m b, 1 ≤ b →
        dist (iter Φ π₀ π₀' (m + 1) b) (iter Φ π₀ π₀' m b) ≤ d m) →
      ∀ m b, dist (iter Φ π₀ π₀' (m + 1) b) (seq Φ π₀ b)
        ≤ F.C₁ * F.τ ^ (L - 1) * d m / (1 - F.τ ^ L))
\end{leancode}
The proposition statement is four parts (lowest identation above) joined by $\wedge$:
\begin{enumerate}\itemsep1pt
\item A family of boundary densities is a fixed point exactly when it is the sequential filter, \ie
  the first half of clause~1 (Theorem~S-12). This part uses neither \texttt{F} nor \texttt{hL} as
  hypothesis, so holds for any block map $\Phi$.
\item For every iteration and every block, the convergence bound of clause~2  (Proposition~S-10).
\item The iteration count of clause~2: from iteration \texttt{iterCount} on, every block is
  within $\delta$ of the sequential filter (Theorem~S-11).
\item The total-variation form of the a posteriori bound (Proposition~S-15): if \texttt{d m}
  bounds the increments between iterations \texttt{m} and \texttt{m + 1}, then the error at
  iteration \texttt{m + 1} is at most $A$ times \texttt{d m}, with $A = C_1\tau^{L-1}/(1-\tau^L)$.
\end{enumerate}
The hypotheses of the theorem are therefore exactly \texttt{F} and \texttt{hL}. 

Clause~3 of P1 and the WRMS form of clause~4 are proved separately.
The stopping iteration is
\texttt{exists\_stop} (Corollary~S-17). The WRMS form follows from part~4 together with
\texttt{dist\_le\_wrms} (Lemma~S-16). 
The bound $|c_i[\mu]| \leq M_e$ used to obtain $\Gamma$ is not formalized. 
We also do not formalize the notion of span complexity, instead simply show the iteration count
does not take as input the trajectory length directly or indirectly, rather just the contraction
data and the tolerance.
\end{leanbox}

\paragraph{Remark (log-likelihood).}
By clause~1 of P1, the computed log-likelihood $\sum_\beta \ell_\beta$ is exact at the fixed
point. 
When the algorithm stops instead on $\delta^{(j)} \leq 1$ , each $\ell_\beta$
has a small error from its initial density, adding up over the $B$ blocks.
The total error is of order $C(\varepsilon)\,B\,\max_\beta E_\beta$, growing linearly with
$N$. 
We do not prove this bound here.
Note that the sequential filter also has extensive-in-$N$ error: its
log-likelihood is a sum of $N$ terms $\log Z_k$, each carrying a rounding error of the order of the
unit roundoff $u$, so its total error is of order $N u$. 
With the stopping tolerances at the floating-point floor, the error of \textsc{PintFilter} is of order $B u \leq N u$, 
the same order in $N$ up to $C(\varepsilon)$.
 Relative to $|\log p|$, which grows linearly in $N$, both errors
are of order $u$ and independent of $N$. 
We report on this relative error in our benchmarks in
the main text.

\subsection{Structure of the proof}

\begin{itemize}\itemsep2pt
\item \emph{Section~\ref{sec:algotheoremsection1}: Mixing property.} From (A3), in whichever
  variant is assumed, we derive the envelope condition~\eqref{eq:E} below.
\item \emph{Section~\ref{sec:algotheoremsection2}: From mixing to block forgetting.}
  From~\eqref{eq:E}, (A1) and (A2), with Le~Gland--Oudjane's Hilbert-metric theory of filter
  stability~\cite{legland2004stability}, we derive the contraction bounds \eqref{eq:Fabs}
  and \eqref{eq:Flip} on the block maps $\Phi_\beta$, with constants depending only on
  $(\varepsilon, L)$, and not on $N$.
\item \emph{Section~\ref{sec:algotheoremsection3}: Algorithmic arguments.} From these two
  bounds and (A4), without further measure theory, we show Proposition~P1.
\end{itemize}

\bigskip

The only property of the model that Section~\ref{sec:algotheoremsection1} passes to the following section
is the \emph{envelope condition} (\ref{eq:E}): a constant $\varepsilon > 0$ with
\begin{equation}\label{eq:E}
  \varepsilon \;\leq\; \kappa_{\Delta t}(x,y) \;\leq\; \varepsilon^{-1}
  \qquad\text{for all } x,y \in \mathcal{X}.
  \tag{E}
\end{equation}
Such an $\varepsilon$ is at most $1$, since $\varepsilon \leq \varepsilon^{-1}$. A different model
that also satisfies~\eqref{eq:E}, by whatever argument, can use Sections~\ref{sec:algotheoremsection2}
and~\ref{sec:algotheoremsection3} as-is.

\begin{leanbox}[Lean: condition (E)]
In \verb|PlanxAlgorithm/Envelope.lean|. 

The conditions on $\varepsilon$ are grouped in the structure
\verb|IsEnvelope|, in the form used by the Le~Gland--Oudjane mixing definition.
\begin{leancode}
structure IsEnvelope (ε : ℝ) (κ : X → X → ℝ) : Prop where
  eps_pos    : 0 < ε
  eps_le_one : ε ≤ 1
  lower      : ∀ x y, ε ≤ κ x y
  upper      : ∀ x y, κ x y ≤ ε⁻¹
\end{leancode}
The field \texttt{eps\_le\_one} follows from the two bounds when \texttt{X} is non-empty, 
kept as a separate property for convenience.
\end{leanbox}

\subsection{Mixing property}\label{sec:algotheoremsection1}

This section derives the envelope condition~\eqref{eq:E} from assumption (A3). Only one of the
two subsections below is needed: \S\ref{sec:compactnessroute} if (A3a) is assumed, and
\S\ref{sec:spectralroute} if (A3b) is assumed. The observations play no role here.

\subsubsection{From compactness alone}\label{sec:compactnessroute}

This route assumes (A3a). We argue that a continuous, strictly positive function on a
compact set has a positive minimum and a finite maximum, and these two numbers give the two sides
of~\eqref{eq:E}.

\begin{propAtwoprime}[Unconditional envelope]
Under (A3a), \eqref{eq:E} holds for every $\Delta t > 0$ with
\[
  \varepsilon_0 \;:=\; \min(m, 1/M),
  \qquad
  m \;:=\; \min_{x,y}\kappa_{\Delta t}(x,y),
  \qquad
  M \;:=\; \max_{x,y}\kappa_{\Delta t}(x,y).
\]
All three depend on $\Delta t$, and we write $\varepsilon_0(\Delta t)$ when relevant.
\end{propAtwoprime}

\begin{proof}
Since $\kappa_{\Delta t}$ is continuous on the compact set $\mathcal{X}\times\mathcal{X}$, the
minimum $m$ and the maximum $M$ are attained, and $m > 0$ by strict positivity. Hence
$\varepsilon_0 > 0$. For all $x,y$,
\[
  \varepsilon_0 \;\leq\; m \;\leq\; \kappa_{\Delta t}(x,y) \;\leq\; M \;\leq\; \varepsilon_0^{-1},
\]
the first inequality because $\varepsilon_0 \leq m$, and the last because $\varepsilon_0 \leq 1/M$.
This is~\eqref{eq:E} with $\varepsilon = \varepsilon_0$.
\end{proof}

\paragraph{Optimality.} The constant $\varepsilon_0$ is the best $\varepsilon$ constant, as a given $\varepsilon > 0$ satisfies~\eqref{eq:E}
exactly when $\varepsilon \leq m$ and $\varepsilon \leq 1/M$.

\paragraph{Limit behavior.} $\varepsilon_0$ behaves as expected in $\Delta t$.
 With $\Delta t \to 0$, the transition density concentrates near the diagonal, so $m \to 0$, $M \to \infty$ and $\varepsilon_0 \to 0$. 
With   $\Delta t \to \infty$, an ergodic process forgets its starting point, $\kappa_{\Delta t} \to 1$, and
$\varepsilon_0 \to 1$. 
The issue of this route is therefore not the size of $\varepsilon_0$, but
that its value is not computable in closed form, contrary to the spectral gap route below.

\medskip

\begin{leanbox}[Lean: Proposition S-2$'$ --- \texttt{KernelData.isEnvelope}]
In \verb|PlanxAlgorithm/Kernel.lean|.
\begin{leancode}
structure KernelData (X : Type*) [TopologicalSpace X] [CompactSpace X]
    [MeasurableSpace X] [OpensMeasurableSpace X] (μ : Measure X)
    [IsProbabilityMeasure μ] where
  kappa  : X → X → ℝ
  cont   : Continuous fun p : X × X => kappa p.1 p.2
  pos    : ∀ x y, 0 < kappa x y
  markov : ∀ x, ∫ y, kappa x y ∂μ = 1

theorem KernelData.isEnvelope [Nonempty X] (K : KernelData X μ) :
    IsEnvelope K.eps K.kappa
\end{leancode}
Here \verb|K.eps| is the constant $\min(m, 1/M)$, defined from the attained extrema. The
hypothesis \verb|[Nonempty X]| ensures these extrema exist. The field \verb|markov|, the mass
condition of (A3a), is used only to prove the field \verb|eps_le_one| of \verb|IsEnvelope|, which
the text instead obtains from $\varepsilon \leq \varepsilon^{-1}$.
\end{leanbox}

\subsubsection{From the spectral gap}\label{sec:spectralroute}

This route assumes (A3b). For our model, the
dynamics is reversible with respect to $p_{\mathrm{eq}}$ by construction, and its generator has a
compact resolvent, thus a discrete spectrum and a spectral gap. 
Whether the sampling is sparse
enough depends on $\Delta t$, which, in the notation introduced below, corresponds to the condition
$\delta(\Delta t) < 1$.
Since the $\kappa_t$ are the transition densities of a Markov process
with invariant measure $P_{\mathrm{eq}}$, they satisfy, for all $s, t > 0$ and $x, y \in
\mathcal{X}$,
\begin{equation}\label{eq:markov}\tag{CK}
  \int_\mathcal{X}\kappa_t(x,y)\,P_{\mathrm{eq}}(dy) \;=\; 1,
  \qquad
  \kappa_{s+t}(x,y) \;=\; \int_\mathcal{X}\kappa_s(x,u)\,\kappa_t(u,y)\,P_{\mathrm{eq}}(du),
\end{equation}
the second identity being the Chapman--Kolmogorov equation. We define the heat-kernel constant
and the envelope
\[
  \Theta(t) \;:=\; \sup_{x}\,\kappa_t(x,x) - 1,
  \qquad
  \delta(\Delta t) \;:=\; \inf_{0 < t_0 < \Delta t/2}\;
    e^{-\lambda_1(\Delta t - 2t_0)}\,\Theta(2t_0).
\]
$\Theta(t)$ is finite because $\kappa_t$ is continuous on the compact set
$\mathcal{X}\times\mathcal{X}$. Heuristically, $\Theta(t)$ measures how far the process, started
at the worst possible point, still is from equilibrium after time $t$, and $\delta(\Delta t)$
bounds how much of its starting point one step of the dynamics still remembers.

\begin{propAone}[Equilibrium envelope]
Under (A3b), $\sup_{x,y}|\kappa_{\Delta t}(x,y) - 1| \leq \delta(\Delta t)$.
\end{propAone}

\begin{proof}
Fix $t_0 \in (0, \Delta t/2)$ and $x, y \in \mathcal{X}$, and write $s := \Delta t - 2t_0$,
which is thus positive.

The spectral gap controls functions in $L^2(P_{\mathrm{eq}})$, but $\kappa_{\Delta t}(x,y)$ is a
pointwise quantity describing the process started exactly at $x$, a point mass not in $L^2$. 
We therefore split the time $\Delta t$ into three segments, $t_0 + s + t_0$.
During the first part, the process started at $x$ spreads into the density $\kappa_{t_0}(x,\cdot)$,
a bounded function and hence in $L^2$.
By reversibility, the same holds at the other end for $y$.
We then apply the spectral gap property to the middle segment, at the cost of a  $\Theta(2t_0)$ factor
and the loss of $2t_0$ in the exponent.

\emph{Step 1: minorization.} For $z \in \mathcal{X}$, let $g_z(u) := \kappa_{t_0}(z,u) - 1$.
By the first identity in~\eqref{eq:markov}, we have
$\int g_z\,dP_{\mathrm{eq}} = 0$. 
We show its norm to be controlled by $\Theta$, by expanding the square:
\begin{align*}
  \|g_z\|^2_{L^2(P_{\mathrm{eq}})}
  &= \int_\mathcal{X}\bigl(\kappa_{t_0}(z,u) - 1\bigr)^2\,P_{\mathrm{eq}}(du) \\
  &= \int_\mathcal{X}\kappa_{t_0}(z,u)^2\,P_{\mathrm{eq}}(du)
     - 2\int_\mathcal{X}\kappa_{t_0}(z,u)\,P_{\mathrm{eq}}(du) + 1
    && \\
  &= \int_\mathcal{X}\kappa_{t_0}(z,u)^2\,P_{\mathrm{eq}}(du) - 1
    && \text{by~\eqref{eq:markov}} \\
  &= \int_\mathcal{X}\kappa_{t_0}(z,u)\,\kappa_{t_0}(u,z)\,P_{\mathrm{eq}}(du) - 1
    && \text{by symmetry} \\
  &= \kappa_{2t_0}(z,z) - 1
    && \text{by~\eqref{eq:markov} with } s = t = t_0 \\
  &\leq \Theta(2t_0).
    &&
\end{align*}

\emph{Step 2: factorization.} We now write $\kappa_{\Delta t}(x,y) - 1$ in terms of $g_x$ and
$g_y$. We split twice  $\Delta t = t_0 + (s +
t_0)$ then  $s + t_0$  using~\eqref{eq:markov},
\begin{align*}
  \kappa_{\Delta t}(x,y)
  &= \int_\mathcal{X}\kappa_{t_0}(x,u)\,\kappa_{s+t_0}(u,y)\,P_{\mathrm{eq}}(du) \\
  &= \int_\mathcal{X}\int_\mathcal{X}
    \kappa_{t_0}(x,u)\,\kappa_s(u,v)\,\kappa_{t_0}(v,y)\,
    P_{\mathrm{eq}}(du)\,P_{\mathrm{eq}}(dv) .
\end{align*}
Substituting $\kappa_{t_0}(x,u) = 1 + g_x(u)$ and, by symmetry, $\kappa_{t_0}(v,y) =
\kappa_{t_0}(y,v) = 1 + g_y(v)$. Expanding the product gives four terms,
\[
  \kappa_{\Delta t}(x,y) \;=\; T_0 + T_x + T_y + T_{xy},
\]
where
\begin{align*}
  T_0 &= \int_\mathcal{X}\int_\mathcal{X} \kappa_s(u,v)\,
    P_{\mathrm{eq}}(du)\,P_{\mathrm{eq}}(dv), \\
  T_x &= \int_\mathcal{X}\int_\mathcal{X} g_x(u)\,\kappa_s(u,v)\,
    P_{\mathrm{eq}}(du)\,P_{\mathrm{eq}}(dv), \\
  T_y &= \int_\mathcal{X}\int_\mathcal{X} \kappa_s(u,v)\,g_y(v)\,
    P_{\mathrm{eq}}(du)\,P_{\mathrm{eq}}(dv), \\
  T_{xy} &= \int_\mathcal{X}\int_\mathcal{X} g_x(u)\,\kappa_s(u,v)\,g_y(v)\,
    P_{\mathrm{eq}}(du)\,P_{\mathrm{eq}}(dv).
\end{align*}
All functions involved are bounded, so by Fubini we evaluate one variable at a time:
\begin{itemize}\itemsep2pt
\item $T_0 = 1$. Integrating over $v$, $\int\kappa_s(u,v)\,P_{\mathrm{eq}}(dv) = 1$ for
  every $u$, by~\eqref{eq:markov}.  The remaining integral over $u$ is then
  $P_{\mathrm{eq}}(\mathcal{X}) = 1$.
\item $T_x = 0$. Integrating over $v$ first again gives $1$. Remaining is $\int
  g_x\,dP_{\mathrm{eq}} = 0$, by Step~1.
\item $T_y = 0$. Integrating over $u$ first, $\int\kappa_s(u,v)\,P_{\mathrm{eq}}(du) = \int
  \kappa_s(v,u)\,P_{\mathrm{eq}}(du) = 1$, by symmetry and~\eqref{eq:markov}. Remaining is
  $\int g_y\,dP_{\mathrm{eq}} = 0$.
\item Finally, for $T_{xy}$, we first integrate over $v$ giving $\int\kappa_s(u,v)\,
  g_y(v)\,P_{\mathrm{eq}}(dv) = P_s g_y(u)$, so
  \[
    T_{xy} \;=\; \int_\mathcal{X} g_x(u)\,P_s g_y(u)\,P_{\mathrm{eq}}(du)
    \;=\; \bigl\langle g_x,\, P_s g_y \bigr\rangle_{L^2(P_{\mathrm{eq}})} .
  \]
\end{itemize}

Reassembling, we get
\[
  \kappa_{\Delta t}(x,y) - 1 \;=\; \bigl\langle g_x,\, P_s g_y \bigr\rangle_{L^2(P_{\mathrm{eq}})} .
\]

\emph{Step 3: spectral gap.} Since $g_y$ has mean zero, the spectral gap of (A3b) gives
$\|P_s g_y\| \leq e^{-\lambda_1 s}\|g_y\|$. We get
\begin{align*}
  |\kappa_{\Delta t}(x,y) - 1|
  &\leq \|g_x\|\,\|P_s g_y\|
    && \text{by Cauchy--Schwarz} \\
  &\leq e^{-\lambda_1(\Delta t - 2t_0)}\,\|g_x\|\,\|g_y\|
    && \text{by the spectral gap} \\
  &\leq e^{-\lambda_1(\Delta t - 2t_0)}\,\Theta(2t_0)
    && \text{by Step~1.}
\end{align*}
This holds for every $t_0 \in (0,\Delta t/2)$ and all $x, y$. 
As the left-hand side does not depend on $t_0$, we can take the infimum over $t_0$ on the right, which is $\delta(\Delta t)$, and
then the supremum over $(x,y)$ on the left, giving $\sup_{x,y}|\kappa_{\Delta t}(x,y) - 1| \leq \delta(\Delta t)$.
\end{proof}

\begin{leanbox}[Lean: Proposition S-1 -- \texttt{HeatData.abs\_kappa\_sub\_one\_le}]
In \verb|PlanxAlgorithm/Envelope.lean|. 
We formalize a special case of (A3b) as the structure \verb|HeatData|.
We notably assumes an eigendecomposition with a discrete spectrum
$0 = \lambda_0 < \lambda_1 \leq \lambda_2 \leq \cdots$, eigenfunctions $\phi_i$ with $\phi_0
\equiv 1$, and a pointwise expansion $\kappa_t(x,y) = \sum_i e^{-\lambda_i t}\phi_i(x)\phi_i(y)$
whose diagonal tails are bounded by $\Theta(t)$. 
Such an expansion implies (A3b), with the same
$\lambda_1$ and $\Theta$. 
(There is not spectral theorem for unbounded operator in Mathlib, which is why we
have to assume the eigendecomposition as member of the structure, rather than obtain it from the operator)
\begin{leancode}
structure HeatData (X : Type*) where
  lam   : ℕ → ℝ                       -- the eigenvalues
  phi   : ℕ → X → ℝ                   -- the eigenfunctions
  kappa : ℝ → X → X → ℝ               -- the ratio kernel
  theta : ℝ → ℝ                       -- the heat-kernel constant
  lam_zero  : lam 0 = 0
  phi_zero  : ∀ x, phi 0 x = 1
  lam_mono  : Monotone lam
  gap       : 0 < lam 1
  hasSum_kappa : ∀ t, 0 < t → ∀ x y,
    HasSum (fun i => Real.exp (-lam i * t) * (phi i x * phi i y)) (kappa t x y)
  diagSummable : ∀ t, 0 < t → ∀ x,
    Summable fun i => Real.exp (-lam (i + 1) * t) * phi (i + 1) x ^ 2
  diag_le_theta : ∀ t, 0 < t → ∀ x,
    ∑' i, Real.exp (-lam (i + 1) * t) * phi (i + 1) x ^ 2 ≤ theta t

theorem HeatData.abs_kappa_sub_one_le (hΔ : 0 < Δt) (ht₀ : 0 < t₀) (h : 2 * t₀ < Δt)
    (x y : X) :
    |H.kappa Δt x y - 1| ≤ Real.exp (-H.lam 1 * (Δt - 2 * t₀)) * H.theta (2 * t₀)
\end{leancode}
The Lean statement is expressed for every admissible $t_0$, with right-hand side $B(t_0) :=
e^{-\lambda_1(\Delta t - 2t_0)}\,\Theta(2t_0)$. 
Since the left-hand side does not depend on $t_0$, the infinum step  $\delta(\Delta t) = \inf_{t_0} B(t_0)$ follows immediately,
but is not formalized.
In the next proposition~S-2, we give the envelope  $\varepsilon = 1 - B(t_0)$  which can be made arbitrarily
close to $1 - \delta(\Delta t)$ but need not equal it if the infimum is not attained. 
Since every later proposition needs only $\varepsilon > 0$, it is fine to not formalize this last step.
\end{leanbox}

\begin{propAtwo}[Envelope $\Rightarrow$ mixing]
Given $\delta(\Delta t) < 1$ by A3b, the
bound~\eqref{eq:E} holds with $\varepsilon := 1 - \delta(\Delta t)$.
 Consequently, for every $x$ and every Borel set $U$,
\[
  \varepsilon\,P_{\mathrm{eq}}(U) \;\leq\; Q_{\Delta t}(x,U) \;\leq\;
  \varepsilon^{-1}P_{\mathrm{eq}}(U),
\]
\ie{} $Q_{\Delta t}$ is mixing in the sense of (Le~Gland--Oudjane, 2004)\cite{legland2004stability} with reference
$P_{\mathrm{eq}}$ and constant $\varepsilon$.
\end{propAtwo}

\begin{proof}
Write $\delta = \delta(\Delta t)$. Proposition~S-1 gives, for all $x,y$,
\[
  1 - \delta \;\leq\; \kappa_{\Delta t}(x,y) \;\leq\; 1 + \delta .
\]
The lower bound is $\varepsilon$ by definition. For the upper bound, since $0 \leq \delta < 1$
we have $(1-\delta)(1+\delta) = 1 - \delta^2 \leq 1$, hence
\[
  1 + \delta \;\leq\; \frac{1}{1-\delta} \;=\; \varepsilon^{-1}.
\]
Together these give~\eqref{eq:E}, with $\varepsilon > 0$ because $\delta < 1$.

For the measure form, fix $x$ and a Borel set $U$. Since $\kappa_{\Delta t}$ is the density of $Q_{\Delta t}$ with respect to $P_{\mathrm{eq}}$,
$Q_{\Delta t}(x,U) = \int_U \kappa_{\Delta t}(x,y)\,P_{\mathrm{eq}}(dy)$. Integrating the
two-sided bound~\eqref{eq:E} over $U$ against $P_{\mathrm{eq}}$ gives
\[
  \varepsilon\,P_{\mathrm{eq}}(U)
  \;\leq\; \int_U \kappa_{\Delta t}(x,y)\,P_{\mathrm{eq}}(dy)
  \;\leq\; \varepsilon^{-1}P_{\mathrm{eq}}(U).
\]
Note that neither bound depends on $x$, because~\eqref{eq:E} is uniform in both arguments.
\end{proof}

\begin{leanbox}[Lean: Proposition S-2, in two pieces]
The algebraic part is in \verb|PlanxAlgorithm/Envelope.lean| and the measure part in \verb|PlanxAlgorithm/Kernel.lean|.
In the code, the set \texttt{A} is the Borel set $U$ of the text.
\begin{leancode}
theorem envelope_of_abs_sub_one_le {δ κ : ℝ} (hδ : δ < 1) (h : |κ - 1| ≤ δ) :
    1 - δ ≤ κ ∧ κ ≤ (1 - δ)⁻¹

theorem HeatData.isEnvelope [Nonempty X] (hΔ : 0 < Δt) (ht₀ : 0 < t₀) (h : 2 * t₀ < Δt)
    (hδ : Real.exp (-H.lam 1 * (Δt - 2 * t₀)) * H.theta (2 * t₀) < 1) :
    IsEnvelope (1 - Real.exp (-H.lam 1 * (Δt - 2 * t₀)) * H.theta (2 * t₀))
      (H.kappa Δt)

theorem setIntegral_le_of_isEnvelope (h : IsEnvelope ε κ)
    (hκ : ∀ x, Integrable (κ x) μ) (x : X) {A : Set X} (hA : MeasurableSet A) :
    ε * μ.real A ≤ ∫ y in A, κ x y ∂μ ∧ ∫ y in A, κ x y ∂μ ≤ ε⁻¹ * μ.real A
\end{leancode}

General condition \verb|IsEnvelope| is shared by both route  to~\eqref{eq:E}.

The Lean statement uses real-valued integrals and measures instead of default measures: \texttt{\(\mu\).real A} is
$P_{\mathrm{eq}}(U)$ as a real number. 
Mathlib measures can take values in $[0,\infty]$ by default,
making two-sided bounds with constant factors awkward to manipulate given $\infty$ is a possible measure.
Since $P_{\mathrm{eq}}$ is a probability measure, every quantity here is finite, so using the real-value measure is equivalent and
avoids this issue.
Both bounds are free of $x$, as necessary
in the proof to have a uniform~\eqref{eq:E}.
\end{leanbox}

\paragraph{Relation to the compactness route.} Under (A3b), the transition density
$\kappa_{\Delta t}$ is continuous, has unit mass by~\eqref{eq:markov}, and is strictly positive by
Proposition~S-1, since $\delta(\Delta t) < 1$. So (A3a) holds as well, and
Proposition~S-2$'$ applies. Since $\varepsilon_0(\Delta t)$ is the largest constant for
which~\eqref{eq:E} holds,
\[
  1 - \delta(\Delta t) \;\leq\; \varepsilon_0(\Delta t).
\]
The spectral route therefore does not give a larger constant. What it gives is a constructive one,
computed from $\lambda_1$ and $\Theta$ rather than from the unknown extrema of the transition
density.

\subsubsection{Remarks}

\paragraph{The implemented kernel must satisfy~\eqref{eq:E} too.}
We have established~\eqref{eq:E} for the exact transition kernel $\kappa_{\Delta t}$.
In the implementation, a discretized kernel $\hat\kappa_{\Delta t}$ is used and must also verify~\eqref{eq:E}.
This requires some care: at small $\Delta t$, the exact kernel is sharply peaked, and
with too few basis functions, the discretized density may show Gibb's phenomenon/ringing, which violates~\eqref{eq:E}.
Therefore, basis function should be increased for small $\Delta t$ and ringing can be checked at runtime.
The change in the $varepsilon$ bound caused by discretization are out of the scope of this proof.

\paragraph{The $\varepsilon = 1$ case.} If~\eqref{eq:E} holds for some $\varepsilon$, it also
holds for any smaller $\varepsilon' \in (0,\varepsilon]$, since the interval $[\varepsilon',
1/\varepsilon']$ only widens. We may therefore assume $\varepsilon \in (0,1)$ from here on, at no
cost: even in the degenerate case $\kappa_{\Delta t} \equiv 1$, where~\eqref{eq:E} holds with
$\varepsilon = 1$, it also holds with any smaller $\varepsilon$. This assumption simplifies several
steps below that divide by $1 - \varepsilon^2$ or similar quantities.

\subsection{From mixing to block forgetting}\label{sec:algotheoremsection2}

The input of this section is the envelope condition~\eqref{eq:E} from
Section~\ref{sec:algotheoremsection1}: a non-random $\varepsilon \in (0,1)$ with $\varepsilon
\leq \kappa_{\Delta t} \leq \varepsilon^{-1}$, the same at every step.

\subsubsection{One-step filter kernel}

Let $R_k(x,dx') := Q_{\Delta t}(x,dx')\,\Psi_k(x')$ be the unnormalized one-step filter kernel,
with $Q_{\Delta t}$ the transition kernel of the latent dynamics over one observation interval and
$\Psi_k$ the likelihood of the $k$-th observation.

\begin{lemAthree}[One-step mixing]
Assume (A1), (A2) and~\eqref{eq:E}, and write
\[
  \nu_k(dx') \;:=\; \Psi_k(x')\,P_{\mathrm{eq}}(dx')
\]
for the observation-weighted reference measure. Then:
\begin{enumerate}\itemsep2pt
\item $R_k$ is mixing  in the sense of (Le~Gland--Oudjane, 2004)\cite{legland2004stability} with the non-random constant $\varepsilon$ and reference $\nu_k$:
  $\varepsilon\nu_k(U) \leq R_k(x,U) \leq \varepsilon^{-1}\nu_k(U)$ for every $x$ and
  every Borel set $U$.
\item $H(R_k) \leq 4\log(1/\varepsilon)$ and $\tau_k := \tau(R_k) \leq \tau :=
  (1-\varepsilon^2)/(1+\varepsilon^2) < 1$, uniformly in $k$ and in the realized observations.
\item For any $r$ consecutive steps, $\tau(R_{k+r-1:k}) \leq \tau^r$.
\end{enumerate}
\end{lemAthree}

\begin{proof}

We use the following three results, for which we refer to (Le~Gland--Oudjane,
2004)~\cite{legland2004stability} for definitions and proofs.
\begin{itemize}\itemsep2pt
\item The \emph{Hilbert projective metric} $h$ between two comparable non-zero finite measures
  $\mu, \mu'$ is the logarithm of the ratio between the supremum and the infimum of
  $\mu(U)/\mu'(U)$ over sets $U$ of positive mass. It is scale-invariant: $h(a\mu, b\mu') =
  h(\mu,\mu')$ for $a, b > 0$.
\item For a non-negative kernel $R$, $H(R) := \sup_{\mu,\mu'} h(R\mu, R\mu')$ is the
  $h$-diameter of its image. Its \emph{Birkhoff contraction coefficient} $\tau(R)$ is the
  smallest constant with $h(R\mu, R\mu') \leq \tau(R)\,h(\mu,\mu')$. Birkhoff's theorem gives
  $\tau(R) = \tanh\bigl(H(R)/4\bigr)$.
\item $\tau$ is submultiplicative under composition. We write $R_{k+r-1:k} := R_{k+r-1}\cdots
  R_k$ for the composite of $r$ consecutive steps.
\end{itemize}

\emph{Part (1).} By (A1), $\Psi_k$ is non-negative and measurable, and since $\kappa_{\Delta t}$ is the density of $Q_{\Delta t}$ with respect to $P_{\mathrm{eq}}$, we have
\[
  R_k(x,U) \;=\; \int_U \Psi_k(x')\,\kappa_{\Delta t}(x,x')\,P_{\mathrm{eq}}(dx') .
\]
Note that $x$ only appears in $\kappa_{\Delta t}(x,\cdot)$.
With the integrant non-negative, we can apply ~\eqref{eq:E} under the integral sign, replacing  $\kappa_{\Delta t}(x,x')$ 
by  $\varepsilon$ and $\varepsilon^{-1}$, yielding

\[
  \varepsilon \int_U \Psi_k\,dP_{\mathrm{eq}}
  \;\leq\; R_k(x,U)
  \;\leq\; \varepsilon^{-1} \int_U \Psi_k\,dP_{\mathrm{eq}} ,
\]

Both the integrals are finite, since $\kappa_{\Delta t} \leq \varepsilon^{-1}$ and $\Psi_k$ is $P_{\mathrm{eq}}$-integrable by (A2).
Substituting $\nu$, we get
\[
  \varepsilon\,\nu_k(U) \;\leq\; R_k(x,U) \;\leq\; \varepsilon^{-1}\nu_k(U) .
\]

The bound in this inequality no longer depend on $x$, \ie,  the constant $\varepsilon$ is a property of the propagator alone, and the observation
$Y_{t_k}$ has been absorbed into the finite, non-zero, reference measure $\nu_k$.

\emph{Part (2).} Let $\mu, \mu'$ be finite non-zero measures. Integrating part~(1) against $\mu$
gives $\varepsilon\,\mu(\mathcal{X})\,\nu_k(U) \leq (R_k\mu)(U) \leq
\varepsilon^{-1}\mu(\mathcal{X})\,\nu_k(U)$, and likewise for $\mu'$. Hence, for every $U$ with
$\nu_k(U) > 0$,
\[
  \varepsilon^2\,\frac{\mu(\mathcal{X})}{\mu'(\mathcal{X})}
  \;\leq\; \frac{(R_k\mu)(U)}{(R_k\mu')(U)}
  \;\leq\; \varepsilon^{-2}\,\frac{\mu(\mathcal{X})}{\mu'(\mathcal{X})} .
\]
Sets with $\nu_k(U) = 0$ receive zero mass under both images, and do not enter $h$. The ratio of
the supremum to the infimum of this quotient is therefore at most $\varepsilon^{-4}$, so
\[
  H(R_k) \;\leq\; \log \varepsilon^{-4} \;=\; 4\log(1/\varepsilon).
\]
By Birkhoff's theorem, and since $\tanh$ is increasing,
\[
  \tau_k \;=\; \tanh\Bigl[\frac{H(R_k)}{4}\Bigr]
  \;\leq\; \tanh\bigl[\log(1/\varepsilon)\bigr]
  \;=\; \frac{\varepsilon^{-1} - \varepsilon}{\varepsilon^{-1} + \varepsilon}
  \;=\; \frac{1-\varepsilon^2}{1+\varepsilon^2} \;=\; \tau ,
\]
and $\tau \in (0,1)$ because $\varepsilon \in (0,1)$. The bound depends on $\varepsilon$ only, hence
neither on $k$ nor on the realized observations.

\emph{Part (3).} The contraction coefficient is submultiplicative under composition, so
\[
  \tau(R_{k+r-1:k}) \;\leq\; \prod_{k'=k}^{k+r-1}\tau_{k'} \;\leq\; \tau^r . \qedhere
\]
\end{proof}

\begin{leanbox}[Lean: Lemma S-3(1) --- \texttt{setIntegral\_mul\_le\_of\_isEnvelope}]
In \verb|PlanxAlgorithm/Kernel.lean|, with the set \texttt{A} being the Borel set $U$ of the text.

Part (1), the integration step, is formalized, with the non-random constant $\varepsilon$, 
the observation-dependent reference measure, and no lower bound on $\Psi$.
\begin{leancode}
theorem setIntegral_mul_le_of_isEnvelope (h : IsEnvelope ε κ) {Ψ : X → ℝ}
    (hΨ : ∀ y, 0 ≤ Ψ y) (hint : ∀ x, Integrable (fun y => Ψ y * κ x y) μ)
    (hΨint : Integrable Ψ μ) (x : X) {A : Set X} (hA : MeasurableSet A) :
    ε * (∫ y in A, Ψ y ∂μ) ≤ ∫ y in A, Ψ y * κ x y ∂μ ∧
      ∫ y in A, Ψ y * κ x y ∂μ ≤ ε⁻¹ * (∫ y in A, Ψ y ∂μ)
\end{leancode}
Part (2) is not formalized, as it depends on Birkhoff's contraction theorem, which Mathlib does not
have.
We instead have the conclusion as member of  the structure \verb|BlockForgetting| below.
\end{leanbox}

\subsubsection{Block operator}

\begin{defblock}[Block operator]
$\Phi_\beta : \mathcal{P}(\mathcal{X}) \to \mathcal{P}(\mathcal{X})$ is the measure-level map
realized by \textsc{BlockFilter} on the segment $Y_{\mathcal{I}_\beta}$: the composition of the $L$
normalized predict--update operators. For $j$ consecutive blocks, $\Phi^{[j]}_\beta := \Phi_{\beta+j-1}
\circ \cdots \circ \Phi_\beta$, a composite of $jL$ one-step kernels, and $\Phi^{[0]}_\beta :=
\mathrm{id}$.
\end{defblock}

We first check that $\Phi_\beta$ is well defined. By Lemma~S-3(1) with $U = \mathcal{X}$, for
every non-zero finite measure $\mu$,
\[
  \varepsilon\,\mu(\mathcal{X})\,\nu_k(\mathcal{X})
  \;\leq\; (R_k\mu)(\mathcal{X})
  \;\leq\; \varepsilon^{-1}\mu(\mathcal{X})\,\nu_k(\mathcal{X}),
\]
and $\nu_k(\mathcal{X}) \in (0,\infty)$ by (A2). So the normalizing constant is positive and
finite at every step, and $\Phi_\beta$ is a well-defined map from $\mathcal{P}(\mathcal{X})$ to
itself.

\begin{lemAfour}[Block forgetting]
Under (A1), (A2), \eqref{eq:E}, for all $\mu,\mu' \in \mathcal{P}(\mathcal{X})$, every $\beta$ and
every $j \geq 1$:
\begin{align*}
  \text{(abs)}\quad & \bigl\|\Phi^{[j]}_\beta(\mu) - \Phi^{[j]}_\beta(\mu')\bigr\|
    \;\leq\; \frac{8\log(1/\varepsilon)}{\log 3}\,\tau^{jL-1} \;=\; C_0\tau^{jL},
    && \text{where } C_0 := \frac{8\log(1/\varepsilon)}{\tau\log 3}, \\
  \text{(lip)}\quad & \bigl\|\Phi^{[j]}_\beta(\mu) - \Phi^{[j]}_\beta(\mu')\bigr\|
    \;\leq\; \frac{2}{\varepsilon^2\log 3}\,\tau^{jL-1}\,\|\mu - \mu'\| .
\end{align*}
\end{lemAfour}

\begin{proof}
Besides the facts recalled before Lemma~S-3, we use two comparisons between $h$ and total variation
from (Le~Gland--Oudjane, 2004)~\cite{legland2004stability}, where both bounds of the lemma appear
as Theorem~4.1 and Corollary~4.2. For a kernel $R$ mixing with constant $\varepsilon$ and
probability measures $\mu, \mu'$,
\[
  h(R\mu, R\mu') \;\leq\; \varepsilon^{-2}\,\|\mu - \mu'\|,
\]
and for non-zero finite measures $\mu, \mu'$ with normalizations $\bar\mu, \bar\mu'$,
\[
  \|\bar\mu - \bar\mu'\| \;\leq\; \frac{2}{\log 3}\,h(\mu, \mu') .
\]

Let $\eta := R_{(\beta+j)L:\beta L+1}\,\mu$ and $\eta' := R_{(\beta+j)L:\beta L+1}\,\mu'$. Since $h$
is scale-invariant, and normalizing at each step only rescales, $\Phi^{[j]}_\beta(\mu)$ and
$\Phi^{[j]}_\beta(\mu')$ are the normalizations of $\eta$ and $\eta'$.

\emph{Step 1: the first kernel.} By the first comparison, and by Lemma~S-3(2),
\[
  h(R_{\beta L+1}\mu, R_{\beta L+1}\mu')
  \;\leq\; \min\bigl\{\varepsilon^{-2}\|\mu - \mu'\|,\; 4\log(1/\varepsilon)\bigr\} .
\]

\emph{Step 2: the remaining $jL-1$ kernels.} By Lemma~S-3(2) and~(3),
\[
  h(\eta, \eta') \;\leq\; \tau^{jL-1}\,h(R_{\beta L+1}\mu, R_{\beta L+1}\mu') .
\]

\emph{Step 3: back to total variation.} By the second comparison,
\[
  \bigl\|\Phi^{[j]}_\beta(\mu) - \Phi^{[j]}_\beta(\mu')\bigr\|
  \;\leq\; \frac{2}{\log 3}\,\tau^{jL-1}\,h(R_{\beta L+1}\mu, R_{\beta L+1}\mu') .
\]
Inserting the second branch of Step~1 gives (abs),
\[
  \frac{8\log(1/\varepsilon)}{\log 3}\,\tau^{jL-1} \;=\; C_0\,\tau^{jL},
\]
and inserting the first branch gives (lip),
\[
  \frac{2}{\varepsilon^2\log 3}\,\tau^{jL-1}\,\|\mu - \mu'\| . \qedhere
\]
\end{proof}

\subsubsection{Contraction bounds}

Section~\ref{sec:algotheoremsection3} uses Sections~\ref{sec:algotheoremsection1}
and~\ref{sec:algotheoremsection2} only through two bounds on the block maps. With $\tau$, $C_0$,
$C_1$ as in Lemma~S-4, for all $\mu,\mu' \in \mathcal{P}(\mathcal{X})$, every $\beta$ and every $j
\geq 0$,
\begin{align}
  \text{(F-abs)}\quad & \bigl\|\Phi^{[j]}_\beta(\mu) - \Phi^{[j]}_\beta(\mu')\bigr\|
    \;\leq\; e(j) := \min\{2,\; C_0\tau^{jL}\}, \tag{F-abs}\label{eq:Fabs}\\
  \text{(F-lip)}\quad & \bigl\|\Phi^{[j]}_\beta(\mu) - \Phi^{[j]}_\beta(\mu')\bigr\|
    \;\leq\; C_1\tau^{jL-1}\,\|\mu - \mu'\|. \tag{F-lip}\label{eq:Flip}
\end{align}
For $j \geq 1$, these are Lemma~S-4; the branch $2$ in $e(j)$ is the trivial bound on probability
measures. The constants depend on $\varepsilon$ alone, so $e(j)$ depends only on $(\varepsilon, L,
j)$, and not on $\beta$, $B$, $N$ or the observations. It decays geometrically in $j$.

\paragraph{The case $j = 0$.}
At $j = 0$, $\Phi^{[0]}_\beta$ is the identity, and both bounds reduce to facts about the
constants. \eqref{eq:Fabs} reads $\|\mu - \mu'\| \leq \min\{2, C_0\}$, which holds if $C_0 \geq 2$.
\eqref{eq:Flip} reads $\|\mu - \mu'\| \leq C_1\tau^{-1}\|\mu - \mu'\|$, which holds if $C_1 \geq
1$, since $\tau < 1$. Both hold:
\begin{itemize}\itemsep2pt
\item $C_1 = 2/(\varepsilon^2\log 3) > 2/\log 3 > 1$, since $\varepsilon < 1$.
\item For $C_0$, let $g(\varepsilon) := \log(1/\varepsilon) - (1-\varepsilon^2)/2$. Then $g(1) =
  0$ and $g'(\varepsilon) = -1/\varepsilon + \varepsilon \leq 0$ on $(0,1]$, so $g \geq 0$, that
  is, $\log(1/\varepsilon) \geq (1-\varepsilon^2)/2$. Hence
  \[
    C_0 \;=\; \frac{8\log(1/\varepsilon)\,(1+\varepsilon^2)}{(1-\varepsilon^2)\log 3}
    \;\geq\; \frac{4\,(1+\varepsilon^2)}{\log 3}
    \;>\; 2 .
  \]
\end{itemize}

\begin{leanbox}[Lean: the contraction bounds (F-abs), (F-lip)]
In \verb|PlanxAlgorithm/Relaxation.lean|. 
Since the  (Le~Gland--Oudjane, 2004)~\cite{legland2004stability} results are not formalized, as discussed
in the previous box, they are taken as hypotehsis as member of the structure \verb|BlockForgetting|.
In the code, \texttt{n} is the number $j$ of consecutive blocks composed, and \texttt{b} the index $\beta$ of the first of them.
\begin{leancode}
structure BlockForgetting (Φ : ℕ → S → S) (L : ℕ) where
  τ  : ℝ
  C₀ : ℝ
  C₁ : ℝ
  τ_pos     : 0 < τ
  τ_lt_one  : τ < 1
  C₀_nonneg : 0 ≤ C₀
  C₁_nonneg : 0 ≤ C₁
  abs : ∀ b n μ μ', dist (comp Φ b n μ) (comp Φ b n μ') ≤ C₀ * τ ^ (n * L)
  lip : ∀ b n μ μ',
    dist (comp Φ b n μ) (comp Φ b n μ') ≤ C₁ * τ ^ (n * L - 1) * dist μ μ'
\end{leancode}
We omit deliberately the trivial branch $\min\{2,\cdot\}$, as it would need a $\mathrm{dist} \leq 2$ field,
 and as it is only a quirk of the total-variation and contributes to no bound in other sections. 
We also do not carry $\varepsilon$ since it enters only through $\tau$, $C_0$, $C_1$.

The fields \verb|abs| and \verb|lip| range over every \verb|n|, including $0$; this is the case
$j = 0$ above. At $j = 0$, Lean's natural-number subtraction gives \verb|n * L - 1 = 0|, so
\verb|lip| reads $\mathrm{dist} \leq C_1\,\mathrm{dist}$, which needs $C_1 \geq 1$. The field
\texttt{$\tau$\_pos} corresponds to $\varepsilon < 1$.
\end{leanbox}

\subsection{Algorithmic arguments}\label{sec:algotheoremsection3}

From previous Sections~\ref{sec:algotheoremsection1} and~\ref{sec:algotheoremsection2}, we take two
results: each block map $\Phi_\beta$ is a well-defined map on $\mathcal{P}(\mathcal{X})$, and the
contraction bounds~\eqref{eq:Fabs} and~\eqref{eq:Flip} hold, with constants depending on
$\varepsilon$ alone, and not on $N$, $B$, $\beta$, the observations or $\bar\pi_0$.
The bounds are needed only for the rates, and nothing else about measures or kernels is used below.

\paragraph{WRMS norm definition.}
The stopping rule measures changes of coefficient vectors in $\mathbb{R}^{N_b}$ with the weighted
root-mean-square norm of the implementation. For a reference vector $y \in \mathbb{R}^{N_b}$,
\[
  \|v\|_{\mathrm{WRMS},y} \;=\; N_b^{-1/2}\,\|W_y v\|_{\ell^2},
  \qquad W_y = \mathrm{diag}(w_1,\dots,w_{N_b}),
  \qquad w_i = (\mathrm{atol} + \mathrm{rtol}\cdot|y_i|)^{-1},
\]
with tolerances $\mathrm{atol} > 0$ and $\mathrm{rtol} \geq 0$, so that $0 < w_i \leq
1/\mathrm{atol}$. Since the weights are strictly positive, this is a norm on $\mathbb{R}^{N_b}$ for each fixed $y$.

\bigskip

We now write the algorithms described in Section~\ref{sec:pint_setting} as recursions on the density at the
start of each block, with one call to \textsc{BlockFilter} on block $\beta$ being one application
of $\Phi_\beta$. \textsc{SequentialFilter} runs the blocks one after another, each starting from
the output of the previous one. \textsc{PintFilter} starts every block $\beta \geq 1$ from the
trial density $\bar\pi_0$ at iteration $0$. At iteration $j+1$, it runs each block from the density
its predecessor produced at iteration $j$. In both algorithms the first block starts from
$\pi_0$:
\[
\begin{aligned}
  p^{\mathrm{seq}}_0 &= \pi_0, & p^{\mathrm{seq}}_{\beta+1} &= \Phi_\beta(p^{\mathrm{seq}}_\beta), \\
  p^{(0)}_0 &= \pi_0, & p^{(0)}_{\beta+1} &= \bar\pi_0, \\
  p^{(j+1)}_0 &= \pi_0, & p^{(j+1)}_{\beta+1} &= \Phi_\beta(p^{(j)}_\beta).
\end{aligned}
\]
Note that the block count $B$ does not appear:
$B$ only truncates the index range, and every statement below is uniform in $\beta$.

\begin{leanbox}[Lean: the three recursions]
In \verb|PlanxAlgorithm/Relaxation.lean|, namespace \verb|WaveformRelax|.
\begin{leancode}
def comp (Φ : ℕ → S → S) (b : ℕ) : ℕ → S → S
  | 0     => id
  | n + 1 => Φ (b + n) ∘ comp Φ b n

def seq (Φ : ℕ → S → S) (π₀ : S) : ℕ → S
  | 0     => π₀
  | b + 1 => Φ b (seq Φ π₀ b)

def iter (Φ : ℕ → S → S) (π₀ π₀' : S) : ℕ → ℕ → S
  | 0,     0     => π₀
  | 0,     _ + 1 => π₀'
  | _ + 1, 0     => π₀
  | n + 1, b + 1 => Φ b (iter Φ π₀ π₀' n b)
\end{leancode}
In the code, \texttt{n} is the iteration index $j$ and \texttt{b} the block index $\beta$.

The state type \verb|S| is an arbitrary type, with \verb|[PseudoMetricSpace S]| added only where
a rate is stated. Mathlib has \verb|SignedMeasure.totalVariation| as a measure but no
total-variation distance on probability measures, so instantiating at $\mathcal{P}(\mathcal{X})$
would mean building that metric first, but we do not need it for the argument, so we do not do so.
 This also avoids the question of the TV normalization convention, \ie, $\|\mu-\mu'\|\leq2$ versus $\leq1$.
  Blocks are indexed by $\mathbb{N}$ rather than \texttt{Fin B}, and all indices are additive, 
  so that no truncated subtraction appears inside an induction.
\end{leanbox}

\subsubsection{Reduction and unrolling}

\begin{lemAfive}[Reduction]
If $p^{(j)}_\beta = p^{\mathrm{seq}}_\beta$ for every $\beta$, then the filtered densities and
log-likelihood at iteration $j$ are exactly those of the sequential filter.
\end{lemAfive}

\begin{proof}
Both algorithms compute each block's filtered densities and normalizers with \textsc{BlockFilter},
from the block's initial condition and its observations. The observations are the same, and by
hypothesis so are the initial conditions.
\end{proof}

\bigskip

\begin{lemAsix}[Staggered chain]
Let $P : \mathbb{N} \to S$ satisfy $P(c+1) = \Phi_c(P(c))$ for all $c$. Then $P(\beta+j) =
\Phi^{[j]}_\beta(P(\beta))$ for all $\beta,j$.
\end{lemAsix}

\begin{proof}
By induction on $j$. For $j = 0$, both sides equal $P(\beta)$, since $\Phi^{[0]}_\beta =
\mathrm{id}$. For the step from $j$ to $j+1$,
\begin{align*}
  P(\beta+j+1)
  &= \Phi_{\beta+j}\bigl(P(\beta+j)\bigr)
    && \text{by the hypothesis on } P \\
  &= \Phi_{\beta+j}\bigl(\Phi^{[j]}_\beta(P(\beta))\bigr)
    && \text{by the induction hypothesis} \\
  &= \Phi^{[j+1]}_\beta\bigl(P(\beta)\bigr)
    && \text{by the definition of } \Phi^{[j+1]}_\beta . \qedhere
\end{align*}
\end{proof}

Applied to $P = p^{\mathrm{seq}}$, whose recursion has exactly this form, Lemma~S-6 gives the
semigroup identity $p^{\mathrm{seq}}_{\beta+j} = \Phi^{[j]}_\beta(p^{\mathrm{seq}}_\beta)$.

\begin{lemAseven}[Diagonal unrolling]
For all $\beta,m,j$: $\;p^{(j+m)}_{\beta+m} = \Phi^{[m]}_\beta\bigl(p^{(j)}_\beta\bigr)$.
\end{lemAseven}

\begin{proof}
Fix $\beta$ and $j$, and argue by induction on $m$. For $m = 0$, both sides equal
$p^{(j)}_\beta$, since $\Phi^{[0]}_\beta = \mathrm{id}$. For the step from $m$ to $m+1$,
\begin{align*}
  p^{(j+m+1)}_{\beta+m+1}
  &= \Phi_{\beta+m}\bigl(p^{(j+m)}_{\beta+m}\bigr)
    && \text{by the recursion for } p^{(j+1)}_{\beta+1} \\
  &= \Phi_{\beta+m}\bigl(\Phi^{[m]}_\beta(p^{(j)}_\beta)\bigr)
    && \text{by the induction hypothesis} \\
  &= \Phi^{[m+1]}_\beta\bigl(p^{(j)}_\beta\bigr)
    && \text{by the definition of } \Phi^{[m+1]}_\beta . \qedhere
\end{align*}
\end{proof}

We can understand the Lemma~S-7 result as follows.
 Taking the starting iteration to be $0$, it gives
$p^{(j)}_\beta = \Phi^{[j]}_{\beta-j}\bigl(p^{(0)}_{\beta-j}\bigr)$ for $\beta \geq j$: the initial
condition of block $\beta$ at iteration $j$ is obtained from the iteration-$0$ state of block
$\beta - j$, followed by $jL$ consecutive filtering steps, using the observations $Y_{(\beta-j)L+1 : \beta
L}$. 

For $\beta > j$, that starting state is the trial density $\bar\pi_0$.
In other words, the only wrong input is the trial density $\bar\pi_0$, and each iteration moves it
one block further upstream of block $\beta$, behind $L$ more filtering steps. Since the filter
forgets its initial condition geometrically (Lemma~S-4), its influence on block $\beta$ shrinks by
a factor $\tau^L$ per iteration; this is the mechanism behind Proposition~S-10. Once $j \geq
\beta$, it has been pushed past the first block, which starts from the correct $\pi_0$, and block
$\beta$ is exact; this is Lemma~S-8 below.

\begin{leanbox}[Lean: Lemmas S-6 and S-7]
In the code, \texttt{n} is the iteration index or the number of blocks composed, both written $j$ in the text, and \texttt{b} is the block index $\beta$.
\begin{leancode}
theorem comp_of_step (P : ℕ → S) (h : ∀ c, P (c + 1) = Φ c (P c)) (b n : ℕ) :
    P (b + n) = comp Φ b n (P b)

theorem seq_add (b n : ℕ) : seq Φ π₀ (b + n) = comp Φ b n (seq Φ π₀ b)

theorem iter_add (b m n : ℕ) :
    iter Φ π₀ π₀' (n + m) (b + m) = comp Φ b m (iter Φ π₀ π₀' n b)

theorem iter_add_zero (b n : ℕ) :
    iter Φ π₀ π₀' n (b + n) = comp Φ b n (iter Φ π₀ π₀' 0 b)
\end{leancode}
\verb|iter_add_zero| is the case $j = 0$ of \verb|iter_add|, as used in Proposition~S-10 proof.
\end{leanbox}

\begin{lemAeight}[Exactness of the leading blocks]
If $\beta \leq j$ then $p^{(j)}_\beta = p^{\mathrm{seq}}_\beta$.
\end{lemAeight}

\begin{proof}
We proceed by induction on $\beta$. For $\beta = 0$, both sides equal $\pi_0$, since the first
block starts from $\pi_0$ at every iteration. For the step, suppose the claim holds for $\beta$ and every
$j \geq \beta$, and let
$\beta + 1 \leq j$. Then $j \geq 1$, so we can write $j = m+1$ with $\beta \leq m$, and
\begin{align*}
  p^{(m+1)}_{\beta+1}
  &= \Phi_\beta\bigl(p^{(m)}_\beta\bigr) && \text{by the recursion} \\
  &= \Phi_\beta\bigl(p^{\mathrm{seq}}_\beta\bigr) && \text{by the induction hypothesis, since } \beta \leq m \\
  &= p^{\mathrm{seq}}_{\beta+1} && \text{by the sequential recursion.} \qedhere
\end{align*}
\end{proof}

\begin{corAnine}[Finite exact termination]
For $j \geq B-1$, $p^{(j)}_\beta = p^{\mathrm{seq}}_\beta$ for every block $\beta \leq B-1$:
\textsc{PintFilter} reproduces the sequential filter exactly after at most $B-1$ iterations.
\end{corAnine}

\begin{proof}
For $\beta \leq B-1 \leq j$, Lemma~S-8 applies.
\end{proof}

This bound depends on $N$ and is therefore irrelevant for our span complexity claim, but it makes
the telescoping sum of Proposition~S-14 finite.

\begin{leanbox}[Lean: Lemma S-8 and Corollary S-9]
In the code, \texttt{n} is the iteration index $j$ and \texttt{b} the block index $\beta$.
\begin{leancode}
theorem iter_eq_seq_of_le {b n : ℕ} (h : b ≤ n) : iter Φ π₀ π₀' n b = seq Φ π₀ b

theorem iter_eq_seq_of_le_of_le {b n B : ℕ} (hb : b ≤ B) (hn : B ≤ n) :
    iter Φ π₀ π₀' n b = seq Φ π₀ b
\end{leancode}
\end{leanbox}

\subsubsection{The fixed point is the sequential filter}

\begin{defblock}[Boundary-update map]
$F$ is the boundary-update map of \textsc{PintFilter}: given $P = (P_\beta)_\beta$,
$F(P)_0 = \pi_0$ and $F(P)_{\beta+1} = \Phi_\beta(P_\beta)$.
\end{defblock}

We write $P^{(j)} := (p^{(j)}_\beta)_\beta$ for the boundary densities at iteration $j$. By the
recursions, $P^{(j)} = F(P^{(j-1)})$ for $j \geq 1$.

\begin{thmAtwelve}[Equivalence]
$P$ is a fixed point of $F$ if and only if $P_\beta = p^{\mathrm{seq}}_\beta$ for every $\beta$. Hence $F$
has a unique fixed point, and at it every filtered density, every normalizer $Z_{t_k}$, and the
total log-likelihood agree with those of \textsc{SequentialFilter} exactly.
\end{thmAtwelve}

\begin{proof}
Componentwise, the equation $F(P) = P$ reads
\[
  P_0 = \pi_0, \qquad P_{\beta+1} = \Phi_\beta(P_\beta) \quad\text{for all } \beta ,
\]
which are exactly the equations defining $p^{\mathrm{seq}}$.

Suppose first that $P$ is a fixed point. We show $P_\beta = p^{\mathrm{seq}}_\beta$ by induction on
$\beta$. The base case is $P_0 = \pi_0 = p^{\mathrm{seq}}_0$. For the step, if $P_\beta =
p^{\mathrm{seq}}_\beta$, then
\[
  P_{\beta+1} \;=\; \Phi_\beta(P_\beta) \;=\; \Phi_\beta(p^{\mathrm{seq}}_\beta) \;=\; p^{\mathrm{seq}}_{\beta+1} .
\]
Conversely, $p^{\mathrm{seq}}$ satisfies the two equations by construction, so it is a fixed
point. Uniqueness follows, since the equations determine $P$ recursively from $\pi_0$.

Finally, at the fixed point $P_\beta = p^{\mathrm{seq}}_\beta$ for every $\beta$, so by
Lemma~S-5 every filtered density, every normalizer and the total log-likelihood agree with those
of \textsc{SequentialFilter}.
\end{proof}

No contraction bound is used here, only that each $\Phi_\beta$ is a well-defined map
(Section~\ref{sec:algotheoremsection2}).

\begin{corAthirteen}[Zero WRMS increment at the fixed point]
Assume (A4). Recall that the increment at iteration $j \geq 1$ is the largest change of a block
initial condition since the previous iteration, measured on coefficient vectors,
\[
  \delta^{(j)} \;=\; \max_{\beta\geq1}\bigl\|\mathbf{c}[p^{(j)}_\beta] -
  \mathbf{c}[p^{(j-1)}_\beta]\bigr\|_{\mathrm{WRMS}} .
\]
If at some iteration $j \geq 1$ the increment vanishes exactly, $\delta^{(j)} = 0$, then
$P^{(j)}$ is the fixed point of $F$, so $p^{(j)}_\beta = p^{\mathrm{seq}}_\beta$ for every $\beta$, and the
algorithm returns the sequential filtered densities and log-likelihood exactly. Conversely, the
current iterate is the fixed point if and only if the next increment is zero.
\end{corAthirteen}

\begin{proof}
Since the weights are strictly positive, $\|\cdot\|_{\mathrm{WRMS},y}$ is a norm on
$\mathbb{R}^{N_b}$, so $\delta^{(j)} = 0$ forces $\mathbf{c}[p^{(j)}_\beta] = \mathbf{c}[p^{(j-1)}_\beta]$ for every $\beta \geq 1$. By (A4), the
coefficient map is injective on the states of the filter, so $p^{(j)}_\beta = p^{(j-1)}_\beta$ for $\beta
\geq 1$; for $\beta = 0$ we have $p^{(j)}_0 = p^{(j-1)}_0 = \pi_0$, since the first block always starts from $\pi_0$. Substituting
$p^{(j-1)}_\beta = p^{(j)}_\beta$ into the recursion $p^{(j)}_{\beta+1} = \Phi_\beta(p^{(j-1)}_\beta)$ gives
\[
  p^{(j)}_0 = \pi_0, \qquad p^{(j)}_{\beta+1} = \Phi_\beta\bigl(p^{(j)}_\beta\bigr),
\]
that is, $F(P^{(j)}) = P^{(j)}$. Theorem~S-12 then gives $p^{(j)}_\beta = p^{\mathrm{seq}}_\beta$ for
every $\beta$. The converse direction follows by reading the same argument backwards: if $P^{(j)}$
is the fixed point, then $P^{(j+1)} = F(P^{(j)}) = P^{(j)}$, so the next increment is zero.
\end{proof}

\begin{leanbox}[Lean: Theorem S-12 and Corollary S-13]
In the code, \texttt{n} is the iteration index $j$ and \texttt{b} the block index $\beta$.
\begin{leancode}
structure IsFixedPoint (Φ : ℕ → S → S) (π₀ : S) (P : ℕ → S) : Prop where
  head : P 0 = π₀
  step : ∀ b, P (b + 1) = Φ b (P b)

theorem isFixedPoint_iff {P : ℕ → S} :
    IsFixedPoint Φ π₀ P ↔ ∀ b, P b = seq Φ π₀ b

theorem IsFixedPoint.unique {P P' : ℕ → S} (h : IsFixedPoint Φ π₀ P)
    (h' : IsFixedPoint Φ π₀ P') : P = P'

theorem increment_eq_zero_iff_isFixedPoint {n : ℕ} :
    (∀ b, iter Φ π₀ π₀' (n + 1) b = iter Φ π₀ π₀' n b) ↔
      IsFixedPoint Φ π₀ (iter Φ π₀ π₀' n)

theorem eq_seq_of_increment_eq_zero {n : ℕ}
    (h : ∀ b, iter Φ π₀ π₀' (n + 1) b = iter Φ π₀ π₀' n b) (b : ℕ) :
    iter Φ π₀ π₀' n b = seq Φ π₀ b
\end{leancode}
The equivalence between the parallel
and the sequential filter is a quasi-trivial function which needs
nothing except \textsc{BlockFilter} being a function.
\end{leanbox}

\begin{leanbox}[Lean: Corollary S-13 for the implementation]
In \verb|PlanxAlgorithm/Wrms.lean|, with \texttt{c} the coefficient map $\mathbf{c}$, \texttt{b} is the block index $\beta$ and \texttt{D} the basis dimension $N_b$. The increment is taken between iterations \texttt{n} and \texttt{n + 1}, that is, $j-1$ and $j$ in the text.
\begin{leancode}
theorem wrms_eq_zero_iff {atol rtol : ℝ} (hatol : 0 < atol) (hrtol : 0 ≤ rtol)
    (y v : Fin D → ℝ) (hD : 0 < D) : wrms atol rtol y v = 0 ↔ v = 0

theorem eq_seq_of_wrms_eq_zero {c : S → Fin D → ℝ} (hc : Function.Injective c)
    {atol rtol : ℝ} (hatol : 0 < atol) (hrtol : 0 ≤ rtol) (hD : 0 < D) {n : ℕ}
    (h : ∀ b, 1 ≤ b → wrms atol rtol (c (iter Φ π₀ π₀' n b))
      (c (iter Φ π₀ π₀' (n + 1) b) - c (iter Φ π₀ π₀' n b)) = 0) (b : ℕ) :
    iter Φ π₀ π₀' n b = seq Φ π₀ b
\end{leancode}
We prove only the property of WRMS that is used: a vector with zero WRMS norm is zero, so a zero
increment means equal coefficient vectors. This needs the weights to be strictly positive.
Assumption~(A4), injectivity of the coefficient map, is the hypothesis \verb|hc|.
\end{leanbox}

\subsubsection{Absolute convergence with \texorpdfstring{$N$}{N}-independent rate}

\begin{propAten}[Absolute convergence]
Under~\eqref{eq:Fabs}, for every $j \geq 1$ and every $\beta$,
$\;\|p^{(j)}_\beta - p^{\mathrm{seq}}_\beta\| \leq e(j) = \min\{2, C_0\tau^{jL}\}$.
\end{propAten}

\begin{proof}
If $\beta \leq j$, then $p^{(j)}_\beta = p^{\mathrm{seq}}_\beta$ by Lemma~S-8, and the bound
holds trivially.

Now let $\beta > j$, and write $\beta = c + j$ with $c \geq 1$, so that block $c$ lies $j$ blocks
before block $\beta$. We express both densities at block $\beta$ in terms of block $c$. By
Lemma~S-7, with starting iteration $0$ and $m = j$,
\[
  p^{(j)}_{c+j} \;=\; \Phi^{[j]}_c\bigl(p^{(0)}_c\bigr) \;=\; \Phi^{[j]}_c(\bar\pi_0),
\]
the last equality because $c \geq 1$. By the semigroup identity following Lemma~S-6,
\[
  p^{\mathrm{seq}}_{c+j} \;=\; \Phi^{[j]}_c\bigl(p^{\mathrm{seq}}_c\bigr).
\]
Both densities are therefore obtained by applying the same map $\Phi^{[j]}_c$, which runs the
filter through blocks $c, \dots, c+j-1$, to two different initial conditions, $\bar\pi_0$ and
$p^{\mathrm{seq}}_c$. Applying~\eqref{eq:Fabs} to this pair gives
\[
  \bigl\|p^{(j)}_{c+j} - p^{\mathrm{seq}}_{c+j}\bigr\|
  \;=\; \bigl\|\Phi^{[j]}_c(\bar\pi_0) - \Phi^{[j]}_c(p^{\mathrm{seq}}_c)\bigr\|
  \;\leq\; e(j) . \qedhere
\]
\end{proof}

\begin{thmAeleven}[$N$-independent iteration count]
Under~\eqref{eq:Fabs}, for any target $\delta_{\mathrm{TV}} \in (0,2)$,
\[
  N_{\mathrm{iter}}(\delta_{\mathrm{TV}}) \;:=\;
  \max\Bigl\{1,\;\Bigl\lceil \tfrac{1}{L}\Bigl(1 + \log_{1/\tau}
  \tfrac{8\log(1/\varepsilon)}{\delta_{\mathrm{TV}}\log3}\Bigr)\Bigr\rceil\Bigr\}
\]
satisfies $\max_\beta \|p^{(j)}_\beta - p^{\mathrm{seq}}_\beta\| \leq \delta_{\mathrm{TV}}$ for every $j
\geq N_{\mathrm{iter}}(\delta_{\mathrm{TV}})$. The quantity $N_{\mathrm{iter}}$ depends only on
$(\varepsilon, L, \delta_{\mathrm{TV}})$, and in particular it is independent of the trajectory
length $N$ and the block count $B$.
\end{thmAeleven}

\begin{proof}
By Proposition~S-10 it suffices to find the $j$ for which $C_0\tau^{jL} \leq
\delta_{\mathrm{TV}}$. Substituting $C_0 = 8\log(1/\varepsilon)/(\tau\log 3)$, this condition
reads
\[
  \frac{8\log(1/\varepsilon)}{\log 3}\,\tau^{jL-1} \;\leq\; \delta_{\mathrm{TV}} .
\]
Taking logarithms and using $\log(1/\tau) > 0$, since $\tau < 1$, this is equivalent to
\[
  (jL-1)\,\log(1/\tau) \;\geq\; \log\frac{8\log(1/\varepsilon)}{\delta_{\mathrm{TV}}\log3},
  \qquad\text{that is,}\qquad
  j \;\geq\; \frac{1}{L}\Bigl(1 + \log_{1/\tau}\frac{8\log(1/\varepsilon)}{\delta_{\mathrm{TV}}\log3}\Bigr).
\]
Since $\tau^{jL}$ decreases in $j$, the inequality holds for every $j$ at or above the right-hand
side, and in particular for every $j \geq N_{\mathrm{iter}}(\delta_{\mathrm{TV}})$. The maximum with
$1$ reflects that at least one iteration is run, which is also the range where
Proposition~S-10 applies. In fact, the ceiling is already at least $1$, since $C_0 > 2 >
\delta_{\mathrm{TV}}$.
\end{proof}

\begin{leanbox}[Lean: Propositions S-10 and S-11]
In the code, \texttt{n} is the iteration index $j$ and \texttt{b} the block index $\beta$.
\begin{leancode}
theorem dist_iter_seq_le (F : BlockForgetting Φ L) (n b : ℕ) :
    dist (iter Φ π₀ π₀' n b) (seq Φ π₀ b) ≤ F.C₀ * F.τ ^ (n * L)

noncomputable def iterCount (C τ : ℝ) (L : ℕ) (δ : ℝ) : ℕ :=
  ⌈Real.log (C / δ) / ((L : ℝ) * (-Real.log τ))⌉₊

theorem dist_iter_seq_le_of_iterCount_le (F : BlockForgetting Φ L) (hL : 0 < L)
    {δ : ℝ} (hδ : 0 < δ) {n : ℕ} (hn : iterCount F.C₀ F.τ L δ ≤ n) (b : ℕ) :
    dist (iter Φ π₀ π₀' n b) (seq Φ π₀ b) ≤ δ

theorem exists_iterCount (F : BlockForgetting Φ L) (hL : 0 < L) {δ : ℝ}
    (hδ : 0 < δ) :
    ∃ N : ℕ, ∀ n, N ≤ n → ∀ b, dist (iter Φ π₀ π₀' n b) (seq Φ π₀ b) ≤ δ
\end{leancode}
\verb|iterCount| is stated in terms of $C_0$ rather than $8\log(1/\varepsilon)/\log3$. 
It solves $C_0\tau^{jL} \leq \delta$ directly, which is the same threshold as $N_{\mathrm{iter}}$
after substituting $C_0$.

The maximum with $1$ in $N_{\mathrm{iter}}$ is absent in Lean, 
because the Lean bound \verb|dist_iter_seq_le| also holds at $j = 0$.

We do not argue $N$-independence in the formalization, as a meta-argument.
Rather, the fact that the statement do not accept either $N$ or $B$ as input proves it.
That is, the arguments of \verb|iterCount| are the contraction
data and the tolerance, no trajectory length and no block count occurs in either statement.
Similarly, in \verb|exists_iterCount|, \verb|N| is produced before any block index is mentioned.
\end{leanbox}

\paragraph{Remark (choice of $L$).} The condition of Theorem~S-11 bounds the product $jL$, the
total number of filtering steps separating a block from the trial density, and not $L$ alone.
Blocks much shorter than the forgetting time are therefore admissible: the algorithm then needs
more iterations, but the span $N_{\mathrm{iter}}\cdot L$ stays of the order of the forgetting time
plus one block. The choice of $L$ trades parallel width $B = N/L$ against per-iteration overhead,
not against correctness.

\subsubsection{A posteriori bound from observable increments}

The error to the sequential filter is not directly computable in practice, since the sequential densities are
exactly what we wish to avoid computing. 
The increment between successive iterations is
our only observable. 

Write $\Delta^{(j)}_\beta := \|p^{(j)}_\beta - p^{(j-1)}_\beta\|$ (the increment) and $E^{(j)}_\beta :=
\|p^{(j)}_\beta - p^{\mathrm{seq}}_\beta\|$ (the error to sequential filter), and let $\delta^{(j)}_{\mathrm{TV}}$ be any bound with
$\Delta^{(j)}_\beta \leq \delta^{(j)}_{\mathrm{TV}}$ for all $\beta \geq 1$.

\begin{propAfourteen}[Telescoping]
For every $j \geq 1$ and every $\beta$,
\[
  E^{(j)}_\beta \;\leq\; \sum_{j'=j+1}^{\beta}\Delta^{(j')}_\beta \;\leq\;
  \sum_{j'=j+1}^{\beta}\delta^{(j')}_{\mathrm{TV}} .
\]
\end{propAfourteen}

\begin{proof}
If $\beta \leq j$, then $E^{(j)}_\beta = 0$ by Lemma~S-8, and the sums are empty. Let $\beta > j$.
By Lemma~S-8, $p^{(\beta)}_\beta = p^{\mathrm{seq}}_\beta$, so the error telescopes over the
iterations $j+1, \dots, \beta$:
\[
  p^{\mathrm{seq}}_\beta - p^{(j)}_\beta \;=\; p^{(\beta)}_\beta - p^{(j)}_\beta
  \;=\; \sum_{j'=j+1}^{\beta}\bigl(p^{(j')}_\beta - p^{(j'-1)}_\beta\bigr).
\]
By the triangle inequality, $E^{(j)}_\beta \leq \sum_{j'=j+1}^{\beta} \Delta^{(j')}_\beta$. Each
term is at most $\delta^{(j')}_{\mathrm{TV}}$, since $\beta > j \geq 1$.
\end{proof}

This bound uses no contraction: it only uses that the iteration reaches $p^{\mathrm{seq}}_\beta$
after finitely many steps (Lemma~S-8). But its sum has up to $\beta$ terms, so it grows with $N$.
Proposition~S-15 uses~\eqref{eq:Flip} to bound the whole sum by the current increment alone.

\begin{propAfifteen}[A posteriori error $\leq$ constant $\times$ observed increment]
Under~\eqref{eq:Flip}, for every $j \geq 1$,
\[
  \max_\beta E^{(j)}_\beta \;\leq\; A\,\delta^{(j)}_{\mathrm{TV}},
  \qquad A := \frac{2}{\varepsilon^2\log3}\cdot\frac{\tau^{L-1}}{1-\tau^L},
\]
with $A$ depending only on $(\varepsilon, L)$, and not on $N$, $B$, $\beta$, $j$, or $\bar\pi_0$.
Setting $\delta^{(j)}_{\mathrm{TV}} = 0$ recovers Corollary~S-13.
\end{propAfifteen}

\begin{proof}
\emph{Step 1: increments decay.} We claim that for all $j, m \geq 1$,
\begin{equation}\label{eq:star}
  \max_{\beta\geq1}\Delta^{(j+m)}_\beta \;\leq\; \rho_m\,\delta^{(j)}_{\mathrm{TV}},
  \qquad \rho_m := \frac{2}{\varepsilon^2\log3}\,\tau^{mL-1}.
  \tag{$*$}
\end{equation}
Fix $\beta \geq 1$. If $\beta \leq j+m-1$, Lemma~S-8 gives $p^{(j+m)}_\beta = p^{\mathrm{seq}}_\beta
= p^{(j+m-1)}_\beta$, so $\Delta^{(j+m)}_\beta = 0$. Otherwise $\beta \geq j+m$, so block $\beta - m
\geq j \geq 1$ exists. By Lemma~S-7, applied over the $m$ blocks from $\beta - m$ to $\beta$, starting
at iterations $j$ and $j-1$,
\[
  p^{(j+m)}_\beta = \Phi^{[m]}_{\beta-m}\bigl(p^{(j)}_{\beta-m}\bigr),
  \qquad
  p^{(j+m-1)}_\beta = \Phi^{[m]}_{\beta-m}\bigl(p^{(j-1)}_{\beta-m}\bigr).
\]
This is the same map $\Phi^{[m]}_{\beta-m}$ applied to two initial conditions, so
\begin{align*}
  \Delta^{(j+m)}_\beta
  &= \bigl\|\Phi^{[m]}_{\beta-m}(p^{(j)}_{\beta-m}) - \Phi^{[m]}_{\beta-m}(p^{(j-1)}_{\beta-m})\bigr\| \\
  &\leq C_1\,\tau^{mL-1}\,\bigl\|p^{(j)}_{\beta-m} - p^{(j-1)}_{\beta-m}\bigr\|
    && \text{by~\eqref{eq:Flip}} \\
  &= \rho_m\,\Delta^{(j)}_{\beta-m}
    && \text{since } C_1 = 2/(\varepsilon^2\log 3) \\
  &\leq \rho_m\,\delta^{(j)}_{\mathrm{TV}}
    && \text{since } \beta-m \geq 1 .
\end{align*}
Taking the maximum over $\beta \geq 1$ gives~\eqref{eq:star}.

\emph{Step 2: summing.} By Proposition~S-14, $E^{(j)}_\beta \leq
\sum_{j'=j+1}^{\beta}\Delta^{(j')}_\beta$ for every $\beta$. Extending this sum of non-negative terms
to all $j' > j$, bounding each term by its maximum over blocks, and writing $j' = j+m$ gives
\[
  \max_\beta E^{(j)}_\beta \;\leq\; \sum_{m\geq1}\,\max_{\beta\geq1}\Delta^{(j+m)}_\beta .
\]
By~\eqref{eq:star}, and summing the geometric series, which converges since $\tau < 1$,
\begin{align*}
  \max_\beta E^{(j)}_\beta
  &\leq \sum_{m\geq1}\rho_m\,\delta^{(j)}_{\mathrm{TV}}
   = \frac{2}{\varepsilon^2\log3}\,\delta^{(j)}_{\mathrm{TV}}\sum_{m\geq1}\tau^{mL-1} \\
  &= \frac{2}{\varepsilon^2\log3}\,\delta^{(j)}_{\mathrm{TV}}\,
     \tau^{L-1}\sum_{m\geq0}\bigl(\tau^{L}\bigr)^{m}
   = \frac{2}{\varepsilon^2\log3}\cdot\frac{\tau^{L-1}}{1-\tau^{L}}\,\delta^{(j)}_{\mathrm{TV}}
   = A\,\delta^{(j)}_{\mathrm{TV}} . \qedhere
\end{align*}
\end{proof}

\begin{leanbox}[Lean: Propositions S-14 and S-15]
In the code, \texttt{n} is the iteration index $j$ and \texttt{b} the block index $\beta$.
\begin{leancode}
theorem dist_iter_seq_le_sum_dist (n b : ℕ) :
    dist (iter Φ π₀ π₀' n b) (seq Φ π₀ b)
      ≤ ∑ j ∈ Finset.Ico n b,
          dist (iter Φ π₀ π₀' j b) (iter Φ π₀ π₀' (j + 1) b)

theorem dist_iter_shift_le (F : BlockForgetting Φ L) {d : ℕ → ℝ}
    (hd : ∀ m b, 1 ≤ b →
      dist (iter Φ π₀ π₀' (m + 1) b) (iter Φ π₀ π₀' m b) ≤ d m)
    (m k b : ℕ) :
    dist (iter Φ π₀ π₀' (m + k + 1) b) (iter Φ π₀ π₀' (m + k) b)
      ≤ F.C₁ * F.τ ^ (k * L - 1) * d m

theorem dist_iter_seq_le_const_mul (F : BlockForgetting Φ L) (hL : 0 < L)
    {d : ℕ → ℝ}
    (hd : ∀ m b, 1 ≤ b →
      dist (iter Φ π₀ π₀' (m + 1) b) (iter Φ π₀ π₀' m b) ≤ d m)
    (m b : ℕ) :
    dist (iter Φ π₀ π₀' (m + 1) b) (seq Φ π₀ b)
      ≤ F.C₁ * F.τ ^ (L - 1) * d m / (1 - F.τ ^ L)
\end{leancode}

There are three differences from the proofs as written above, mostly to avoid awkward subtraction of indices.

Proposition~S-14 is reindexed to \verb|Finset.Ico n b| with increments $(j', j'+1)$, where the
summation index $j'$ is Lean's \verb|j|.
This is the same as $\sum_{j'=j+1}^\beta$, but avoids the subtraction $j'-1$.

 Indexing blocks by $\mathbb{N}$ rather than \texttt{Fin B} lets the telescoping sum end at $j' = \beta$ rather than at
$j' = B-1$.

In Proposition~S-15, \verb|d| is an abstract bounding sequence rather than
$\sup_\beta$, which would need \verb|ciSup| and a boundedness side condition, 
which we ultimately do not use later.

Finally, \verb|dist_iter_shift_le| drops the condition $m \geq 1$ of~\eqref{eq:star}: it
is unnecessary, because~\eqref{eq:Flip} at $j = 0$ already asserts $\|\mu - \mu'\| \leq
C_1\|\mu-\mu'\|$ and so forces $1 \leq C_1$.
\end{leanbox}

\subsubsection{From TV to WRMS}

So far, we measured increments and errors in total variation (TV).
However, the stopping rule of \textsc{PintFilter} is based on the WRMS norm of coefficient differences. 
We now translate TV bounds to WRMS.

\begin{lemAsixteen}[TV controls WRMS]
Assume (A4). For any two states $p, p'$ of the filter and every weighting reference $y \in
\mathbb{R}^{N_b}$:
\begin{enumerate}\itemsep2pt
\item
  \[
    \bigl\|\mathbf{c}[p] - \mathbf{c}[p']\bigr\|_{\mathrm{WRMS},y}
    \;\leq\; \frac{M_e}{\mathrm{atol}}\,\|p - p'\|_{\mathrm{TV}} ;
  \]
\item conversely,
  \[
    \|p - p'\|_{\mathrm{TV}} \;\leq\; {N_b}\,\max_{i \leq {N_b}}\bigl((\mathrm{atol} + \mathrm{rtol}\,|y_i|)\,
      \|e_i\|_{L^1}\bigr)\,\bigl\|\mathbf{c}[p] - \mathbf{c}[p']\bigr\|_{\mathrm{WRMS},y} .
  \]
\end{enumerate}
\end{lemAsixteen}

\begin{proof}
Write $v := \mathbf{c}[p] - \mathbf{c}[p']$. Since $\mathbf{c}$ is linear, $v = \mathbf{c}[p -
p']$, and $p - p'$ lies in $V$.

\emph{Part (1).} By the definition of $M_e$ in (A4), for each $i$,
\[
  |v_i| \;=\; \bigl|c_i[p - p']\bigr| \;\leq\; M_e\,\|p - p'\|_{\mathrm{TV}} .
\]
The weights satisfy $w_i \leq 1/\mathrm{atol}$, whatever $y$ is. A root mean square is at most the
maximum of the entries, so
\[
  \|v\|_{\mathrm{WRMS},y}
  \;=\; \Bigl(\frac{1}{{N_b}}\sum_i (w_i v_i)^2\Bigr)^{1/2}
  \;\leq\; \max_i w_i|v_i|
  \;\leq\; \frac{M_e}{\mathrm{atol}}\,\|p - p'\|_{\mathrm{TV}} .
\]

\emph{Part (2).} By (A4), $p = \sum_i c_i[p]\,e_i$ and likewise for $p'$, so $p - p' =
\sum_i v_i e_i$. Using this expansion, and inserting
$w_i w_i^{-1} = 1$ in each term,
\begin{align*}
  \|p - p'\|_{\mathrm{TV}} \;=\; \int|p - p'|
  &\leq \sum_i |v_i|\,\|e_i\|_{L^1}
   = \sum_i w_i|v_i|\cdot w_i^{-1}\|e_i\|_{L^1} \\
  &\leq \max_i\bigl(w_i^{-1}\|e_i\|_{L^1}\bigr)\sum_i w_i|v_i| .
\end{align*}
By the Cauchy--Schwarz inequality on $\mathbb{R}^{N_b}$,
\[
  \sum_i w_i|v_i| \;\leq\; \sqrt{N_b}\,\|W_y v\|_{\ell^2}
  \;=\; \sqrt{N_b}\cdot\sqrt{N_b}\,\|v\|_{\mathrm{WRMS},y}
  \;=\; N_b\,\|v\|_{\mathrm{WRMS},y} .
\]
Substituting $w_i^{-1} = \mathrm{atol} + \mathrm{rtol}|y_i|$ gives the claim.
\end{proof}

\begin{corAseventeen}[Implementation stops at $N$-independent iteration count]
Under~\eqref{eq:Fabs} and (A4), for every $j \geq 1$,
\[
  \delta^{(j)} \;\leq\; \frac{2M_eC_0}{\mathrm{atol}}\,\tau^{(j-1)L}.
\]
Hence the rule $\delta^{(j)} \leq 1$ is met at the latest at iteration
\[
  N_{\mathrm{stop}} := 1 + \max\Bigl\{0,\; \Bigl\lceil
    \frac{\log(2M_eC_0/\mathrm{atol})}{L\log(1/\tau)}\Bigr\rceil\Bigr\},
\]
which depends on $(\varepsilon, L, M_e, \mathrm{atol})$ only, not on $N$ or $B$.
\end{corAseventeen}

\begin{proof}
We first note that $E^{(m)}_\beta \leq C_0\tau^{mL}$ holds for every $m \geq 0$, not only for
$m \geq 1$. For $m \geq 1$ this is Proposition~S-10. For $m = 0$ it reads $E^{(0)}_\beta \leq C_0$,
which holds because total variation distances between probability measures are at most $2$,
and $C_0 > 2$ (Section~\ref{sec:algotheoremsection2}, the case $j = 0$).

Now fix $j \geq 1$ and $\beta \geq 1$. By the triangle inequality through $p^{\mathrm{seq}}_\beta$, then
the bound just established at $j$ and at $j-1 \geq 0$, and finally $\tau^{jL} \leq
\tau^{(j-1)L}$,
\[
  \bigl\|p^{(j)}_\beta - p^{(j-1)}_\beta\bigr\|_{\mathrm{TV}}
  \;\leq\; E^{(j)}_\beta + E^{(j-1)}_\beta
  \;\leq\; C_0\tau^{jL} + C_0\tau^{(j-1)L}
  \;\leq\; 2C_0\,\tau^{(j-1)L}.
\]
Now apply Lemma~S-16(1) with the reference $y = \mathbf{c}[p^{(j-1)}_\beta]$ that the
implementation uses:
\[
  \bigl\|\mathbf{c}[p^{(j)}_\beta] - \mathbf{c}[p^{(j-1)}_\beta]\bigr\|_{\mathrm{WRMS},y}
  \;\leq\; \frac{M_e}{\mathrm{atol}}\,\bigl\|p^{(j)}_\beta - p^{(j-1)}_\beta\bigr\|_{\mathrm{TV}}
  \;\leq\; \frac{2M_eC_0}{\mathrm{atol}}\,\tau^{(j-1)L}.
\]
The bound is uniform in $y$, so it does not matter that the reference changes from block to
block and from iteration to iteration; taking the maximum over $\beta \geq 1$ gives the first
claim. For the second, the right-hand side is at most $1$ as soon as $(j-1)L\log(1/\tau) \geq
\log(2M_eC_0/\mathrm{atol})$, that is, for $j \geq 1 + \lceil\log(2M_eC_0/\mathrm{atol}) /
(L\log(1/\tau))\rceil$; the maximum with $0$ ensures $j \geq 1$, as the first claim requires.
\end{proof}

\begin{leanbox}[Lean: Lemma S-16 and Corollary S-17]
In \verb|PlanxAlgorithm/Wrms.lean|. The Lean state space is abstract, with no densities or basis
functions, so Assumption~(A4) enters as two hypotheses on the coefficient map \verb|c|:
\begin{itemize}\itemsep2pt
\item \verb|hc|: $|c_i[x] - c_i[x']| \leq M\,\mathrm{dist}(x,x')$, the bound defining $M_e$ in
  (A4), with \verb|M| $= M_e$;
\item \verb|hsyn|: $\mathrm{dist}(x,x') \leq \sum_i a_i\,|c_i[x] - c_i[x']|$, the consequence of
  the expansion $p = \sum_i c_i[p]\,e_i$ used in Lemma~S-16(2), with $a_i = \|e_i\|_{L^1}$.
\end{itemize}
In the code, \texttt{b} is the block index $\beta$ and \texttt{D} the basis dimension $N_b$.
\texttt{A} is not a Borel set here, but a bound on $(\mathrm{atol} + \mathrm{rtol}\,|y_i|)\,a_i$, so
that \texttt{D * A} is the constant of Lemma~S-16(2). The increment is taken between iterations
\texttt{n} and \texttt{n + 1}, so Lean's \texttt{n} is the paper's $j-1$.
\begin{leancode}
theorem wrms_le_of_lipschitz (hM : 0 ≤ M)
    (hc : ∀ x x' i, |c x i - c x' i| ≤ M * dist x x')
    (hatol : 0 < atol) (hrtol : 0 ≤ rtol) (y : Fin D → ℝ) (x x' : S) :
    wrms atol rtol y (c x - c x') ≤ M / atol * dist x x'

theorem dist_le_wrms (hA0 : 0 ≤ A) (hA : ∀ i, (atol + rtol * |y i|) * a i ≤ A)
    (hatol : 0 < atol) (hrtol : 0 ≤ rtol)
    (hsyn : dist x x' ≤ ∑ i, a i * |c x i - c x' i|) :
    dist x x' ≤ (D : ℝ) * A * wrms atol rtol y (c x - c x')

theorem wrms_increment_le (F : BlockForgetting Φ L) (hM : 0 ≤ M)
    (hc : ∀ x x' i, |c x i - c x' i| ≤ M * dist x x')
    (hatol : 0 < atol) (hrtol : 0 ≤ rtol) (n b : ℕ) (y : Fin D → ℝ) :
    wrms atol rtol y (c (iter Φ π₀ π₀' (n + 1) b) - c (iter Φ π₀ π₀' n b))
      ≤ 2 * M * F.C₀ / atol * F.τ ^ (n * L)

theorem exists_stop (F : BlockForgetting Φ L) (hL : 0 < L) (hM : 0 ≤ M)
    (hc : ∀ x x' i, |c x i - c x' i| ≤ M * dist x x')
    (hatol : 0 < atol) (hrtol : 0 ≤ rtol) :
    ∃ N : ℕ, ∀ n, N ≤ n → ∀ b (y : Fin D → ℝ),
      wrms atol rtol y (c (iter Φ π₀ π₀' (n + 1) b) - c (iter Φ π₀ π₀' n b)) ≤ 1
\end{leancode}
The four theorems are, in order:
\begin{itemize}\itemsep2pt
\item \verb|wrms_le_of_lipschitz| is Lemma~S-16(1);
\item \verb|dist_le_wrms| is Lemma~S-16(2);
\item \verb|wrms_increment_le| is the first claim of Corollary~S-17, with $\tau^{nL} =
  \tau^{(j-1)L}$. The case \verb|n = 0| needs no separate argument in Lean, because
  \verb|BlockForgetting| already includes the case $j = 0$ of~\eqref{eq:Fabs};
\item \verb|exists_stop| is the second claim. Its witness is \verb|iterCount| applied to
  $(2MC_0/\mathrm{atol}, \tau, L, 1)$, so, as for Theorem~S-11, the stopping iteration visibly
  depends on none of $N$, $B$ or the observations.
\end{itemize}
\end{leanbox}

\subsubsection{Proof of Proposition P1}

We recall the statement.

\PoneStatement*

\begin{proof}[Proof of Proposition~P1]
By Section~\ref{sec:algotheoremsection2}, which uses (A1), (A2) and~\eqref{eq:E}, the block maps
$\Phi_\beta$ are well defined, and the contraction bounds~\eqref{eq:Fabs} and~\eqref{eq:Flip} hold
with the constants of Lemma~S-4.

\emph{Clause 1 (equivalence).} By Theorem~S-12, $F$ has a unique fixed point, it is
$p^{\mathrm{seq}}$, and, at it, the filtered densities, normalizers and log-likelihood coincide
with those of \textsc{SequentialFilter}. By Corollary~S-13, which uses (A4), an exactly zero
increment implies that the current iterate is that fixed point.

\emph{Clause 2 (convergence).} Proposition~S-10 gives $\max_\beta\|p^{(j)}_\beta -
p^{\mathrm{seq}}_\beta\|_{\mathrm{TV}} \leq \min\{2, C_0\tau^{jL}\}$. Theorem~S-11 gives
the iteration count $N_{\mathrm{iter}}(\delta_{\mathrm{TV}})$, which depends only on
$(\varepsilon, L, \delta_{\mathrm{TV}})$.

\emph{Clause 3 (span).} Each iteration runs the $B$ block filters in parallel, each of $L$
steps, so the span of one iteration is $\mathcal{O}(L)$. By Corollary~S-17, which uses (A4),
the rule $\delta^{(j)} \leq 1$ is met at the latest at iteration $N_{\mathrm{stop}}$, which
depends only on $(\varepsilon, L, M_e, \mathrm{atol})$. The span is therefore
$\mathcal{O}(N_{\mathrm{stop}}\cdot L)$, independent of $N$.

\emph{Clause 4 (a posteriori certificate).} Proposition~S-15 gives $\max_\beta E^{(j)}_\beta \leq
A\,\delta^{(j)}_{\mathrm{TV}}$ with $A = A(\varepsilon, L)$, for any
$\delta^{(j)}_{\mathrm{TV}}$ bounding the total-variation increments at iteration $j$. The
algorithm observes the WRMS increment $\delta^{(j)}$, not the total-variation one, and
Lemma~S-16(2) converts one into the other. For each $\beta \geq 1$, with the reference $y =
\mathbf{c}[p^{(j-1)}_\beta]$ that the implementation uses, it gives
\[
  \bigl\|p^{(j)}_\beta - p^{(j-1)}_\beta\bigr\|_{\mathrm{TV}}
  \;\leq\; {N_b}\max_i\bigl((\mathrm{atol} + \mathrm{rtol}\,|y_i|)\,\|e_i\|_{L^1}\bigr)\,
  \delta^{(j)} .
\]
The constant still depends on the coefficients $y_i$ of the current iterate, but it is bounded
uniformly. The reference $y$ is the coefficient vector of a state $p = p^{(j-1)}_\beta$ of the filter, a
probability density, so $\|p\|_{\mathrm{TV}} = 1$ and, by (A4),
\[
  |y_i| \;=\; |c_i[p]| \;\leq\; M_e\,\|p\|_{\mathrm{TV}} \;=\; M_e .
\]
Hence the constant is at most $\Gamma = {N_b}\,(\mathrm{atol} +
\mathrm{rtol}\,M_e)\,\max_i\|e_i\|_{L^1}$. Therefore $\delta^{(j)}_{\mathrm{TV}} := \Gamma\,
\delta^{(j)}$ bounds the total-variation increments, and Proposition~S-15 gives $\max_\beta
E^{(j)}_\beta \leq A\,\Gamma\,\delta^{(j)}$. At termination $\delta^{(j)} \leq 1$, which gives the
bound $A\,\Gamma$. The constant $\Gamma$, and hence the error bound, grows linearly with the
basis size $N_b$, which is chosen together with $\Delta t$.
\end{proof}

\subsection{Limitations}

\paragraph{Loose constants.}
The proof establishes the $\mathcal{O}(1)$ span complexity, but the convergence rate obtained is very loose compared
to that obtained in practice, for instance in our benchmarks.
This is because the  Le~Gland--Oudjane contraction  ignores the observation, which can contribute greatly to 
forgetting.
 We are not aware of results that account for the observations \textit{and} give a contraction rate computable from the actual data,
with the possible exception of the recent work of Kim \textit{et al.}~\cite{kim2024variance}. 
The proof therefore gives no insight into the actual mechanism of
convergence, apart from the sparse observation regime, where we expect the bound to be close to sharp.

\paragraph{Mixing assumptions.}
(A3) makes the transition density uniformly bounded above and below, which in
practice requires a compact latent state space. 
While our model satisfies this, many other state-space models do not (such as even the simple  Kalman filter on $\mathbb{R}^n$).
A more general proof would accept weaker forms of mixing, and consequently worse span complexity.
We expect $\mathcal{O}(\log N)$ in the general case, and still useful speed-up in practical cases.

\paragraph{Finite-dimensional representation.}
Assumption (A4) matches the implementation, which works with a finite basis and the WRMS norm.
This makes the total variation to WRMS conversion simple. 
A proof that stops at convergence in
$L^2$, for instance, without assuming a discretization, would be more elegant.

\paragraph{Partial formalization.}
The Lean formalization covers Section~\ref{sec:algotheoremsection3} and parts of
Sections~\ref{sec:algotheoremsection1} and~\ref{sec:algotheoremsection2}, but two gaps remain.
First, Birkhoff's contraction theorem and the Hilbert projective metric on
measures are absent from Mathlib, as is the measure-level block operator of
Le~Gland--Oudjane, so Lemmas~S-3(2) and~S-4 are not formalized, rather enter as hypothesis in 
structure \verb|BlockForgetting|.
Second, Mathlib has no theory of stochastic differential equations, so the latent dynamics
cannot be constructed, and we instead directly assume eigendecomposition results.

\section{Implementation details}
\label{sec:si_impl}

We use the biorthogonal 3.3 synthesis wavelet as the parametrization basis throughout, owing to its compact support and regularity, its synthesis functions being quadratic B-splines, continuously differentiable with bounded second derivative.
For the computation, we use Gaussians on a nD grid, with equal spacing and $\sigma$ corresponding to a 65 percent overlap between nearby basis, and $N=96$ basis per dimension by default.

For the prediction step of Bayesian filtering, we proceed by diagonalization of the Fokker-Planck operator. 
To facilitate this, we internally use the self-adjoint operator $\tilde{\mathcal{L}} = p_{\mathrm{eq}}^{-1/2} \mathcal{L} \, p_{\mathrm{eq}}^{1/2}$ and convert back to probability densities when needed, as this form is more amenable to differentiation (lack of gauge freedom in the eigenvectors).
We then assemble the matrix exponential propagator $ \mathbf{P} =  \mathbf{V} \cdot  \textrm{diag}_{i}(\exp(D \cdot  \Delta t \cdot \lambda_i)) \cdot  \mathbf{V} ^\mathrm{T} \cdot  \mathbf{S}$ with $\mathbf{V}$ the eigenvectors, $\lambda_i$ the eigenvalues, $ \mathbf{S} $ the computation basis Gram Matrix. Diagonalization is obtained with the built-in jax \texttt{eigh} routine.
For the update step, we project onto a grid comprised of  Gauss-Hermite quadrature nodes, then perform the multiplication of density pointwise at those nodes. We then reduce back to densities-as-vector using the  Gauss-Hermite quadrature weights.
We use a mixed precision scheme, where operator assembly, diagonalization and propagator assembly occur in double precision, while the filtering loop takes place in single precision. Final loglikelihood reduction is then conducted in double precision. All computation, including the Hamiltonian Monte Carlo sampling, take place on GPU.

Because the observation likelihood depends on the data $Y_{0:K}$ but not on the model parameters $\theta$, we precompute them as a function of the emission parameter $s$, and project them in a Gaussian basis, the so-called data basis, at program start-up. For each evaluation of the likelihood, we project from data basis to latent coordinate Gauss-hermite node according to the emission function  $f_{\theta_{\mathrm{em}}}(x)$ in a single large matrix computation. 
This reduce the filtering step, which is the critical path, to only the pointwise multiplication described below for the update step.

We use confining, exponential wall potential: 

\begin{equation}
G_{\mathrm{wall}}(\mathbf{x}) = k_w \sum_{i=1}^{n} \left[ e^{-a_w (x_i - a_i)} + e^{\,a_w (x_i - b_i)} \right]
\end{equation}

where $k_w=20\, k_\mathrm{b}T$ sets the wall strength, and $a_w=8$ controls the steepness. 
Of note, this particular $G_{\mathrm{wall}}(\mathbf{x})$ form is not essential to the method.

Since the landscape and emission function can only be inferred up to isometry, \ie, the physics of the reflected (in 1D) or rotated system is indistinguishable,  we use an isometry breaking prior, which can be of either of two type.
For the first type, the center of probability mass prior,  we compute the center of mass of the equilibrium distribution, and penalize loglikelihood  when this point lies either to the left of the 1D landscape, or away from the lower coordinate quadrant in 2D.
Similar quadrant convention can be used in dimension above 2.
This penalty is the distance $d$ to the allowed boundary, softened with the $log(1 + e^d)$ softplus function to allow for gradient-based optimization.
For the second type, the emission maximum prior, we compute the soft maximum of the emission function, and apply a penalty as described above to ensure it lies to the right of the 1D landscape , or away from the lower coordinate quadrant in 2D.
The second type of prior is preferable when applicable (\ie, when the emission is close to linear), as it imparts no bias in the inferred landscape, whereas the first has a small but non-negligible impact when both side of the landscape have similar probability density (\ie, it forces slightly deeper wells on the right).

\section{Additional benchmark results}
\label{sec:si_benchmarks}

We quantify the agreement between the PinT algorithm and sequential
filters by the relative log-likelihood error
\begin{equation}
  \varepsilon \;=\;
  \frac{\bigl|\log p_{\mathrm{WR}} - \log p_{\mathrm{seq}}\bigr|}
       {\max\bigl(|\log p_{\mathrm{WR}}|,\,|\log p_{\mathrm{seq}}|\bigr)},
  \label{eq:rel_logp_error}
\end{equation}
normalised symmetrically so that $\varepsilon \in [0, 2]$ and neither
evaluation is privileged as the denominator.

The single-GPU runtime of a likelihood-and-gradient evaluation is well described by a linear cost model in the trajectory length $N$, $T(N) = T_0 + c\,N$, fitted to replicated benchmark runtimes.
The intercept $T_0 \approx 3.9$~ms is a fixed cost per evaluation, comprising operator assembly, prior evaluation and, above all, operator diagonalization.
The slope $c \approx 0.63\,\mu$s is the average cost per time point of PinT filtering, including its repeated filtering of each block.
It is not comparable with the 36--42~$\mu$s per time point of the sequential filter (main text, Figure~3b), which runs as a latency-bound chain of matrix-vector products.
The two terms are equal at $N^* = T_0/c \approx 6300$ time points, below which the fixed cost dominates.
The model fits closely beyond a few thousand time points, and slightly underestimates the runtime of the shortest trajectories.
Since additional GPUs reduce only the per-time-point term, this fixed cost caps strong scaling at these lengths.

\includegraphics[width=0.5\textwidth]{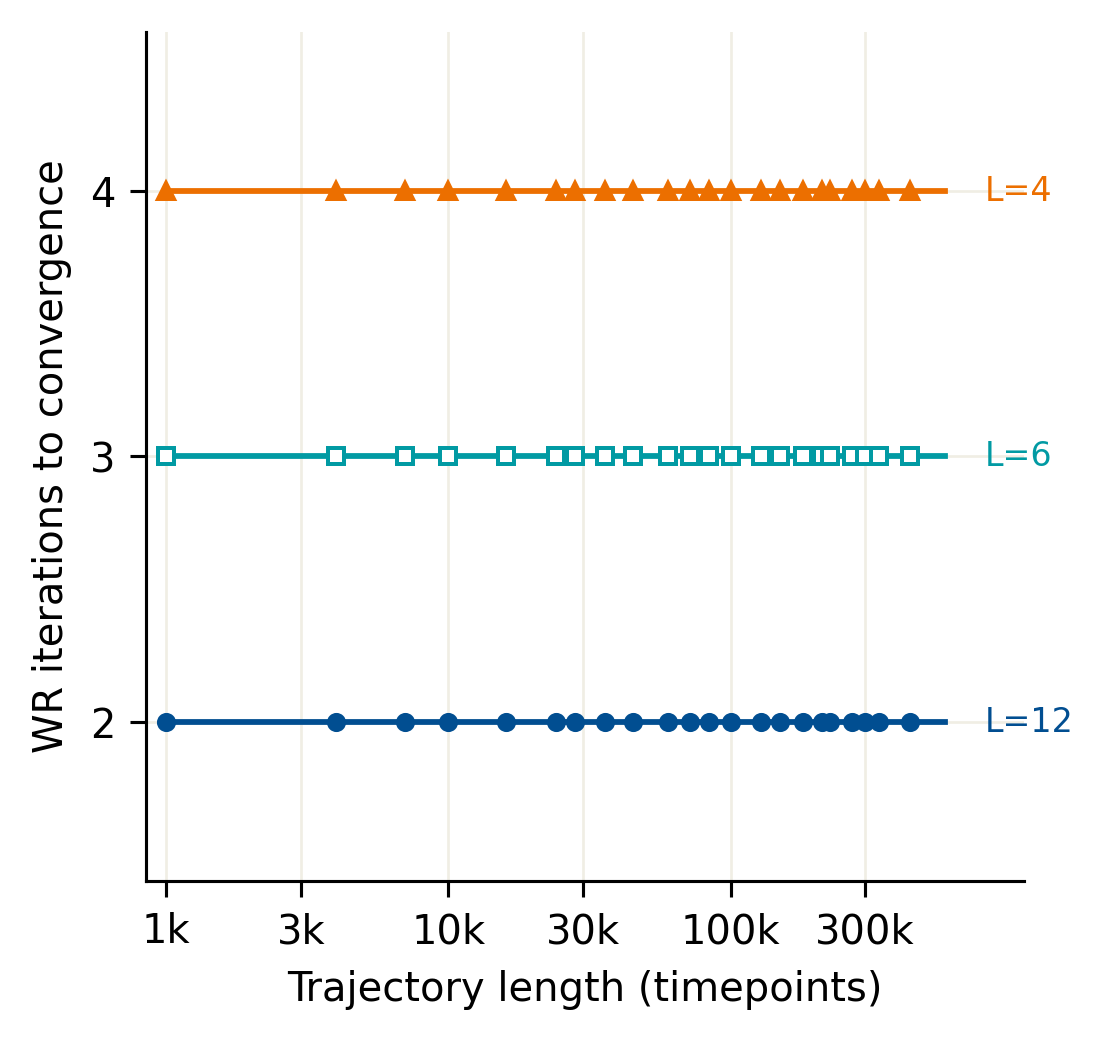}
\captionof{figure}{%
    \textbf{PinT iteration count as a function of trajectory length.} Number of iterations required for the PinT filter to converge, at parameters near the maximum a posteriori estimate, as a function of total trajectory length, for three block lengths $L$. The count is constant over close to three orders of magnitude of trajectory length, consistent with a constant span, and shows no run-to-run variation. It is set by $L$ and by the number of time steps the filter needs to forget its initial condition, here 12, and reaches its floor of two iterations, the minimum to compute inter-iteration differences, for $L \geq 12$.
    \label{fig:si_benchmark_iterations}
}

\includegraphics[width=0.5\textwidth]{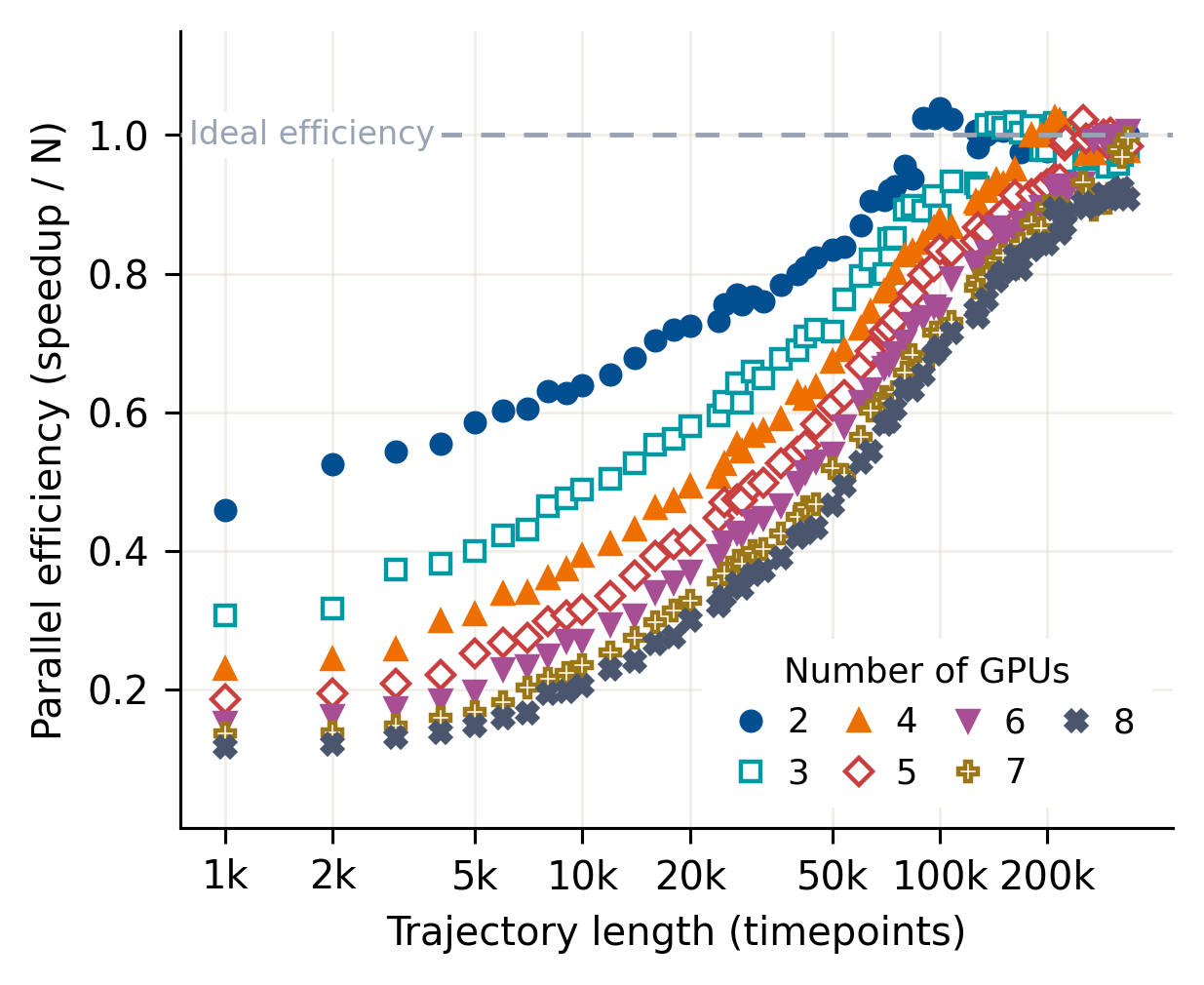}
\captionof{figure}{%
    \textbf{Weak scaling parallel efficiency as a function of trajectory length.} and number of GPU (symbol and color), with the ideal linear scaling as dashed line. Note that the noise, as well as the slight overshoot beyond perfect scaling, is due to JAX code generation, as the run-ton-run variability at exactly the same settings is below 1 percent. This is a known downside of high level frameworks where code generation is not manually controlled but tuned by heuristics.
    \label{fig:si_benchmark_weak_efficiency}
}

\includegraphics[width=0.5\textwidth]{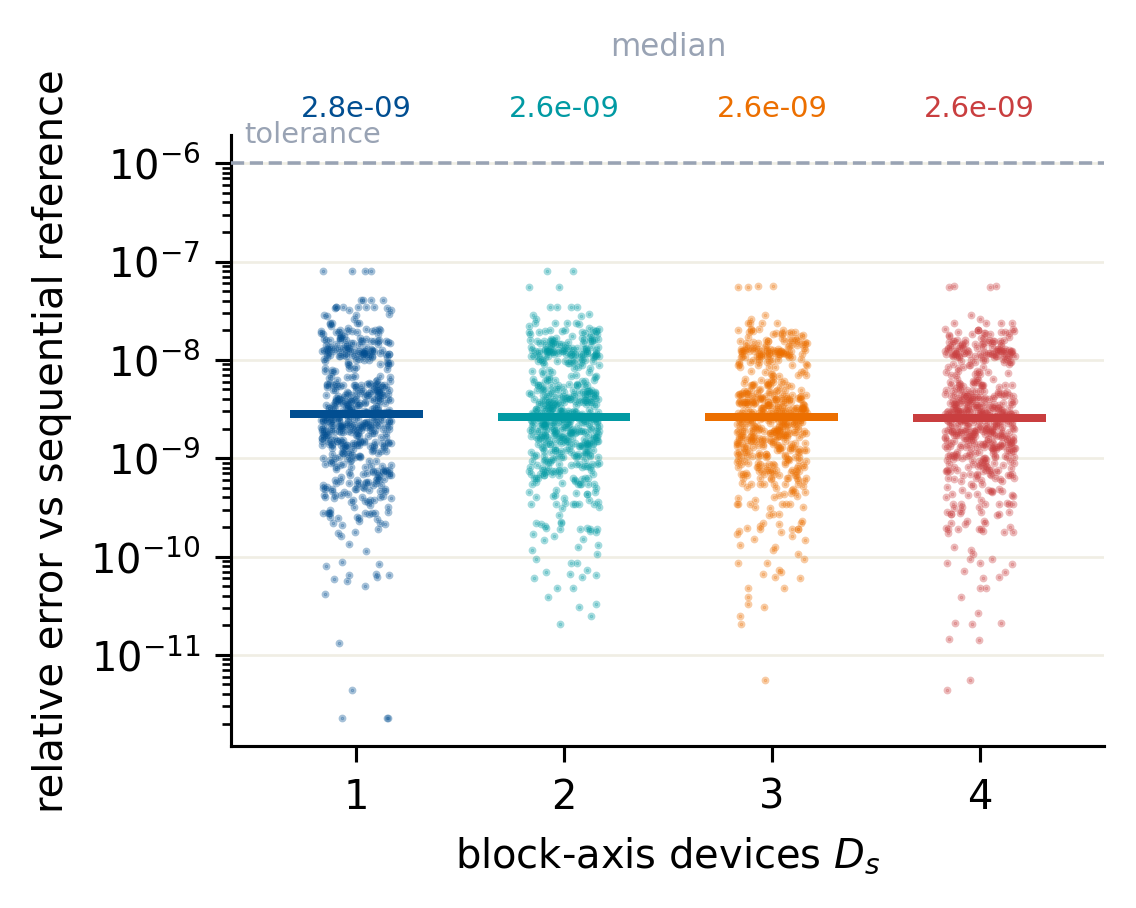}
\captionof{figure}{%
    \textbf{PinT loglikelihood  relative error as a function of the number of GPU} compared to the sequential reference loglikelihood. The slight increase with GPU count reflects the changed order of floating-point summations, which are not associative, when the work is split across more devices.
    \label{fig:si_benchmark_pint_accuracy_numgpu}
}

\includegraphics[width=0.5\textwidth]{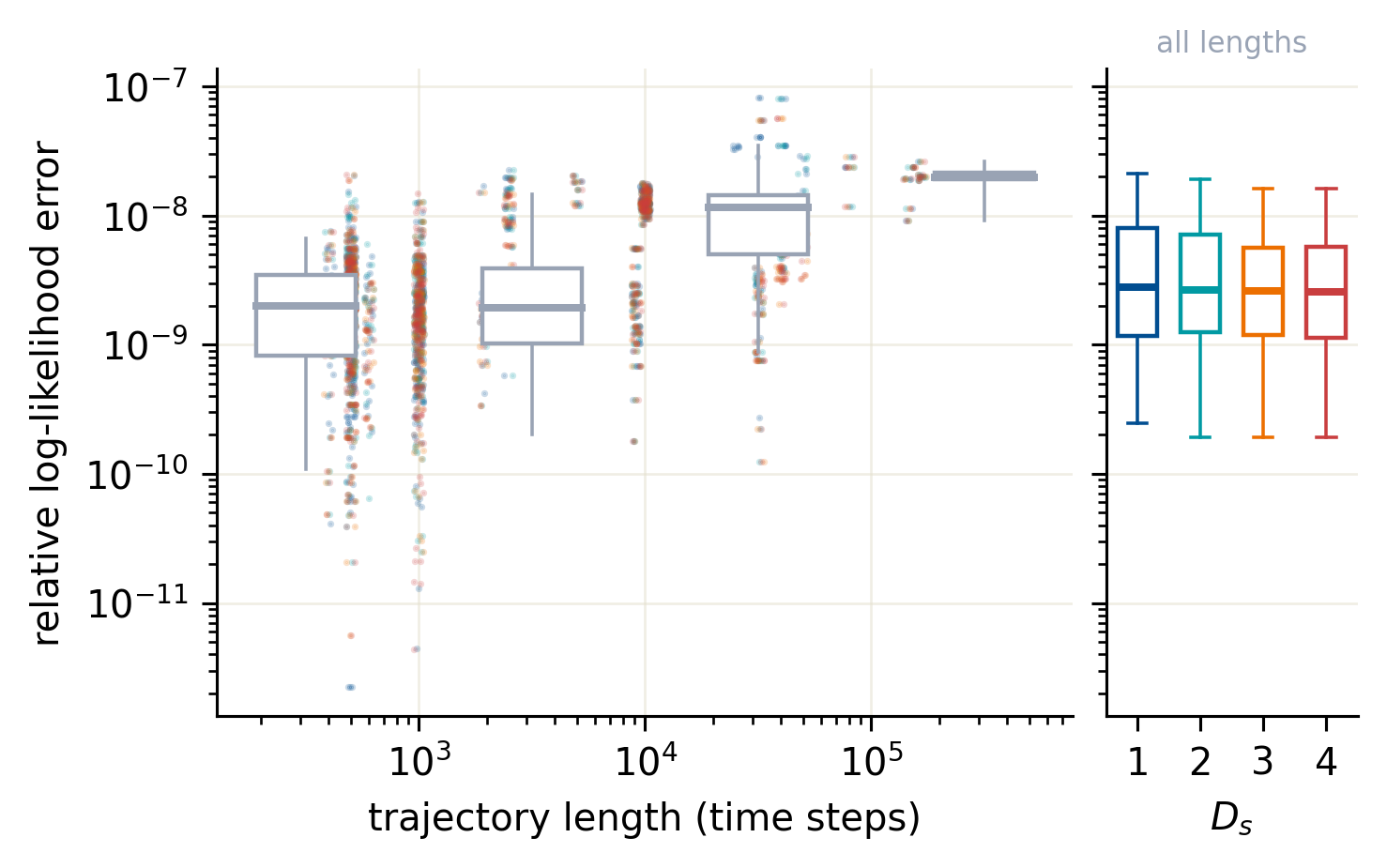}
\captionof{figure}{%
    \textbf{PinT loglikelihood  relative error as a function of the trajectory length} (and number of GPU) compared to the sequential reference loglikelihood.
    \label{fig:si_benchmark_pint_accuracy_trajlenght}
}

\section{Gauge freedom and gauge fixing in the diffusion D coefficient}
\label{sec:si_gauge_d}

 When $D$ is inferred alongside the landscape and emission function, an affine freedom is present: for any $\alpha \neq 0$ and $c \in \mathbb{R}$,
 \begin{equation}
 D \to \alpha^2 D, \qquad G(x) \to G\!\left(\frac{x - c}{\alpha}\right), \qquad f(x) \to f\!\left(\frac{x - c}{\alpha}\right)
 \end{equation}
 leaves the observed process $\{Y_{t_k}\}$ invariant.
 The fixed wall potential formally breaks this symmetry.
 Where $G_{\mathrm{interior}}$ confines the dynamics to an effective domain $[a', b']^n \subset [a, b]^n$, however, the walls are barely explored and the degeneracy remains severe in practice.

 We therefore fix $D$ rather than infer it, removing the scaling freedom $\alpha$.
 Choosing a value for which the latent trajectory spans the whole domain $[a,b]^n$ removes the translation freedom $c$ as well, leaving only isometries: reflections in one dimension, rotations in two.
 These we eliminate with a prior that lightly biases the centre of mass of the equilibrium distribution toward a conventional location (Section~\ref{sec:si_impl}).

 To choose the value of $D$, we first treat it as a free parameter rather than a constant, inferring it alongside $G$ and $f$.
 A short posterior ascent followed by a brief HMC run then locates the range of $D$ the posterior supports, in which the latent landscape spans the whole domain.
 We fix a value in that range and report it alongside the posterior distributions.

Note that this is a separate issue from whether a constant diffusion $D$, \ie, independent of latent location, is appropriate.
Even with a location-dependent $D$, there is a gauge between the length of the latent dimensions and a multiplicative prefactor 
for $D$, such that gauge fixing must still be done.
Note also that in 1 dimension, a location-dependent  $D$ can be absorbed into the landscape shape (stretching when $D$ is  locally low, compressing when
$D$ is locally high).
However, beyond 1 dimensions, our choice of a global $D$ is a modelling choice (as stretching of the landscape cannot account for local $D$).

 \section{Wavelet representation and the Besov prior}
\label{sec:si_wavelet_besov}

 \includegraphics[width=\textwidth]{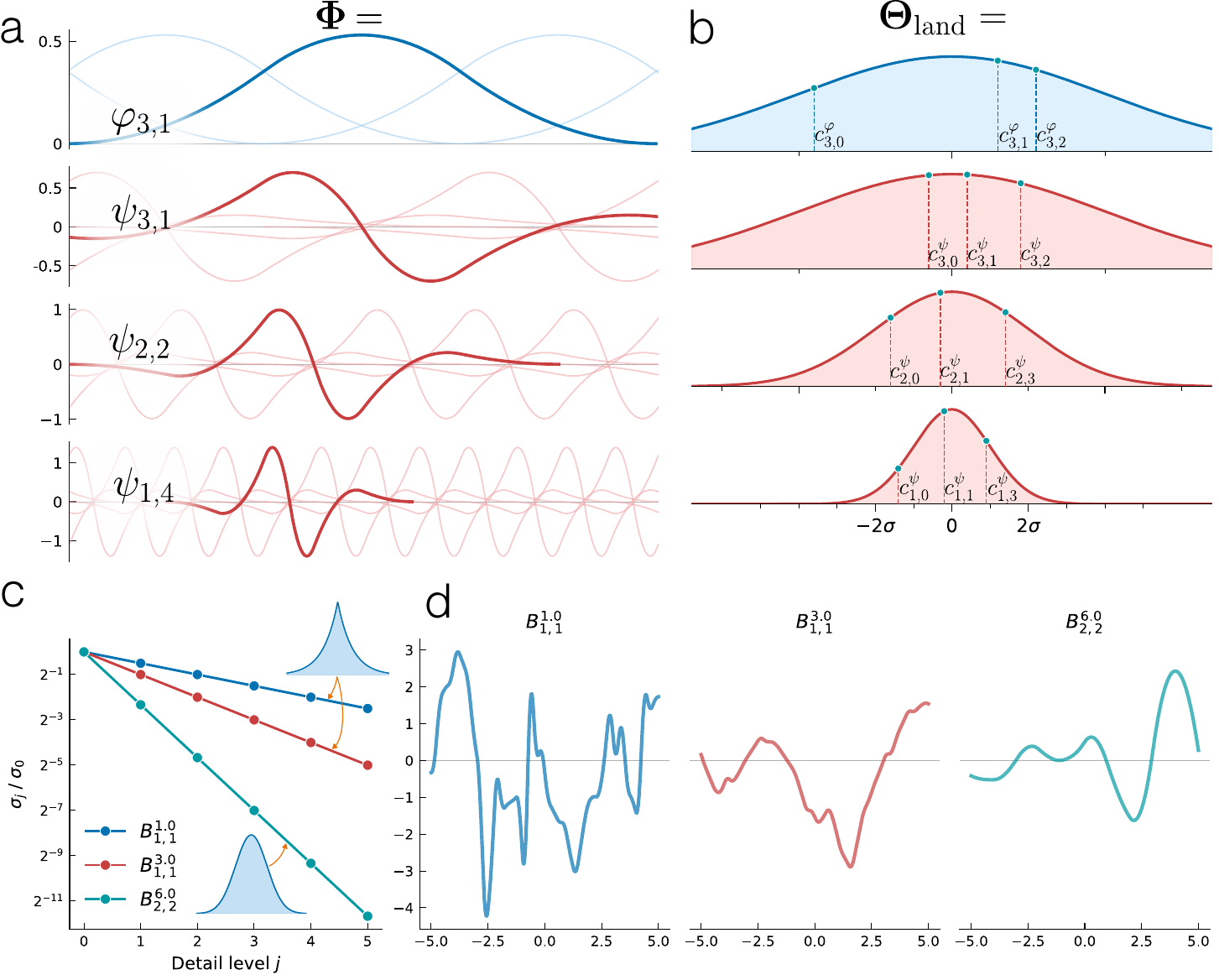}
 \captionof{figure}{\textbf{Wavelet representation and Besov prior}.
 \label{fig:besov_prior}
 \textbf{a}, Biorthogonal 3.3 synthesis basis functions: the scaling functions $\varphi_{3}$ (blue) capture coarse shape, while wavelet functions $\psi_{3}$, $\psi_{2}$, $\psi_{1}$ (red) encode progressively finer detail. Translates at each level shown faded.
 \textbf{b}, A sample draw of coefficients from the prior, illustrating how each coefficient weights the corresponding basis functions. Coefficients at finer levels are shrunk toward zero, suppressing high-frequency detail unless supported by data.
 \textbf{c}, Prior standard deviation $\sigma_j/\sigma_0$ as a function of detail level $j$ for three Besov classes $B^s_{p,p}$. Higher smoothness $s$ and lower $p$ produce steeper decay, more aggressively regularizing fine-scale structure. Insets show the corresponding coefficient prior density (Laplace for $p{=}1$, Gaussian for $p{=}2$).
 \textbf{d}, Representative functions drawn from each prior, showing how $s$ controls smoothness (rougher at left, smoother at right) and $p$ controls sparsity structure (piecewise features for $p{=}1$ versus globally smooth variation for $p{=}2$).}

 We regularize using Besov function spaces $B^s_{p,p}$, whose members have a regularity set by the smoothness order $s$ and the sparsity type $p$.\cite{meyer1992wavelets,mallat2009wavelet}
 Besov regularity can be enforced directly by expressing landscape and emission function in a wavelet basis (Figure~\ref{fig:besov_prior}):
 \begin{equation}
   g = \sum_k c^{\varphi}_{k}\,\varphi_k + \sum_{j=1}^{J}\sum_k c^{\psi}_{j,k}\,\psi_{j,k},
 \end{equation}
 where $\varphi_k$ are scaling functions capturing the coarse shape and $\psi_{j,k}$ are wavelet functions at successively finer resolution levels $j=1,\dots,J$.
 Membership in $B^s_{p,p}$ is enforced by assigning each detail coefficient a level-dependent prior
 \begin{equation}
   \pi(c^{\psi}_{j,k}) \propto \exp\!\bigl(-|c^{\psi}_{j,k}/\sigma_j|^p\bigr),
 \end{equation}
 with scale decaying geometrically across levels:
 \begin{equation}\label{eq:besov_scale}
   \sigma_j = \sigma_0 \cdot 2^{-j\alpha}, \qquad \alpha = s + \tfrac{1}{2} - \tfrac{1}{p},
 \end{equation}
 where $\sigma_0$ sets the overall prior scale and the exponent $\alpha$ controls how aggressively fine-level coefficients are shrunk toward zero.
 For $p=2$ the prior is Gaussian; for $p=1$ it is Laplace, thus sparsity-enforcing.
 This prior is discretization-invariant in the sense of Lassas et al.,\cite{lassas2009discretization} so the expansion can be truncated at a finite level $J$ without altering the prior at resolved scales.

 The scaling coefficients $c^{\varphi}_k$ determine the coarse shape of the function (overall depth and broad structure), which is dictated by the physical system rather than by smoothness assumptions: their scale $\sigma_0^{\varphi}$ is accordingly set independently and more permissively.

\FloatBarrier

 \section{Inference protocol parameters}
\label{sec:si_inference_protocol}

We now list the parameters common to each validation or application case:

{\small
\begin{longtable}{@{}p{0.34\linewidth}p{0.62\linewidth}@{}}
\caption{Settings common to all inference runs. Per-experiment settings are given in
Tables~\ref{tab:si-enso}--\ref{tab:si-l010e008-sweep}. RT is evaluated at the run temperature.}
\label{tab:si-common}\\
\toprule
Setting & Value \\
\endfirsthead
\multicolumn{2}{@{}l}{\textit{Table~\ref{tab:si-common} (continued)}}\\
\toprule
Setting & Value \\
\endhead
\midrule
\multicolumn{2}{@{}r}{\textit{continued on next page}}\\
\endfoot
\bottomrule
\endlastfoot
\sisec{Latent space and confinement}
Latent domain $\mathcal{X}$ & $[-5,\,5]$, one dimension, dimensionless \\
Wall potential & $U(x) = k_{\mathrm{w}}\,[e^{-a(x+5)} + e^{a(x-5)}]$, soft-capped at $k_{\mathrm{w}}$:
  $G_{\mathrm{wall}} = U - \tau\,\mathrm{softplus}((U-k_{\mathrm{w}})/\tau)$ \\
\quad wall parameters & $a = 8$, $k_{\mathrm{w}} = 20\,RT$, $\tau = 1\ \mathrm{kJ\,mol^{-1}}$, no standoff \\
Diffusion coefficient $D$ & Fixed (not sampled); value per experiment \\
\sisec{Representation}
Inference basis & Biorthogonal 3.3 (synthesis) wavelets, 4 levels with dilations $8, 4, 2, 1$;
  functions kept if $\geq 30\%$ of their area lies in $\mathcal{X}$ \\
\quad coefficients & 24 for $G$ and 24 for $f$ (shared basis): 2 scaling + $2 + 4 + 6 + 10$ wavelet \\
Computation basis (latent) & $N_g$ Gaussians on a regular grid over $\mathcal{X}$,
  width $\sigma_g = 0.65 \cdot 10 / N_g$; 16-point Gauss--Hermite quadrature \\
Data basis (observation) & $N_g$ Gaussians over $[y_{\min} - 2,\ y_{\max} + 2]$,
  width $0.65 \cdot (y_{\max} - y_{\min} + 4) / N_g$ \\
Precision & Filtering in float32; operator assembly, diagonalization and
  log-likelihood accumulation in float64 \\
\sisec{Priors}
Besov prior, detail coefficients & level $\ell = 0,1,2,3$ (coarse to fine):
  scale $\sigma_\ell = \sigma_0\,2^{-\ell\alpha}$, $\alpha = s + \tfrac12 - \tfrac1p$;
  $p=2$: $\mathcal{N}(0,\sigma_\ell^2)$, $p=1$: $\mathrm{Laplace}(0,\sigma_\ell)$ \\
Besov prior, scaling coefficients & Same family, scale $\sigma_0^{\mathrm{sc}}$ \\
\quad landscape $G$ & $\sigma_0 = 20$, $\sigma_0^{\mathrm{sc}} = 100\ \mathrm{kJ\,mol^{-1}}$
  ($8.0$ and $40\,RT$ at 300 K);
  $s = 2$, $p = 2$ (except the sensitivity sweep, Table~\ref{tab:si-l010e008-sweep}) \\
\quad emission $f$ & $\sigma_0 = 50$, $\sigma_0^{\mathrm{sc}} = 100$ (observation units); $s = 2$, $p = 2$ \\
Isometry breaking & Weight 8; mode (landscape or emission) per experiment \\
\sisec{Initialization (fan-out)}
Candidate library & Flat landscape; piecewise-linear emission through a node grid
  (1--5 segments, 4 levels spanning the observed range), simplest first; the first
  $K$ candidates are used for each chain \\
Local optimization & L-BFGS with line search, preconditioned by the Besov prior scales,
  2500 steps, on the tempered posterior \\
Fan-out HMC phases & T-tail: $\beta$ raised from $\beta_0$ to 1 on the tempered likelihood;
  F-tail: $\beta = 1$, tempered likelihood; F-ramp: likelihood interpolated from tempered
  to standard; F-full: standard likelihood \\
\quad adaptation & Step size and diagonal mass adapted every 50 samples; $\beta$ (T-tail)
  and the interpolation weight (F-ramp) raised in equal steps at the same interval \\
Candidate selection & One candidate per chain, drawn with probability
  $\propto \exp(\overline{\log \pi})$ over F-full \\
Pre-warmup & HMC, dual-averaging step size (target acceptance 0.8), mass matrix fixed
  from the fan-out; adaptive stop once the log-posterior gain falls below
  $0.3\times$ its spread over the last 10 chunks \\
\sisec{Production sampling}
Warmup & Stan window adaptation (initial / terminal fast buffers 75 / 50,
  first slow window 25, doubling); diagonal mass matrix \\
Sampler & Static HMC, fixed number of leapfrog steps, target acceptance 0.8;
  each chain runs the full protocol independently from its own fan-out \\
\sisec{Parallel-in-time filtering}
Boundary initialization & $p_{\mathrm{eq}}$ for all blocks \\
Convergence & WRMS change of block-boundary densities $\leq 1$,
  $\mathrm{atol} = \mathrm{rtol} = 10^{-6}$; at most $B$ iterations \\
\end{longtable}
}

\FloatBarrier

We now list the parameters for each system:

\subsection{Validation system 1}

\begin{table}[ht]
\centering
\caption{Synthetic validation, landscape 9 / emission 7. Experiment
\texttt{newlandscan\_l009\_e007\_A02}. Ground truth: $D = 1$, Gaussian observation noise
$\sigma = 0.25$, Euler--Maruyama step $5 \times 10^{-4}$, trajectory duration 1200.}
\label{tab:si-l009e007}
\small
\begin{tabular}{@{}p{0.34\linewidth}p{0.62\linewidth}@{}}
\toprule
Setting & Value \\
\sisec{Data}
Trajectories $\times$ points & $32 \times 1000$, $\Delta t = 1.2$ \\
Observed range & $[-8.51,\ 16.44]$ \\
\sisec{Model}
$D$ (fixed) & 1 (ground truth) \\
Temperature & 300 K \\
Observation model & $0.995\,\mathcal{N}(\sigma = 0.25) + 0.005\,\mathrm{Tri}$ \\
Tempered observation model & $0.70\,\mathcal{N}(\sigma = 0.25) + 0.30\,\mathrm{Tri}$ \\
$N_g$ (computation / data) & 96 / 96; $\sigma_g = 0.068$ / $0.196$ \\
Isometry breaking & Landscape mode \\
\sisec{Fan-out}
Starts per chain $K$ & 150 \\
$\beta_0$; leapfrog steps & 0.1 (10 levels); 64 \\
Samples T-tail / F-tail / F-ramp / F-full & 500 / 500 / -- / 500 \\
Pre-warmup (min / max steps; leapfrog) & 1000 / 2000; 64 \\
\sisec{Production}
Chains & 4 \\
Warmup / samples per chain & 4000 / 50\,000 \\
Leapfrog steps & 128 \\
Adapted step size & $1.22$--$1.78 \times 10^{-3}$ \\
Metric & Diagonal \\
\sisec{Parallel-in-time filtering}
Block length $L$ / blocks $B$ & 10 / 100 per trajectory \\
GPUs & 1 \\
WRMS threshold; rtol & 1.0; $10^{-6}$ \\
\bottomrule
\end{tabular}
\end{table}

\FloatBarrier
\subsection{Validation system 2}

\begin{table}[ht]
\centering
\caption{Synthetic validation, landscape 10 / emission 8. Experiment
\texttt{newlandscan\_l010\_e008\_A02}. Ground truth as in Table~\ref{tab:si-l009e007}.}
\label{tab:si-l010e008}
\small
\begin{tabular}{@{}p{0.34\linewidth}p{0.62\linewidth}@{}}
\toprule
Setting & Value \\
\sisec{Data}
Trajectories $\times$ points & $32 \times 1000$, $\Delta t = 1.2$ \\
Observed range & $[-7.80,\ 11.82]$ \\
\sisec{Model}
$D$ (fixed) & 1 (ground truth) \\
Temperature & 300 K \\
Observation model & $0.995\,\mathcal{N}(\sigma = 0.25) + 0.005\,\mathrm{Tri}$ \\
Tempered observation model & $0.70\,\mathcal{N}(\sigma = 0.25) + 0.30\,\mathrm{Tri}$ \\
$N_g$ (computation / data) & 96 / 96; $\sigma_g = 0.068$ / $0.160$ \\
Isometry breaking & Landscape mode \\
\sisec{Fan-out}
Starts per chain $K$ & 150 \\
$\beta_0$; leapfrog steps & 0.1 (10 levels); 64 \\
Samples T-tail / F-tail / F-ramp / F-full & 500 / 500 / -- / 500 \\
Pre-warmup (min / max steps; leapfrog) & 8000 / 8000; 64 \\
\sisec{Production}
Chains & 4 \\
Warmup / samples per chain & 4000 / 30\,000 \\
Leapfrog steps & 256 \\
Adapted step size & $0.85$--$1.02 \times 10^{-3}$ \\
Metric & Diagonal \\
\sisec{Parallel-in-time filtering}
Block length $L$ / blocks $B$ & 10 / 100 per trajectory \\
GPUs & 1 \\
WRMS threshold; rtol & 1.0; $10^{-6}$ \\
\bottomrule
\end{tabular}
\end{table}

\FloatBarrier

\subsection{ENSO application example}

\begin{table}[ht]
\centering
\caption{ENSO (Southern Oscillation Index). Experiment \texttt{enso\_A71}.
$\dagger$\,Step size in Gram-whitened coordinates.}
\label{tab:si-enso}
\small
\begin{tabular}{@{}p{0.34\linewidth}p{0.62\linewidth}@{}}
\toprule
Setting & Value \\
\sisec{Data}
Observable & SOI (standardized, z-score), monthly, Jan 1866 -- Dec 2025 \\
Trajectories $\times$ points & $1 \times 1920$, $\Delta t = 1$ month \\
Observed range & $[-4.34,\ 4.07]$ \\
\sisec{Model}
$D$ (fixed) & $0.35\ \mathrm{month^{-1}}$ (latent units$^2$) \\
Landscape units & $k_BT$ (dimensionless effective potential) \\
Observation model & $0.995\,t_{\nu=9.5}(\sigma = 0.60) + 0.005\,\mathrm{Tri}$ \\
Tempered observation model & $0.70\,t_{\nu=9.5}(\sigma = 0.60) + 0.30\,\mathrm{Tri}$ \\
$N_g$ (computation / data) & 96 / 96; $\sigma_g = 0.068$ / $0.084$ \\
Isometry breaking & Emission mode \\
\sisec{Fan-out}
Starts per chain $K$ & 25 \\
$\beta_0$; leapfrog steps & 0.8 (30 levels); 128 \\
Samples T-tail / F-tail / F-ramp / F-full & 1500 / 1500 / 500 / 3000 \\
Pre-warmup (min / max steps; leapfrog) & 2000 / 4000; 256 \\
\sisec{Production}
Chains & 4 \\
Warmup / samples per chain & 10\,000 / 60\,000 \\
Leapfrog steps & 256 \\
Adapted step size & $3.3$--$4.0 \times 10^{-3}$\,$\dagger$ \\
Metric & Diagonal, Gram-whitened coefficients \\
\sisec{Parallel-in-time filtering}
Block length $L$ / blocks $B$ & 8 / 240 \\
GPUs & 1 \\
WRMS threshold; rtol & 1.0; $10^{-6}$ \\
\bottomrule
\end{tabular}
\end{table}

\paragraph{Domain boundary.} The credible intervals of the landscape widen toward the edges of the domain, where extreme observations are rare. There, the stiff exponential wall imposed at the domain boundary compresses the landscape, and the latent coordinate saturates for the strongest events of either sign (main text, Figure~5b). A flexible, inferable wall stiffness is a natural extension of the model.

\FloatBarrier
\subsection{Nano-aperture optical tweezer application example}

\begin{table}[ht]
\centering
\caption{Nano-aperture optical tweezer, BSA at 37\,$^\circ$C. Experiment
\texttt{not\_1\_37deg\_A31}. $\ddagger$\,Chains 3 and 5 stopped at 10\,000 samples.}
\label{tab:si-not}
\small
\begin{tabular}{@{}p{0.34\linewidth}p{0.62\linewidth}@{}}
\toprule
Setting & Value \\
\sisec{Data}
Observable & Transmitted laser intensity (APD voltage), offset and scaled $\times 30$ \\
Trajectories $\times$ points & $1 \times 24\,000$ (first 75\% of the 32\,000-point record),
  $\Delta t = 2500\ \mu\mathrm{s}$ \\
Observed range & $[0.34,\ 5.90]$ \\
\sisec{Model}
$D$ (fixed) & $5.0 \times 10^{-5}\ \mu\mathrm{s}^{-1}$ (latent units$^2$) \\
Temperature & 310.15 K \\
Observation model & $0.99\,\mathcal{N}(\sigma = 0.10) + 0.01\,\mathcal{U}$ \\
Tempered observation model & $0.70\,\mathcal{N}(\sigma = 0.10) + 0.30\,\mathcal{U}$ \\
$N_g$ (computation / data) & 96 / 96; $\sigma_g = 0.068$ / $0.065$ \\
Isometry breaking & Landscape mode \\
\sisec{Fan-out}
Starts per chain $K$ & 35 \\
$\beta_0$; leapfrog steps & 0.8 (8 levels); 128 \\
Samples T-tail / F-tail / F-ramp / F-full & 400 / 300 / 500 / 1200 \\
Pre-warmup (min / max steps; leapfrog) & 1000 / 1000; 256 \\
\sisec{Production}
Chains & 6 \\
Warmup / samples per chain & 5000 / 18\,000\,$\ddagger$ \\
Leapfrog steps & 256 \\
Adapted step size & $1.47$--$1.64 \times 10^{-3}$ \\
Metric & Diagonal \\
\sisec{Parallel-in-time filtering}
Block length $L$ / blocks $B$ & 16 / 1500 \\
GPUs & 4 (blocks sharded over devices) \\
WRMS threshold; rtol & 1.0; $10^{-6}$ \\
\bottomrule
\end{tabular}
\end{table}

\paragraph{Model checks.} The observed autocorrelation exceeds the posterior predictive credible interval from roughly 20 to 135~ms of lag (main text, Figure~4b). As the excess is absent at the shortest lags, frame averaging and particle inertia, which would act at those timescales, are unlikely causes. Its timescale, slower than any relaxation within the wells ($\le 20.8$~ms) but faster than inter-well transitions (325~ms), indicates a slow degree of freedom acting within the wells rather than on barrier crossings, such as conformational substates hidden by the one-dimensional projection. Slow drift of the experimental apparatus could produce a similar excess. The ten frame-to-frame jumps of 4 to 5 standard deviations (main text, Figure~4c) may be rare instrumental artifacts, or entropically disfavored but faster crossings of the landscape, another artifact of the one-dimensional projection.

\paragraph{Conformer assignment.} The two wells could correspond to the F and E conformers of BSA, respectively, following the assignment of Peters \textit{et al.}\cite{peters2024energy} Because the two analyses use different models, the correspondence between their states and our wells is uncertain.

\FloatBarrier
\subsection{LogVIX application example}

\begin{table}[ht]
\centering
\caption{Implied volatility (logVIX). Experiment \texttt{logvix\_A61}.}
\label{tab:si-logvix}
\small
\begin{tabular}{@{}p{0.34\linewidth}p{0.62\linewidth}@{}}
\toprule
Setting & Value \\
\sisec{Data}
Observable & $\log(\mathrm{VIX})$, daily close, 2 Jan 1990 -- 13 Mar 2026 \\
Trajectories $\times$ points & $1 \times 9142$, $\Delta t = 1$ trading day \\
Observed range & $[2.21,\ 4.42]$ \\
\sisec{Model}
$D$ (fixed) & $0.020\ \mathrm{day^{-1}}$ (latent units$^2$) \\
Landscape units & $k_BT$ (dimensionless effective potential) \\
Observation model & $0.995\,t_{\nu=2.6}(\sigma = 0.023) + 0.005\,\mathcal{U}$ \\
Tempered observation model & $0.95\,t_{\nu=2.6}(\sigma = 0.023) + 0.05\,\mathcal{U}$ \\
$N_g$ (computation / data) & 256 / 256; $\sigma_g = 0.025$ / $0.016$ \\
Isometry breaking & Emission mode \\
\sisec{Fan-out}
Starts per chain $K$ & 25 \\
$\beta_0$; leapfrog steps & 0.8 (10 levels); 128 \\
Samples T-tail / F-tail / F-ramp / F-full & 500 / 500 / 500 / 1500 \\
Pre-warmup (min / max steps; leapfrog) & 2000 / 2000; 256 \\
\sisec{Production}
Chains & 4 \\
Warmup / samples per chain & 2000 / 14\,000 \\
Leapfrog steps & 256 \\
Adapted step size & $0.98$--$1.27 \times 10^{-3}$ \\
Metric & Diagonal \\
\sisec{Parallel-in-time filtering}
Block length $L$ / blocks $B$ & 16 / 572 \\
GPUs & 2 (blocks sharded over devices) \\
WRMS threshold; rtol & 1.0; $10^{-6}$ \\
\bottomrule
\end{tabular}
\end{table}

\FloatBarrier

\subsection{Prior sensitivity scans}

Figure~\ref{fig:si_prior} shows how the inferred landscape depends on the prior hyperparameters and on the amount of data, all other settings being at their defaults.
The prior scale $\sigma_0$ and multiplier $C_s$, which together set the expected free-energy range, have virtually no effect over the scanned range (a), the data overriding the prior.
The weight $w_{\mathrm{sym}}$ of the isometry-breaking prior likewise has no noticeable effect between 1 and 20 (c), as expected for a weak prior whose only role is to select one orientation of the latent coordinate.
The posterior is more sensitive to the smoothness order $s$ of the Besov class $B^s_{2,2}$ targeted by the prior (b, Methods in the main text, and Supplementary Information~S5).
For $p = 2$, the negative log-prior is, up to constants, the squared $H^s$ norm of the landscape, so that $s = 2$ is the lowest order penalizing its curvature.
We want curvature regularized because the second derivative of the landscape enters the Fokker--Planck operator, through the effective potential of its symmetrized form.
Strictly, samples of this prior lie almost surely in $B^t_{2,2}$ only for $t < s - 1/2$,\cite{lassas2009discretization} so that $s = 2$ sets a penalty on curvature rather than the regularity of every sample, which the finite quadratic-spline expansion guarantees instead.
At $s = 1$, the curvature is unpenalized and the posterior is far from converged, its chains remaining in widely different basins even after about 240\,000 draws per chain.
At $s = 1.5$, the landscape is recovered, though the posterior is slightly unconverged after about 256\,000 draws per chain.
At $s = 2.5$ and $3$, higher derivatives are penalized too, and the prior can no longer accommodate the sharp features of the true landscape.
At $s = 2.5$, the bands exclude the ground truth in places, and at $s = 3$ the main barrier is underestimated and the two small wells within the rightmost basin are no longer resolved.
The validation ground truths were themselves drawn from an $s = 2$ prior before modification (Methods in the main text), so this scan favors the default by construction, and $s$ should be relaxed where finer detail is expected.
Finally, the posterior concentrates on the ground truth as the trajectory lengthens (d), its credible intervals narrowing from $N = 10\,000$ to $50\,000$ time steps.

\includegraphics[width=\textwidth]{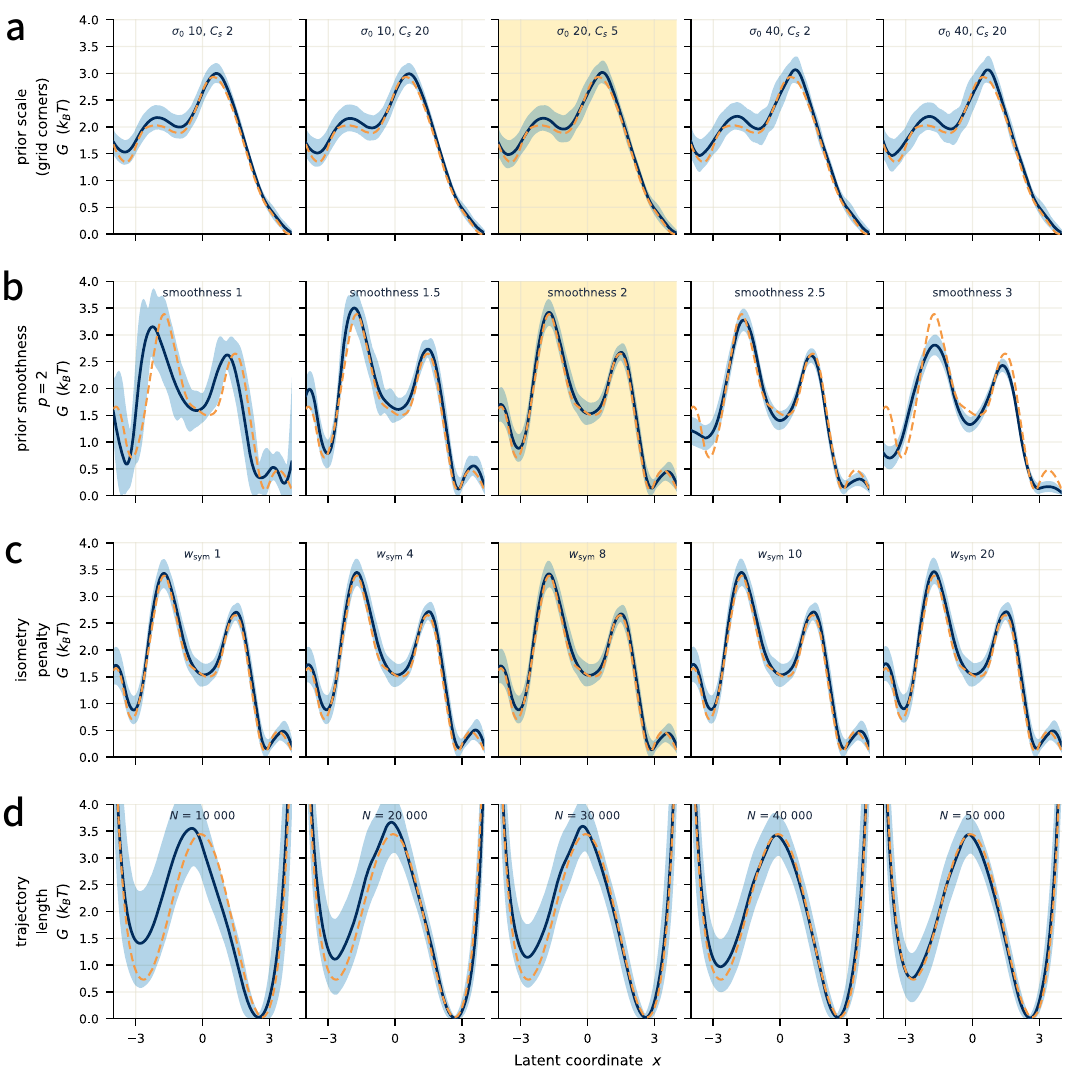}
\captionof{figure}{%
    \textbf{Prior and data sensitivity of the inferred landscape.} Posterior mean (solid line), 95\% credible interval (band) and ground truth (dashed), as in main text Figure~2. Highlighted panels mark the defaults used throughout this work.
    (\textbf{a})~Prior scale $\sigma_0$ ($k_BT$) of the coarsest wavelet level and multiplier $C_s$ of the coarsest scale function, on validation system~1.
    (\textbf{b})~Smoothness order $s$ of the Besov prior, at $p = 2$, on validation system~2.
    (\textbf{c})~Weight $w_{\mathrm{sym}}$ of the isometry-breaking prior, on validation system~2.
    (\textbf{d})~Trajectory length $N$, in time steps, on a double-well landscape with linear emission, chosen to isolate the effect of the amount of data.
    \label{fig:si_prior}
}

\begin{table}[ht]
\centering
\caption{Prior sensitivity sweep on landscape 10 / emission 8. Experiment
\texttt{newlandscan\_l010\_e008\_A07}: eight runs differing only in the landscape Besov
parameters (Table~\ref{tab:si-l010e008-sweep}). Ground truth as in Table~\ref{tab:si-l009e007}.}
\label{tab:si-l010e008-a07}
\small
\begin{tabular}{@{}p{0.34\linewidth}p{0.62\linewidth}@{}}
\toprule
Setting & Value \\
\sisec{Data}
Trajectories $\times$ points & $32 \times 1000$, $\Delta t = 1.2$ \\
Observed range & $[-7.80,\ 11.82]$ \\
\sisec{Model}
$D$ (fixed) & 1 (ground truth) \\
Temperature & 300 K \\
Observation model & $0.99\,\mathcal{N}(\sigma = 0.25) + 0.005\,\mathrm{Tri} + 0.005\,\mathcal{U}$ \\
Tempered observation model & $0.70\,\mathcal{N}(\sigma = 0.25) + 0.25\,\mathrm{Tri} + 0.05\,\mathcal{U}$ \\
$N_g$ (computation / data) & 96 / 96; $\sigma_g = 0.068$ / $0.160$ \\
Landscape Besov $s$, $p$ & Swept, Table~\ref{tab:si-l010e008-sweep} \\
Isometry breaking & Landscape mode \\
\sisec{Fan-out}
Starts per chain $K$ & 150 \\
$\beta_0$; leapfrog steps & 0.1 (10 levels); 32 \\
Samples T-tail / F-tail / F-ramp / F-full & 500 / 500 / -- / 500 \\
Pre-warmup (min / max steps; leapfrog) & 3000 / 3000; 128 \\
\sisec{Production}
Chains & 4 \\
Warmup / samples per chain & 4000 / 152\,000, except 240\,000 ($p = 2$, $s = 1$) and 256\,000 ($p = 2$, $s = 1.5$) \\
Leapfrog steps & 128 \\
Adapted step size & $0.69$--$1.57 \times 10^{-3}$ (all runs) \\
Metric & Diagonal \\
\sisec{Parallel-in-time filtering}
Block length $L$ / blocks $B$ & 10 / 100 per trajectory \\
GPUs & 1 \\
WRMS threshold; rtol & 1.0; $10^{-6}$ \\
\bottomrule
\end{tabular}
\end{table}

\begin{table}[ht]
\centering
\caption{Landscape Besov parameters of the sensitivity sweep (Table~\ref{tab:si-l010e008-a07}).
$\alpha = s + \tfrac12 - \tfrac1p$ is the level-decay exponent of the prior scale.
Step sizes are ranges over the four chains.}
\label{tab:si-l010e008-sweep}
\small
\begin{tabular}{@{}cccc@{}}
\toprule
$p$ & $s$ & $\alpha$ & Adapted step size ($10^{-3}$) \\
\midrule
1 & 1.0 & 0.5 & 0.88--1.16 \\
1 & 1.5 & 1.0 & 0.78--1.08 \\
1 & 2.5 & 2.0 & 0.93--1.24 \\
1 & 3.0 & 2.5 & 0.72--1.09 \\
\midrule
2 & 1.0 & 1.0 & 0.73--1.23 \\
2 & 1.5 & 1.5 & 0.80--1.15 \\
2 & 2.5 & 2.5 & 0.86--1.04 \\
2 & 3.0 & 3.0 & 0.69--1.57 \\
\bottomrule
\end{tabular}
\end{table}

\FloatBarrier

\bibliography{bayesian_filtering}